\documentstyle[prd,aps]{revtex}
\begin{document}
\input epsf
\draft
\renewcommand{\topfraction}{0.8}
\twocolumn[\hsize\textwidth\columnwidth\hsize\csname
@twocolumnfalse\endcsname
\preprint{SU-ITP-97-18, hep-ph/9704452, December 1, 1996}
\title { \bf   Towards the Theory of Reheating After Inflation}
\author{Lev Kofman}
\address{ Institute for Astronomy, University of Hawaii,
2680 Woodlawn Dr., Honolulu, HI 96822, USA }
\author{
Andrei Linde}
\address{ Department of Physics, Stanford University, Stanford CA
94305-4060, USA}
\author {Alexei A. Starobinsky}
\address{ Landau Institute for Theoretical Physics,
Kosygina St. 2, Moscow 117334, Russia}
\date {April 30, 1997}
\maketitle
\begin{abstract}
Reheating after inflation occurs due to particle production by the oscillating inflaton field. In this paper we briefly describe the perturbative approach to reheating, and then concentrate on effects beyond the perturbation theory. They are related to the stage of parametric resonance, which we called {\it preheating}. It may occur in an expanding universe if the initial amplitude of oscillations of the inflaton field is large enough. We investigate a simple model of a massive inflaton field $\phi$ coupled to another scalar field $\chi$ with the interaction term $g^2\phi^2\chi^2$. Parametric resonance in this model is very broad. It occurs in a very unusual stochastic manner, which is quite different from   parametric resonance in the case when the expansion of the universe is neglected. Quantum fields interacting with the oscillating inflaton field experience a series of kicks which, because of the rapid expansion of the universe, occur with phases uncorrelated to each other. Despite the stochastic nature of the process, it leads to exponential growth of fluctuations of the field $\chi$. We call this process {\it stochastic resonance}. We develop the theory of preheating taking into account the expansion of the universe and backreaction of produced particles, including the effects of rescattering. This investigation extends our previous study of reheating after inflation \cite{KLS}. We show that the contribution of the produced particles to the effective potential $V(\phi)$ is proportional not to $\phi^2$, as is usually the case, but to $|\phi|$. The process of preheating can be divided into several distinct stages. In the first stage the backreaction of created particles is not important. In the second stage backreaction increases the frequency of oscillations of the inflaton field, which makes the process even more efficient than before. Then the effects related to scattering of $\chi$-particles on the
oscillating inflaton field terminate the resonance. We calculate the number density of particles $n_\chi$ produced during preheating and their quantum fluctuations $\langle\chi^2\rangle$ with all backreaction effects taken into account. This allows us to find the range of masses and coupling constants for which one can have efficient preheating. In particular, under certain conditions this process may produce particles with a mass much greater than the mass of the inflaton field.

\end{abstract}
\pacs{PACS: 98.80.Cq  \hskip 2.5 cm IfA-97-28~~~~~~~SU-ITP-97-18 \hskip 2.5 cm 
hep-ph/9704452}
 \vskip2pc]

\section {Introduction}

 According to inflationary theory, (almost) all elementary particles
populating the
universe were
created during the process of reheating of the universe after inflation
\cite{book}.
It makes this process extremely important. However, for many years the theory
of reheating remained the least developed part of inflationary theory. Even
now, when many features of the mechanism of reheating are understood, the
literature on
this subject is still full of contradictory statements.

The basic idea of reheating after inflation was proposed in the first paper
on
new inflation \cite{New}: reheating occurs due to particle production by
the oscillating scalar field $\phi$.
In the
simplest inflationary models, this field is the same inflaton field $\phi$
that drives inflation at the early stages of the evolution of
the universe.
After inflation, the scalar field $\phi$ (which we will call inflaton)
oscillates near the minimum of its
effective potential and produces elementary particles. These
particles interact with each other  and eventually they come to a
state of thermal equilibrium at some temperature $T$. This process
completes when all (or almost all) the energy of the classical scalar
field $\phi$ transfers
to the thermal energy of elementary particles. The temperature of the
universe at this stage is called the reheating temperature, $T_r$.

A first attempt at a phenomenological description of this process was made in
ref. \cite{ASTW}. The authors added various friction terms to the equation of
motion of the scalar field in order to imitate energy transfer from the
inflaton field to matter. However, it remained unclear
what kind of terms should be added and whether one should add
them at the stage of slow rolling of the inflaton field, or only at the stage
of rapid oscillations of the inflaton
field.

The theory of reheating in application to the new inflation scenario
was first developed in refs. \cite{DL,AFW}, and, in application to 
$R^2$ inflation, in ref. \cite{st81}. It was based on perturbation theory,
which was quite sufficient for obtaining the
reheating temperature, $T_r$, in many
realistic models. We will give a detailed description of this theory and
develop it even further in a forthcoming publication \cite{PERT}. However,
 perturbation theory has certain limitations, which
have been realized only very recently. In particular, the mechanism of decay of
the
inflaton field to
the vector fields discussed in \cite{DL} is efficient only at an intermediate
stage of reheating in the new inflation model considered.
The
decay of the inflaton field to fermions described in \cite{AFW} typically
is
important only at very late stages of reheating. In many inflationary
models neither of these mechanisms gives a correct description of the first
stages of the process.

Indeed, recently it was understood \cite{KLS} that in many inflationary
models
the first stages of reheating occur in a regime of a broad parametric
resonance. To distinguish this stage from the subsequent stages of slow
reheating and thermalization, we called it {\it preheating}. The energy
transfer from the inflaton field to other bose fields and particles during
preheating is extremely efficient. As we pointed out in \cite{KLS}, reheating
never completes at the
stage of parametric resonance; eventually the resonance becomes narrow and
inefficient, and the final stages of the decay of the inflaton field and
thermalization of its decay products can be described by the elementary
theory
of reheating \cite{DL,AFW,PERT}. Thus,   the elementary
theory
of reheating proves to be very useful even in the theories where reheating
begins at the stage of parametric resonance. However, it should be applied not
to
the
original coherently oscillating inflaton field, but to the products of its
decay, as well as to the
part of the inflaton field which survived preheating. The short stage of
explosively
rapid preheating
 in the broad resonance regime may have long-lasting effects on
the subsequent evolution of the universe.
 It may lead to specific nonthermal
phase transitions in the early universe \cite{KLSSR,tkachev} and to
topological
defect production, it may make possible novel mechanisms of baryogenesis
\cite{Kolb,Riotto}, and it may change the final value of the reheating
temperature $T_r$.

The theory of parametric resonance in application to
particle production by
oscillating external fields was developed more than 20 years ago, see e.g.
\cite{24}. The methods used in this theory were developed mainly for the case
of narrow parametric resonance. A first attempt to apply this theory  to
reheating after inflation was made by Dolgov and Kirilova \cite{9} and by
Traschen and Brandenberger \cite{Brand1} for the narrow
resonance
regime in the context of the new inflation. In \cite{9} it was
conjectured that the parametric resonance in an expanding universe cannot lead
to efficient reheating.  The authors of Ref. \cite{Brand1} came to an important
  conclusion that parametric resonance in new inflation can be  efficient.
However, their investigation of    parametric resonance
was not quite correct,
see Sec. \ref{LIMITS} of this paper.

In any case, at the moment we do not have any consistent inflationary models
based on the new inflation scenario. The step towards the general theory of
reheating in chaotic inflation was rather nontrivial. Indeed, the
effective potential in new inflation is anomalously flat near $\phi = 0$. As
a result of this fine-tuned property of the effective potential, the Hubble
constant
at the end of inflation in this scenario is much smaller
than the mass of the oscillating scalar field.
Therefore the effects related to the expansion of the universe are not very
destructive for the
development of the resonance, which may be rather efficient even if the
resonance is narrow.
Narrow resonance can be rather efficient in chaotic inflation as well, in the
context of conformally-invariant theories
 of the type
of $\lambda\phi^4$. In such
theories
the expansion of the universe does not interfere with the development of the resonance, and therefore preheating may be efficient even if the resonance is
rather   narrow \cite{KLS,Shtanov,Boyan1,Kaiser,GKLS}. However,   generally the
effective
potential is
quadratic with respect to $\phi$ near the minimum of the potential, which
breaks the conformal invariance.
As we will show in this paper, for
the simplest models of inflation, such as the theory of a
massive inflaton field $\phi$ with quadratic effective potential and
interaction
$ g^2\phi^2\chi^2$, preheating is efficient only if the
resonance is extremely broad. The theory of a broad
parametric resonance in an expanding universe is dramatically different from
the theory of a narrow resonance.

The basic features of the theory of a broad parametric resonance were
outlined in \cite{KLS}, where
the theory of preheating was developed in the context of the chaotic
inflation scenario, taking into account backreaction of created particles and
the expansion of the universe. This issue was studied later by many other
authors,
and a lot of very
interesting results on parametric resonance and particle production have
been
obtained
\cite{Shtanov} -\cite{GRAVWAVES}. Of all these papers one is especially
relevant to our investigation. Khlebnikov and Tkachev \cite{Khleb} performed a
detailed three-dimensional numerical lattice simulation of  broad parametric
resonance in an expanding universe, taking into account the backreaction of
produced
particles, including, in particular, their rescattering. Their method (see
also \cite{KhTk,KhTk2,Prokopec}) is based on solving numerically the  classical
equations
for fluctuations of all interacting fields. It is presumably the best way to perform
computer simulations of preheating.

 From the point of view of analytical investigation of preheating in the broad
resonance regime we should
 mention   ref. \cite{Fujisaki}, where this  regime was investigated for the
case of a non-expanding
universe, and some of the results of ref. \cite{KLS} concerning this regime
were  obtained by
a different method. However, after our paper \cite{KLS} there was not much 
progress in analytical investigation of the broad resonance regime in an expanding
universe. This is not very surprising, because the
analytical investigation of
preheating including backreaction is very difficult; one must describe a
system of particles far away from equilibrium in the regime where effective
coupling becomes strong because of anomalously large occupation numbers of
bose particles produced by parametric resonance. But the main problem was
related to the very unusual nature of  broad parametric resonance in
an expanding universe. As we will show in this paper, instead of staying in a
particular resonance band, each growing mode
scans many  stability/instability bands within a single oscillation of the
inflaton field, so the usual concept of separate resonance bands becomes
inadequate. It is
a stochastic process,
during
which the number of produced particles changes in a chaotic way.   On average,
the number of produced particles grows exponentially, but at some moments
their
number may decrease; a process which would be impossible
at
the classical level. We call this process {\it stochastic resonance}.
The
standard methods developed for investigation of parametric resonance simply
do
not apply here, so it was necessary to develop a new, more general approach.

 The main purpose of the present paper is to develop the theory
of preheating with an account taken of the expansion of the universe and the
backreaction of created particles, including the effects of their rescattering.
 We will give here a
detailed derivation of the results of Ref. \cite{KLS}, and describe
recent progress in the understanding of physical processes which
occur soon after the end of inflation.

We will begin our paper with discussion of the evolution of the scalar
fields after inflation neglecting the effects of reheating, see Sec.
\ref{EVOLUTION}. Sec. \ref{OSCILLATIONS} contains an introduction to
the elementary theory of
reheating \cite{DL,AFW,PERT}. We will then develop the theory of particle
production
due to parametric resonance following \cite{KLS}. First of all, in Sec.
\ref{LIMITS} we introduce the theory of reheating due to parametric
resonance
and discuss the relation between this theory and the elementary theory of
reheating. Both theories are
very simple, but the transition from one to the other is quite nontrivial; it
is
very difficult to understand the theory of parametric resonance using the
elementary
theory of reheating as a starting point, and, conversely, perturbation
theory is not simply a limiting case of a weak parametric resonance.
 A more detailed discussion of all these issues will be contained in our
forthcoming paper \cite{PERT}.

In Sec. \ref{BROAD} we discuss the difference between the narrow and  
broad
resonance regimes. Sec. \ref{STOCHASTIC} is devoted to a qualitative
description
of the development of   broad resonance in an expanding universe. We
describe the effect of stochastic resonance and illustrate this effect by
solving the resonance equations numerically, taking into account the expansion
of
the universe. We find that it is much easier to perform the investigation in
terms
of the number of created particles, which is an adiabatic invariant, rather
than in terms of wildly oscillating quantities such as $\langle\chi^2\rangle$
which are studied in many publications on preheating. In particular, in some
cases $\langle\chi^2\rangle$ continues to grow even after the resonance ceases
to exist and the number of $\chi$ particles remains constant.
 In Sec. \ref{ANALYTIC} we develop analytic methods for the description of
broad resonance. These methods are especially appropriate for the
investigation
of  stochastic resonance. They are applicable in those cases where the
standard approach based on the investigation of Mathieu or Lame equations
fails.

Sec. \ref{BACKREACTION} contains a discussion of the backreaction of the
$\chi$-particles created by parametric resonance on the effective potential
of
the inflaton field.  In Sec. \ref{FIRST} we describe the process of
reheating
in the broad resonance regime with an account taken of the change of the
frequency of oscillations of the inflaton field due to its interaction with the
$\chi$-particles produced during preheating. In Sec. \ref{RESCATTERING} we
discuss the
process of rescattering of $\chi$-particles and the production of
$\phi$-particles
in this process. We also consider some modifications of the picture
of
the second stage of reheating with an account taken of rescattering. We
calculate the number of particles produced during reheating and the amplitude
of perturbations $\langle\chi^2\rangle$. In Sec. \ref{SUPERMASSIVE} we
investigate the possibility of a copious production of particles with mass much
greater than the inflaton mass. Finally,
in Sec. \ref{DISCUSSION} we give a summary of our results and discuss their
possible implications.

\section{\label{EVOLUTION} Evolution of the Inflaton Field}

During inflation the leading contribution to the energy-momentum tensor is
given by the inflaton scalar field $\phi$ with the Lagrangian
 \begin{equation}
{ L}(\phi) = {\textstyle {1 \over 2}} \phi_{,i} \phi^{,i}
 -V(\phi)\ ,
 \label{grav}
\end{equation}
where $V(\phi)$ is the effective potential of the scalar field $\phi$.
The evolution of the (flat) FRW universe is described by the Einstein
equation
\begin{equation}
H^2={{ 8\pi} \over {3 M_p^2}}\biggl( {\textstyle {1 \over 2}}\dot \phi^2 +
V( \phi) \biggr)\ ,
 \label{E0}
\end{equation}
where $H={\dot a / a}$.
The Klein-Gordon equation for $\phi(t)$
 is
\begin{equation}
\ddot \phi + 3H\, \dot \phi + V_{, \phi}=0\ .
 \label{KG0}
\end{equation}
For  sufficiently large initial values of $\phi > M_p$,
the ``friction'' term $3H \dot \phi$ in (\ref{KG0})
dominates over $\ddot \phi$ and
the potential term in (\ref{E0}) dominates over the kinetic term.
This is the inflationary stage,
 where the universe expands
quasi-exponentially, $a(t)=a_0 \exp \bigl( \int dt H(t) \bigr)$.
For definiteness, we will consider here the simplest models of chaotic
inflation: $V(\phi) = {\textstyle {1 \over 2}} m \phi^2$
 \cite{book}. In these models inflation
occurs at $\phi {\
\lower-1.2pt\vbox{\hbox{\rlap{$>$}\lower5pt\vbox{\hbox{$\sim$}}}}\ } M_p$.
Density perturbations responsible for large-scale structure formation in
these models are produced at $\phi \sim 3 - 4~ M_p$. With a decrease of the
field $\phi$ below $M_p$ the ``friction'' term $3H\, \dot \phi$ becomes less
and less important, and inflation terminates at $\phi \sim M_p/2$.

When making numerical estimates one should take into account that at the
last
stages of inflation the friction term is still non-negligible, and
therefore during the first oscillation the amplitude of the field rapidly
drops down.

\begin{figure}[t]
\centering
\leavevmode\epsfysize=5.2cm \epsfbox{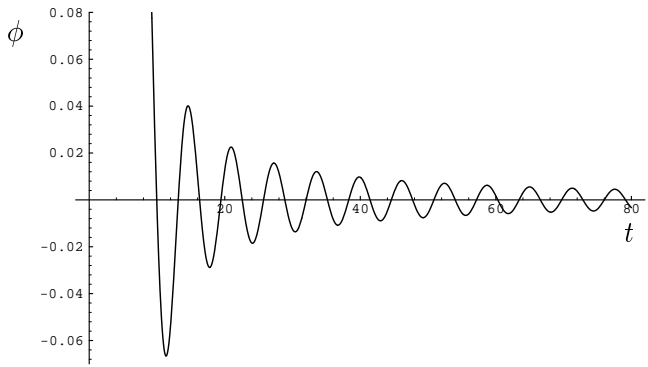}\\

\caption[fig1]{\label{fig1} Oscillations of the field $\phi$ after inflation
in the theory ${m^2\phi^2\over 2}$. The value of the scalar field here and in
all other figures in this paper is measured in units of $M_p$, time is
measured
in units of $m^{-1}$.}
\end{figure}

For the quadratic potential
$V(\phi)={\textstyle {1 \over 2}} m \phi^2$ the amplitude after the first
oscillation becomes only $0.04 M_p$, i.e. it drops by a factor of ten during
the
first
oscillation, see Fig. 1. Later on the solution for the scalar field $\phi$
asymptotically
approaches the
 regime
\begin{eqnarray}
\phi(t)&=& \Phi(t) \cdot
 \sin{ m  t  }\ , \nonumber\\
 \Phi(t) &=& {M_p \over \sqrt{3\pi} mt} \sim {M_p \over 2\pi\sqrt{3\pi}
N}.
\label{870}
\end{eqnarray}
Here $\Phi(t)$ is the amplitude of oscillations, $N$ is the number of
oscillations since the end of inflation.
For simple estimates which we will make later one may use
\begin{equation}\label{870aa}
 \Phi(t) \approx {M_p \over 3 mt} \approx {M_p \over 20 N}.
\end{equation}
The scale factor
averaged over several oscillations grows as
$ a(t) \approx a_0 \bigl({t\over t_0}\bigr)^{2/3}$.
Oscillations of $\phi$ in this theory are
% approximately
 sinusoidal, with  the decreasing amplitude
$\Phi(t) = {M_p \over {3 }} \bigl({a_0\over a(t) }\bigr)^{3/2}$.
The energy density
of the field $\phi$ decreases in the same way as the density of
nonrelativistic
particles of mass $ m$:
 $\rho_{\phi} = {\textstyle {1 \over 2}} \dot \phi^2 +
 { 1\over2 } m^2 \phi^2 \sim a^{ - 3}$.
Hence the coherent oscillations of the homogeneous scalar field
correspond to the matter dominated effective equation of state with
vanishing
pressure.

Reheating occurs when the amplitude of oscillations of the inflaton field
$\phi$ decreases much faster than in (\ref{870}), and its
energy density is transferred to the energy density of other particles and
fields.

\section {\label{OSCILLATIONS} Oscillations and decay of the scalar field}
In the present section, we will discuss the elementary theory of reheating
developed in \cite{DL,AFW}; see also \cite{book}. A more detailed discussion
of
this theory will be contained in \cite{PERT}. We will consider a basic
model describing the inflaton scalar field $\phi$  interacting with a scalar field $\chi$ and a
spinor field
$\psi$:
\begin{eqnarray}\label{1}
L &=& {\textstyle {1 \over 2}} \phi_{,i} \phi^{,i} - V(\phi) +
{\textstyle {1 \over 2}}
\chi_{,i}
\chi^{,i} - {\textstyle {1 \over 2}} m_\chi^2(0) \chi^2 + {\textstyle {1
\over 2}} \xi R \chi^2 \nonumber \\
 &+& \bar \psi \bigl( i \gamma^i
\partial_i
 - m_{\psi}(0)\bigr) \psi - {\textstyle {1 \over 2}} g^2 \phi^2 \chi^2
 - h \bar \psi \psi \phi\ .
\end{eqnarray}
Here $ g$, $h$ and $\xi$ are small coupling constants,
$R$ is the space-time curvature, and $V(\phi)$ is the effective potential
of the field $\phi$. We will suppose here, for generality, that the effective
potential has a minimum at $\phi = \sigma$, and near the minimum it is
quadratic with respect to the field $\phi$: $V(\phi) \sim {\textstyle {1 \over
2}} m^2
(\phi-\sigma)^2 $.
 Here $m^2$ is the effective mass squared of the field
$\phi$. After the shift $\phi
-\sigma \to \phi$, the effective potential acquires the familiar form $
{\textstyle {1 \over 2}} m^2 \phi^2$, and the Lagrangian acquires an
interaction term
which is linear with respect to the field $\phi$:\ \ $\Delta L =
-g^2
\sigma\phi\chi^2$. This term vanishes in the case without spontaneous
symmetry
breaking, where $\sigma = 0$. The masses of the $\chi$-particles and $\psi$
after the shift become $m_\chi = \sqrt{m^2_\chi(0) + g^2\sigma^2}$ and $m_\psi
= m_\psi(0)
+
h\sigma$.
In this section we will consider the case $m \gg
m_{\chi}$, $m_{\psi}$. We will assume that after inflation $H \ll m$. This
condition is always satisfied during the last, most important stages of
reheating.

We will study now the  oscillation  of the scalar field
near the minimum of its effective potential. The energy density of
the oscillating field (after the shift $\phi - \sigma \to \phi$)
is $\rho_{\phi} = {\textstyle {1 \over 2}} \dot \phi^2 +
 { 1\over2 } m^2 \phi^2$. If
we ignore for a moment the effects associated with particle
creation, the field
$\phi$ after inflation oscillates near the point $\phi = 0$
with the frequency $k_0 = m$. The amplitude of oscillation 
decreases as $ a^{ - {3\over2}}$ due to the expansion of the universe,
and the energy
of the field $\phi$ decreases in the same way as the density of
nonrelativistic
particles of mass $m$: $\rho_{\phi} = {\textstyle {1 \over 2}} \dot \phi^2 +
 { 1\over2 } m^2 \phi^2 \sim a^{ - 3}$. A homogeneous scalar
field oscillating with frequency $m$ can be
considered as a coherent wave of $\phi$-particles with zero
momenta and with   particle density $n_{\phi} = \rho_{\phi}/m$. In other
words, $n_{\phi}$ oscillators of the same
frequency $m$, oscillating coherently with the same phase, can be described
as a single homogeneous wave $\phi(t)$.
Note that if we consider time intervals larger than the typical
oscillation time $m^{-1}$, the energy density of the
oscillating field, and
the number density of the particles $n_{\phi}$ will be related to its
amplitude
 $\Phi$ in a simple way:
\begin{equation}\label{1aa}
\rho_{\phi} ={\textstyle {1 \over 2}} m^2 \Phi^2 \ ,
\end{equation}
\begin{equation}\label{1aaa}
n_{\phi} = {\textstyle {1 \over 2}}m \Phi^2\ .
\end{equation}

Now we will consider effects related to the expansion of the
universe and to particle production. For a homogeneous scalar field
in a universe with a Hubble constant $H$, the equation of motion
with non-gra\-vi\-ta\-tio\-nal quantum corrections is
\begin{equation}\label{2}
\ddot \phi + 3 H(t)\, \dot \phi + \left( m^2 + \Pi
(\omega) \right) \phi = 0\ .
\end{equation}
Here $\Pi (\omega)$ is the flat space polarization operator  for the
field $\phi$ with   four-momentum $ k_i = (\omega , 0,0,0),\, \omega =
m$.

The real part of $\Pi(\omega)$ gives only a small correction
to $m^2$, but when $ \omega \ge min(2m_{\chi},2m_{\psi})$, the
polarization operator $\Pi(\omega)$ acquires an imaginary part $
\mbox{Im}\,
\Pi(\omega)$. We will assume that
 $m^2 \gg H^2$, $m^2 \gg \mbox{Im}\, \Pi$. The first
condition
is
automatically satisfied after the end of inflation; the second
is usually also
true. We have $\Phi(t) = \Phi_0 a^{-3/2}(t) = \Phi_0 \exp (-{3\over 2}\int dt
H(t))$. Neglecting for simplicity the time-dependence of $H$ and $\mbox{Im}\,
\Pi$
due to
the expansion of the universe, we obtain a solution of (\ref{2}) that
describes
damped oscillations of the field near the
point $\phi = 0$:
\begin{eqnarray}\label{3}
\phi &=& \Phi (t)\, \exp ( i mt) \nonumber \\
&\approx& \phi_0\, \exp ( i
mt)\cdot \exp {\Bigl[ -
{1 \over 2}
{\left(3H + {{\mbox{Im}} \, \Pi(m) \over m} \right)}
t
\Bigr]}\
{}.
\end{eqnarray}
 From unitarity it follows that \cite{Peskin}
\begin{equation}\label{4}
{\mbox{Im}}\, \Pi = m \Gamma \ ,
\end{equation}
where $\Gamma = \Gamma ( \phi \to \chi \chi) +\Gamma( \phi \to \psi \psi )$
is
the total decay rate of $\phi$-particles. (In a more general case one should
calculate not only the imaginary part of the polarization operator, but the
imaginary part of the effective action \cite{DL}.)
Thus, Eq. (\ref{3}) implies that the amplitude of oscillations of
the field $\phi$ decreases as $\exp {\big [ - {\textstyle {1 \over 2}}
{\left(
3H + \Gamma \right)} t \big]}$ due to particle production which
occurs during the decay of the inflaton field.

Note that under the condition $m \gg H$, the polarization
operator $\Pi$
and the decay rates $\Gamma$ do not depend on the curvature of the
universe
(and thus on time) and coincide with their flat-space limits. In
particular, the
probability of decay of a $\phi$-particle into a pair of scalar
$\chi$-particles or spinor $\psi$-particles for $m \gg
m_{\chi},
m_{\psi}$ is given by the following expressions \cite{book}:
\begin{equation}\label{7}
 \Gamma ( \phi \to \chi \chi) = { g^4 \sigma^2\over 8
\pi m}\ , \ \ \ \ \
\Gamma( \phi \to \psi \psi ) = { h^2 m\over 8 \pi}\ .
\end{equation}

 For a phenomenological
description of the damping of oscillations of the scalar field $\phi$
(\ref{3})
one may add an extra friction term
$\Gamma\dot\phi$ to the classical equation of motion of the field
$\phi$,
instead of adding the term proportional to the imaginary part of the
polarization operator,
\begin{equation}\label{5}
\ddot\phi + 3 H(t)\, \dot\phi + \Gamma \dot\phi +
m^2\phi = 0\ .
\end{equation}
This phenomenological equation together with relation (\ref{4}) for
$\Gamma
$ reproduces the damped oscillator solution (\ref{3}) of Eq. (\ref{2}).
The idea that one can describe effects of reheating by adding friction terms
to the equation of motion goes back to one of the first papers on reheating
\cite{ASTW}. At first the physical origin of such terms as
well as their value remained obscure. Some authors added
various auxiliary friction terms to the equations of the inflaton field in
order to
slow down its motion and make inflation longer, see e.g. \cite{ASTW,6}. From
the derivation of
expression (\ref{4}) for $\Gamma $ it follows, however, that the simple
phenomenological equation (\ref{5}) is valid only at the stage of rapid
oscillations of the field $\phi$ near
the minimum of $V(\phi)$. This equation cannot be used to
investigate the stage of
 slow rolling of the field $\phi$ during
inflation.

According to
(\ref{3}), the field amplitude $\Phi(t)$ obeys the equation
\begin{equation}\label{ogogo}
{1 \over a^3}\ {d\over dt}(a^3 \Phi^2) = - \Gamma \Phi^2\
{}.
\end{equation}
 If one
multiplies it by
$m$, one obtains the following equation for the number density
(\ref{1aaa}) of the coherently oscillating $\phi$-particles:
\begin{equation}\label{ogogogo}
 {d\over dt}(a^3 n_\phi) = - \Gamma\cdot a^3\, n_\phi\ .
\end{equation}
This equation has a simple interpretation. It shows that the total comoving
number density
 of particles $\sim a^3 n_\phi$
exponentially decreases with the decay rate $\Gamma$. Similarly, one
obtains
the following equation for the total energy of the
oscillating
field $\phi$:
\begin{equation}\label{ogogogogo}
 {d\over dt}(a^3 \rho_\phi) = - \Gamma\cdot a^3\, \rho_\phi\ .
\end{equation}

The decay products of the scalar field $\phi$ are ultrarelativistic (for $m
\gg
m_\chi, m_\psi$), and their energy density decreases due to the expansion of
the
universe much faster than the energy of the oscillating field $\phi$.
Therefore, reheating in our model ends only when the Hubble constant $H \sim
{2\over 3t}$ becomes smaller than $\Gamma$, because otherwise the main
portion
of energy remains stored in the field $\phi$. Therefore the age of the
universe when reheating completes is given by $t_r
\sim {\textstyle {2\over3}}
\Gamma^{-1}$. At that stage the main part of the matter in the universe becomes
ultrarelativistic. The age of the universe with the energy density
$\rho$ is $t = \sqrt M_p/\sqrt {6\pi \rho}$ \cite{book}. This,
together with the
condition $t_r \sim {\textstyle {2\over3}} \Gamma^{-1}$, gives the energy
density at   time
$t_r$:
\begin{equation}\label{10}
\rho(t_r) \simeq {3\Gamma^2M_p^2 \over 8\pi}\ .
\end{equation}

If thermodynamic equilibrium sets in quickly after the
decay of the inflaton field, then the matter acquires a temperature
$T_r$, which is defined by the equation
\begin{equation}\label{11}
\rho(t_r)
\simeq {3 \Gamma^2M_p^2 \over 8 \pi} \simeq {\pi^2 N(T_r)
\over
30}T_r^4\ .
\end{equation}
Here $N(T)$ is the number of relativistic degrees of freedom at the temperature
$T$; one should take 1 for each scalar, two for each massless vector particle,
etc. \cite{book}. In realistic models one may expect $N(T_r) \sim 10^2 -
10^3$, which gives the
following estimate of the reheating temperature:
\begin{equation}\label{12}
T_r \simeq 0.2 \sqrt{\Gamma M_p}\ .
\end{equation}
Note that $T_r$ does not depend on the initial value
of the field $\phi$; it is completely
determined by the parameters of the underlying elementary particle
theory.

Here we should make an important comment. In the absence of fermions, the
only
contribution to the decay rate would be $\Gamma ( \phi \to \chi \chi) = {
g^4
\sigma^2\over 8 \pi m}$. Note that this term disappears in the theories
without
spontaneous symmetry breaking, where $\sigma = 0$. This does not necessarily
mean that there is no reheating at all in such theories. Indeed, decay is
possible not only in the presence of a constant field $\sigma$ but in the
presence of a large oscillating field $\phi(t)$ as well. In what follows we
will study parametric resonance and reheating in  models with $\sigma = 0$,
or $\sigma \ll \Phi$, where $\Phi$ is the amplitude of the oscillations.
However, when reheating proceeds and $\Phi$ becomes small one may expect
perturbation theory to work well. To get an estimate for the decay rate at
$\sigma = 0$ let us simply write $\Phi$ instead of $\sigma$ in Eq. (\ref{7}):
$\Gamma ( \phi\phi \to \chi \chi) \sim { g^4 \Phi^2\over 8 \pi m}$. The
problem
with this term is that $\Phi^2$ decreases as $t^{-2}$ in the expanding
universe, whereas the Hubble constant decreases only as $t^{-1}$. Therefore
the
decay rate never catches up with the expansion of the universe, and reheating
never completes. Reheating can be complete only if $\Gamma$ decreases more
slowly than $t^{-1}$. Typically this requires either spontaneous symmetry
breaking ($\sigma \not = 0$) or coupling of the inflaton field to fermions
with
$m_\psi < m/2$. If both of these conditions are violated, the inflaton field
$\phi$ never decays completely. Such fields may be responsible
 for the dark matter of the universe, but it requires certain fine-tuning of
the parameters. Normally, an incomplete decay of the inflaton field implies
that
the universe at the age of 10 billion years is cold, empty and unsuitable
for
life. We should emphasize that this may happen even if the coupling constant
$g^2$ is very large. Thus the requirement that reheating is complete
imposes important constraints on the structure of the theory.

The elementary theory of reheating described above is simple and
intuitively
appealing. It proves to be very successful in describing reheating after
inflation in many realistic inflationary models. That is why we are going to
develop this theory even further in \cite{PERT}. However, in some cases where
the amplitude of the oscillating field is sufficiently large, reheating
occurs
in a different way, in the regime of parametric resonance.

\section{\label{LIMITS} Parametric resonance and limits of applicability of
perturbation
theory}

\subsection{\label{NARROW} Perturbation theory versus narrow resonance}

In the investigation performed above we made a natural assumption that the
decay probability $\Gamma$ of the scalar field $\phi$ can be calculated by
ordinary
methods of quantum field theory describing the decay $ \phi \to \chi \chi$.
However, if many $\chi$-particles have already been produced,
$n_k >1$, then the
probability of decay becomes greatly enhanced due to effects related to
Bose-statistics. This may lead to explosive particle production.

For simplicity, we
 consider here the interaction between
 the {\it classical } inflaton field $\phi$
and the {\it quantum} scalar field
$\hat\chi$ with the Lagrangian (\ref{1}).
 The Heisenberg representation of the quantum scalar field $\hat \chi$ is
\begin{equation}
\hat \chi(t, {\bf x}) =
{1\over{(2\pi)^{3/2}}} \int d^3k\ {\Bigl( \hat a_{{k}} \chi_{
k}(t)\, e^{ -
i{{\bf k}}{{\bf x}}}
+ \hat a_{{k}}^ + \chi^*_k(t)\, e^{i{{\bf k}}{{\bf x}}}
\Bigr)}\ ,
\label{37}
\end{equation}
where $\hat a_{{k}}$ and $\hat a_{{k}}^ + $ are annihilation and creation operators. For a flat Friedmann background with scale
factor $a(t)$ the temporal
part of the eigenfunction with comoving momentum ${\bf k}$ obeys the
following equation:
\begin{equation}
\ddot \chi_k + 3{{\dot a}\over a}\dot \chi_k + {\left(
{{\bf k^2}\over a^2}
 + m^2_{\chi}(0) - \xi R + g^2\phi^2 \right)} \chi_k = 0 \ .
\label{38}
\end{equation}
(The physical momentum ${\bf p} = {\bf k \over a(t)}$ coincides with ${\bf k}$
for
Minkowski space, where $a = 1$.) Eq. (\ref{38}) describes an
oscillator with a variable
frequency $\omega$ due to the time-dependence of   $a(t)$ and the
background field $\phi(t)$. Until the last section of this paper we
will suppose that the effective mass of the field $\chi$ vanishes for $\phi =
0$: $m_\chi(0) = 0$. In Sec. \ref{SUPERMASSIVE} we will investigate the
opposite case, $m_\chi(0) \gg m$.

As in the previous section, consider the simplest potential $V(\phi) \sim
{\textstyle {1 \over 2}} m^2
(\phi-\sigma)^2 $ (to mimic the situation with spontaneous symmetry breaking)
and make the shift $\phi
-\sigma \to \phi$, after which the effective potential becomes $
{\textstyle {1 \over 2}} m^2 \phi^2$, and the interaction term
 $-{\textstyle {1 \over 2}} g^2\phi^2\chi^2$ transforms to $ -{\textstyle {1
\over 2}} g^2\phi^2\chi^2
-g^2
\sigma\phi\chi^2 - {\textstyle {1 \over 2}} g^2\sigma^2\chi^2$.
We shall analyze the general equation (\ref{38}) in different regimes.

Suppose first that the amplitude of oscillations $\phi$ is much smaller than
$\sigma$, and neglect for a moment the expansion of
the universe, taking $a = 1$ in Eq. (\ref{38}). Then one can write the
equation
for modes (quantum fluctuations)
of
the field $\chi$
with   physical momentum ${\bf k}$ in the following form:
\begin{equation}\label{Mux}
\ddot \chi_k + \left( k^2
+ g^2\sigma^2 +2g^2\sigma \Phi \, \sin mt \right) \chi_k = 0 \ ,
\end{equation}
where $k = \sqrt {{\bf k}^2 }$, and $\Phi$ stands for the amplitude of
oscillations of the inflaton field.
This equation
describes an oscillator with a periodically changing
frequency $\omega_k^2(t)= k^2
+ g^2\sigma^2 +2g^2\sigma \Phi \, \sin mt$.
This periodicity may lead to parametric resonance for modes with certain
values
of $k$. The simplest way to describe this important effect is to make a change
of variables
$mt = 2z -\pi /2$, which reduces
Eq. (\ref{Mux}) to the well-known Mathieu equation \cite{ML}:
\begin{equation}\label{M1ux}
\chi_k'' + \left(A_k - 2q \cos 2z \right) \chi_k = 0 \ .
\end{equation}
Here $A_k
= 4{k^2+g^2\sigma^2 \over m^2 }$, $q = {4g^2\sigma\Phi \over
 m^2} $, $z
= {mt\over 2}$, and prime denotes differentiation with respect to $z$.
The properties of the solutions of the Mathieu equation
are well represented by its
stability/instability chart which can be found, e.g., in \cite{ML}.
An important feature of solutions of Eq. (\ref{M1ux}) is the
existence of an exponential instability $\chi_k \propto \exp
(\mu_k^{(n)}z)$ within the set of resonance bands of frequencies
$\Delta k^{(n)}$ labeled by an integer index $n$.
This instability corresponds to exponential growth of occupation
numbers of quantum fluctuations
$n_{k}(t) \propto \exp (2\mu_k^{(n)} z)$
 that may be interpreted as particle
production. In a state with a large number of Bose particles the estimates
for $\Gamma$ obtained in the previous subsection do not apply, and one should
use much more elaborate methods of investigation based on the theory of
parametric resonance.

In the case under consideration, $g\Phi \ll g\sigma \ll m$, the theory of  
parametric resonance is well known \cite{LLMechanics}. Indeed, in this case
one has $q \ll 1$, and the resonance occurs only in some narrow bands near
$
A_k \simeq l^2, \ l=1, 2, ... $. Each band in momentum space has  width of
 order  $\Delta k \sim q^l$, so for $q< 1$ the widest and  most
important instability band is the first one, $A_k \sim 1 \pm q =1 \pm
{4g^2\sigma\Phi \over
 m^2}$.

The factor $\mu_k$ which describes the rate of exponential growth for the
first
instability band for $m^2 \gg g^2\sigma^2$ is given by \cite{ML}
\begin{equation}\label{epsilon}
\mu_k =\sqrt{\left({ q \over 2} \right)^2 -
\left( {2k\over m} - {1 } \right)^2 } \ .
\end{equation}
Thus  resonance occurs for ${k } = {m\over 2}(1 \pm {q \over 2})$. The
index
$\mu_k$ vanishes at the edges of
the
resonance band and takes its maximal value $\mu_k = {q\over 2} =
{2g^2\sigma\Phi \over m^2}$ at $k={m \over 2}$. The corresponding modes
$\chi_k$ grow at a maximal rate $\exp({qz\over 2})$, which in our case is
given
by $\exp ({qmt\over 4}) = \exp({g^2\sigma\Phi t\over m})$.

The growth of the modes $\chi_k$ leads to the growth of the occupation
numbers
of the
 created particles $n_k(t)$. Indeed, the
 number density $n_k$ of particles with  momentum $\bf k$
 can be evaluated as the energy of that mode
${\textstyle {1 \over 2}} |\dot \chi_k|^2 +
{\textstyle {1 \over 2}}\omega_k^2 |\chi_k|^2$ divided by the energy $\omega_k$
of each
particle:
\begin{equation}\label{number}
n_k={\omega_k\over 2} \left( { |\dot \chi_k|^2 \over \omega_k^2}
+|\chi_k|^2 \right)-{1\over 2}.
\end{equation}

When the modes $\chi_k$ grow as $\exp({qz\over 2})$, the number of
$\chi$-particles grows as $\exp({qz })$, which in our case is equal to $\exp
({qmt\over 2}) = \exp({2g^2\sigma\Phi t\over m})$.

The fact that the resonance occurs near $k={m \over 2}$ has a simple
interpretation: In the limit $g\sigma \ll m$ the effective mass of the
$\chi$-particles is much smaller than $m$. Therefore one decaying
$\phi$-particle
 creates two $\chi$-particles with momentum $k \sim m/2$. This picture
is
very similar to the process of decay $\phi \to \chi\chi$ discussed in the
previous section, but the results are absolutely different. Indeed, in
perturbation theory the amount of produced particles did not depend on the
number of particles produced earlier, and the rate of production for our
model
was given by $\Gamma (\phi \to \chi \chi) = { g^4 \sigma^2\over 8 \pi m}$.
Thus the decay rate $\Gamma^{-1}$ was
suppressed by the factor $g^{4}$, which made the decay very slow in the weak
coupling limit. By contrast, in the regime of parametric resonance the rate of
production of $\chi$-particles is proportional to the amount of particles
produced earlier (which is why we have exponential growth), and the rate of
the
process is given by an absolutely different expression $\mu_k m \sim
{g^2\sigma\Phi \over m}$, which is greater than $\Gamma$ for $\Phi >
{g^2\over
8\pi}\sigma$.

Thus, before going any further we should understand how these two processes
are related to each other, and why we did not find the effect of parametric
resonance in the investigation performed in the previous section. Is the
perturbation
theory discussed there a limiting case of the narrow resonance regime or is
it
something else?

The reason we missed the effect of parametric resonance is rather
delicate.
In our calculations of the imaginary part of the polarization operator we
assumed that the $\chi$-particles produced by the oscillating scalar field
$\phi$ are normal particles on the mass shell, $k^2_\chi = m^2_\chi$. This is
what one would get solving Eq.  (\ref{Mux}) in any finite order of
perturbation theory with respect to the
interaction term $2g^2 \chi\sigma \Phi \sin mt$. However, if one solves the
equation
for
the fluctuations of the field $\chi$ (\ref{Mux}) {\it exactly}, one finds
exponentially growing
modes $\chi_k$. This creates a new channel of decay of the scalar field
$\phi$.

Note that exponentially growing modes occupy a very small portion of momentum
space in the narrow resonance limit. This means that the
fluctuations of the field $\chi$ for almost all $k$ are normal fluctuations
with $k^2_\chi = m^2_\chi$. In this case our calculation of the imaginary
part
of the polarization operator does apply. If the resulting value of $\Gamma$
appears to be smaller than $2\mu_k m \sim qm$, then the perturbative decay of
the scalar
field may coexist with the parametric resonance. One may consider several
different possibilities. In the beginning the scalar field $\phi$ can be
expected to oscillate with   amplitude $ \Phi > {g^2\over 8\pi}\sigma $. In
this regime parametric resonance leads to the exponential growth of modes
$\chi_k$, as we discussed above. However, gradually the field $\phi$
loses its energy, and its amplitude $\Phi$ becomes smaller than ${g^2\over
8\pi}\sigma$. In this regime the amplitude of the field $\Phi$ decays
exponentially within a time $\Gamma^{-1}$ which is smaller than the typical
time  necessary for   parametric resonance to occur. One may
say
that the perturbative decay
makes the energy eigenstate (the mass) of the field $\phi$ ``wide," with  width $\Gamma$, and when this width exceeds the width of the resonance band
$\sim q m/2$, the resonance terminates. Starting from this moment
perturbation
theory takes over, and the description of reheating should be given along the
line
of the elementary theory described in the previous section.

Thus, the standard effect of  scalar field decay described by the elementary
theory of reheating \cite{DL,AFW,PERT} and preheating due to parametric
resonance
are two {\it different} effects. In an expanding universe there exist other
reasons for evolving from  parametric resonance to  perturbative
decay.

First of all, during the expansion of the universe the field $\phi$ decreases
not
only because of its decay, but because of the ``friction term'' $3H\dot \phi$
in the equation of motion for the field $\phi$. Thus one should compare $q m$
with the effective decay rate $3H + \Gamma$: Parametric resonance occurs
only
for $q m {\
\lower-1.2pt\vbox{\hbox{\rlap{$>$}\lower5pt\vbox{\hbox{$\sim$}}}}\
} 3H + \Gamma$. Note that for $\Gamma > H$ perturbative decay leads to
reheating even neglecting parametric resonance. Therefore to check whether
parametric resonance appears at the time when perturbative decay is
inefficient, i.e. in the case $\Gamma < H$, it is enough to consider the
condition $q m {\
\lower-1.2pt\vbox{\hbox{\rlap{$>$}\lower5pt\vbox{\hbox{$\sim$}}}}\ } 3H$.

Another important mechanism which can prevent parametric resonance from being
efficient is the redshift of momenta $k$ away from the resonance band. The
total width of the first band is given by ${q m}$; the width of the
part
where the resonance is efficient is somewhat smaller; one can estimate it as
${q
m\over 2}$. The time $\Delta t$ during which
a given mode remains within this band depends on the equation of state of
matter, and typically can be estimated by
$q H^{-1}$. During this time the number of particles in
 growing modes increases as $\exp \Bigl({q^2 m\over 2H}\Bigr)$. This leads
to
efficient decay of
inflatons only if $q^2 m {\
\lower-1.2pt\vbox{\hbox{\rlap{$>$}\lower5pt\vbox{\hbox{$\sim$}}}}\ } H$. In
the
narrow resonance limit $q \ll 1$ this is a stronger condition than the
condition $q m {\
\lower-1.2pt\vbox{\hbox{\rlap{$>$}\lower5pt\vbox{\hbox{$\sim$}}}}\ } 3H$.

In general, it is possible that exponential growth during the time $\Delta t$
is small, but $\Delta t \ll H^{-1}$ and therefore resonance still plays some
role in reheating. However, this is a rather exceptional situation.
Therefore
typically the set of conditions for the resonance to be efficient can be
formulated as follows:
\begin{eqnarray}\label{cond}
 q m &\gtrsim&
 \Gamma \ ,\\
q^2 m &\gtrsim&
H
 \ .
\end{eqnarray}\label{LL}

In the model considered above these conditions yield:
\begin{eqnarray}\label{cond2a}
\Phi &>& {g^2\over 32\pi}\sigma\ , \\
{\Phi} &\gtrsim
& {m\sqrt {m H}\over 4 g^2 \sigma} \ .
\end{eqnarray}

Thus parametric resonance can be efficient at a sufficiently large $\Phi$,
but
reheating never ends in the regime of parametric resonance. As
soon as the amplitude of oscillations becomes sufficiently small, parametric
resonance terminates, and reheating can be described by the elementary theory
developed in \cite{DL,AFW,PERT}. Typically the reheating temperature is
determined
by
these last stages of this process. Therefore one should not calculate
the reheating
temperature simply by finding the endpoint of the stage of parametric
resonance, as many authors do. The role of the stage of preheating is to
prepare a different setup for the last stage of reheating. It changes
the reheating
temperature, and it may lead to interesting effects such as nonthermal
symmetry restoration and new mechanisms of baryogenesis. However, reheating
never ends in the regime of parametric resonance; it does not
make
much sense to calculate the reheating temperature at the end of the stage of
preheating.

The expansion of the universe and the inflaton decay are not the only
mechanisms
which could prevent the development of resonance. As we will show,
backreaction of created particles may change the parameters $A_k$ and $q$.
There will be no resonance if the $\chi$-particles decay with decay rate
$\Gamma > \mu_k m$, or if within the time $\sim (\mu_k m)^{-1}$ they are
taken
away from the resonance band because of their interactions. Also, there is no
explosive
reheating if the decay products include fermions since the fermion occupation
numbers cannot be large because of the Pauli principle.
This happens, for example, in many inflationary models based on supergravity
where  inflaton decay is often accompanied by gravitino production
\cite{Primordial}.

If reheating never {\it ends} in a state of narrow parametric resonance, one
may wonder whether reheating may {\it begin} in a state of narrow resonance.
As
we are going to show, in most cases inflation begins in a state of  broad
parametric resonance; the resonance typically ceases to exist as soon as it
becomes narrow. But before analyzing this issue, we will take one last look
at
the model which we studied above.

 \subsection{\label{SPONTSYMMBR}Processes at $\phi \sim \sigma$}

In our investigation of the simple model with spontaneous symmetry breaking
($\sigma \not = 0$) we assumed that the amplitude of oscillations of the
scalar field $\phi$ is very small, $\Phi \ll \sigma$. Therefore we retained
only the quadratic part of the effective potential, $V(\phi) \sim (\phi -
\sigma)^2$. However, in realistic models of spontaneous symmetry breaking this
condition is
satisfied only at the end of  parametric resonance. Indeed, let us
consider a theory with spontaneous symmetry breaking with the usual potential
${ \lambda\over 4}(\phi^2 -\sigma^2)^2$. Then after spontaneous symmetry
breaking and the corresponding shift $\phi-\sigma \to \phi$ the theory at
$\phi
\ll \sigma$ can be represented as a theory of a massive scalar field with a
mass $m^2 = 2\lambda \sigma^2$ interacting with the field $\chi$ which
acquires mass $m^2_\chi = g^2 \sigma^2$. In this respect, it coincides with the
toy
model studied in the previous subsection.
However, there are some important differences.

First of all, the process $\phi \to \chi \chi$ is possible only if $m > 2
m_\chi$. This was one of the conditions which we used in our investigation:
we
assumed that $m \gg m_\chi$, i.e. $\lambda \gg g^2$. However, in this case
the
interaction ${\lambda\over 4} \phi^4$ which we did not take into account so
far may become more important than the interaction ${g^2\over 2}
\phi^2\chi^2$
which we considered. As a result, the production of $\phi$-particles may be
more efficient than the production of $\chi$-particles.

In order to investigate this possibility let us study for a moment a model
with
the
effective potential ${ \lambda\over 4}(\phi^2 -\sigma^2)^2$ in the limit
$\lambda \gg g^2$, i.e. neglecting the interaction ${g^2\over 2}
\phi^2\chi^2$.
We will assume here that in the beginning the field $\phi$ was at the top of
the effective potential. At that time its effective mass squared was
negative,
$m^2(0) = -\lambda\sigma^2$. This fact alone, independent of any parametric
resonance, leads to the production of particles of the field $\phi$. The main
point
here is that all modes with $k < \sqrt\lambda \sigma$ grow exponentially,
which
breaks the homogeneity of the oscillating scalar field. This is an
interesting
effect, which has some nontrivial features, especially if one takes the
expansion
of the universe into account. We will return to its discussion elsewhere.
However, this effect does not last long because away from the maximum of the
effective potential its curvature becomes positive.

When the amplitude of the oscillations of the field $\phi$ near $\phi =
\sigma$
becomes   smaller than $\sigma$, the field begins
oscillating
near its minimum with a frequency $m \approx \sqrt{2\lambda} \sigma$. The
parametric resonance with $\phi$-particle production in this regime can be
qualitatively understood if the equation
for
the fluctuations   $\delta \phi_k$ is approximately
represented
as a Mathieu equation.  The modes $\phi_k$   grow in essentially  the same way
as the modes in the second instability band of the Mathieu equation with
 $A_k = {4k^2\over m^2} + 4$, $q = 6{\Phi\over \sigma} $.  For
$q\gtrsim 1$, we are in the broad resonance regime, and there is a significant
production of
$\phi$-particles.
However, for $q \ll1$, i.e. for $\Phi \ll \sigma/6$, the parametric resonance
in the second band   becomes very inefficient. (One
can obtain the same   result by a more accurate investigation of parametric
resonance in this situation
in terms of the Lame equation,  but this is not our purpose here.)

Thus, we are coming to the following picture of   parametric resonance in
this  model.
In the beginning of the rolling of the field $\phi$ down to the minimum of
the
effective potential, the leading source of particle production is associated
with the tachyonic mass of the field $\phi$. Soon after that, the leading
mechanism is the decay of a coherently oscillating field $\phi$ into
$\phi$-particles. This mechanism remains dominant until the amplitude of the
field $\Phi$ becomes much smaller than $\sigma$, after which the decay $\phi
\to \chi\chi$ studied in the previous section becomes more important. (This process becomes somewhat more complicated if the backreaction of the produced particles it taken into account.) 
Finally,
when the amplitude of the oscillations $\Phi$ becomes smaller than ${g^2\over
32\pi} \sigma$, or when
it becomes smaller than ${m\sqrt {m H}\over 4 g^2 \sigma}$,
whichever
comes first, the parametric resonance ceases to exist, and the decay
$\phi \to \chi\chi$ is described by the elementary theory of reheating
based on perturbation theory.

 We should note that the $\chi$-particle production in this model for $
\lambda\gg g^2$ was first studied in
\cite{Brand1}. However, as we just mentioned, at $\Phi \sim \sigma$ this
process is
subdominant as compared to the $\phi$-particle production, which was not
studied in \cite{Brand1}. The process of $\chi$-particle production is more
efficient than $\phi$-particle production only for $\Phi \ll \sigma$. In
this regime our results differ from those obtained in
\cite{Brand1} by the factor ${\Phi\over \sigma}$ in the exponent. This
difference is very significant because it leads to a much less efficient
reheating, which shuts down as soon as $\Phi$ becomes sufficiently small.

The models studied in the last two subsections can be considered as a good
laboratory where one can study different features of parametric resonance.
However, in our investigation so far we did not discuss the question of
initial
conditions for  resonance in these models. Indeed, after 15 years of
investigation we still have not found any simple mechanism which will put
the
inflaton field on the top of the potential at $\phi = 0$ in the new inflation
scenario. Also,
the shape of the potential required for new inflation (extremely flat near
the
origin) is rather artificial. As soon as we consider generic initial
conditions
for the scalar field $\phi$ in more realistic inflationary models, such as
chaotic inflation in the theory with a simple potential ${m^2\phi^2\over 2}$,
the theory of parametric resonance becomes different in many respects from
the simple theory described above.

\section{\label{BROAD} Broad resonance versus narrow resonance in Minkowski
space}

In the chaotic inflation scenario one does not impose any {\it a priori}
conditions on the initial value of the scalar field. In many models of
chaotic
inflation the initial amplitude of oscillations of the
 field $\phi$ can be as large as $M_p$, i.e. much greater than any other
dimensional parameters such as $\sigma$. Therefore in the remaining part of
the paper we will concentrate on the simplest chaotic inflation model without
symmetry breaking with the effective potential $V(\phi) = {m^2\over
2}\phi^2$, and the interaction term $-{\textstyle {1 \over 2}}
g^2\phi^2\chi^2$.
In this case instead of Eq. (\ref{Mux})
one has
\begin{equation}\label{Mu}
\ddot \chi_k + \left( k^2
+ g^2 \Phi^2\, \sin^2(mt) \right) \chi_k = 0 \ .
\end{equation}
 This equation
describes an oscillator with a periodically changing
frequency $\omega^2(t)=
 k^2 + g^2\Phi^2\, \sin^2 mt $. One can write it as a Mathieu equation (Eq.
(\ref{M1ux}))
with $A_k
= {k^2 \over m^2 }+2q$, $q = {g^2\Phi^2\over
4m^2} $, $z
= mt$.

For $g\Phi < m$ we have a narrow resonance with $q \ll 1$. In this regime the
resonance is more pronounced in the first resonance band, for modes with
${k^2
} \sim m^2(1 - 2q \pm q)$. The modes $\chi_k$ with momenta corresponding to
the center of
the resonance at $k \sim m$ grow as $e^{q z/2}$, which in our case equals  $e^{\mu_k mt} \sim \exp \Bigl({g^2\Phi^2\, t\over
8m}\Bigr)$, and the number of $\chi$-particles grows as
$e^{2\mu_k
z} \sim e^{q z} \sim \exp \Bigl({g^2\Phi^2\, t\over
4m}\Bigr)$. This process can be interpreted as a resonance with decay of two
$\phi$-particles  with mass $m$ to two $\chi$-particles with momenta $k \sim
m$. We show the results of the numerical solution of Eq. (\ref{Mu}) for the fastest growing mode $\chi_k$ in the narrow resonance regime in   Fig. 2.  Typically, the rate of  development of the parametric resonance does not differ much from the rate of the growth of the leading mode $\chi_k$, see a discussion of this issue in the next section.

 \begin{figure}[t]
 \centering
 \leavevmode\epsfysize=9.1cm \epsfbox{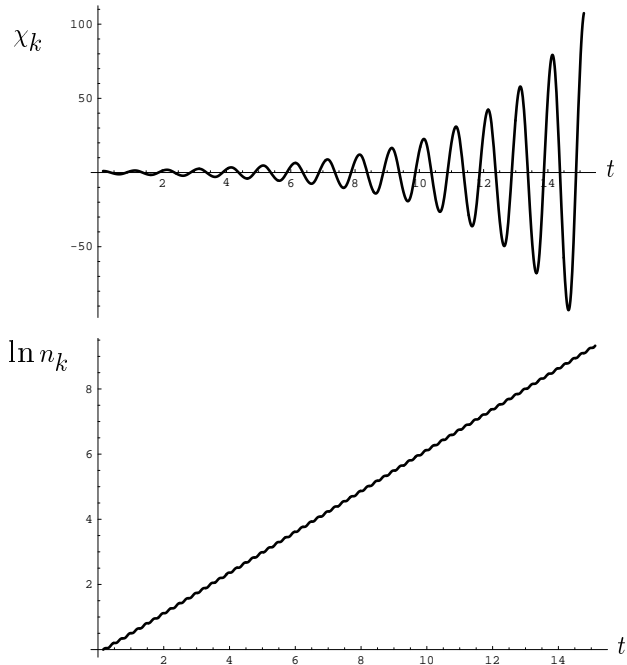}\\
\

 \caption[fig2]{\label{fig2} Narrow parametric resonance for the field $\chi$
in the theory ${m^2\phi^2\over 2}$ in Minkowski space for $q \sim 0.1$.
Time
is shown in units of $m/2\pi$, which is equal to the number of oscillations
of
the inflaton field $\phi$. For
each oscillation of the field $\phi(t)$ the growing modes of the field $\chi$
oscillate one time. The upper figure shows the growth of the mode $\chi_k$
for
the momentum $k$ corresponding to the maximal speed of growth. The lower
figure
shows the logarithm of the occupation number of particles $n_k$ in this mode,
see Eq. (\ref{number}). As we see, the number of particles grows
exponentially,
and $\ln n_k$ in the narrow resonance regime looks like a straight line with
a
constant slope. This slope divided by $4\pi$ gives the value of the parameter
$\mu_k$. In this particular case $\mu_k \sim 0.05$, exactly as it should be
in
accordance with the relation $\mu_k \sim q/2$ for this model.}
 \end{figure}

On the other hand, for oscillations with a large amplitude $\Phi$
 the parameter
 $q={g^2\Phi^2\over 4m^2} $ can be very large.
 In this regime the
resonance occurs for a broad range of values of $k$, the parameter $\mu_k$
can be rather large, and reheating becomes extremely efficient. The resonance
occurs for modes with ${k^2 \over m^2 } = A - 2q$, i.e. above the line $A =
2q$  on the stability/instability chart for the Mathieu equation \cite{KLS}.
The
standard methods of investigation of narrow parametric resonance do not work
here. The difference between these two regimes can be easily grasped by
comparing solutions of Eq. (\ref{Mu}) for small and for large $q$, see
Figs.
2 and 3.

 \begin{figure}[t]
\centering
\leavevmode\epsfysize=9.2cm \epsfbox{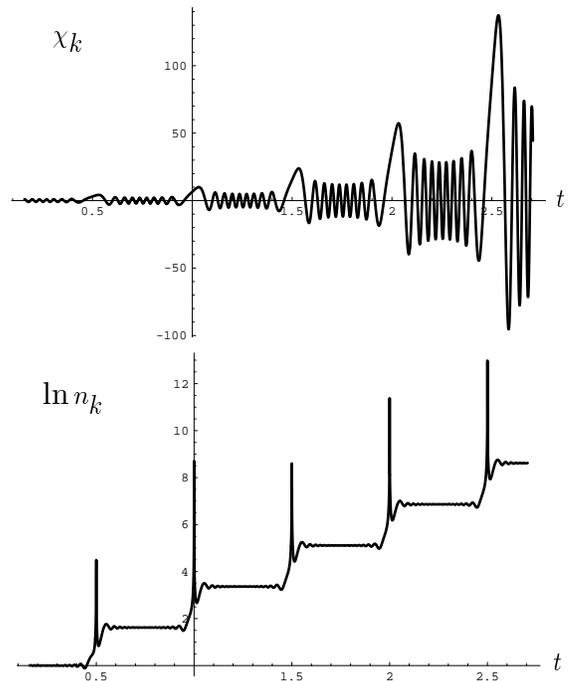}\\
\

\caption[fig3]{\label{fig3} Broad parametric resonance for the field $\chi$
in
Minkowski space for $q \sim 2\times10^2$
in the theory ${m^2\phi^2\over 2}$. For each oscillation of the field
$\phi(t)$
the field $\chi_k$ oscillates many times. Each peak in the amplitude of the
oscillations of the field $\chi$ corresponds to a place where $\phi(t) =
0$.
At this time the occupation number $n_k$ is not well defined, but soon after
that time it stabilizes at a new, higher level, and remains constant until
the
next jump. A comparison of the two parts of this figure demonstrates
the importance
of using proper variables for the description of preheating. Both $\chi_k$
and
the integrated dispersion $\langle\chi^2\rangle$ behave erratically in the
process of parametric resonance. Meanwhile $n_k$ is an adiabatic invariant.
Therefore, the behavior of $n_k$ is relatively simple and predictable
everywhere
except  the short intervals of time when $\phi(t)$ is very small and the particle production occurs. In our
particular case, the average rate of   growth of $n_k$ is close to the
maximal
possible rate for our model, $\mu_k \sim 0.3$.}
\end{figure}

The time evolution is shown in units $m/2\pi$, which corresponds to the
number of oscillations $N$ of the inflaton field $\phi$. The oscillating
field $\phi(t) \sim \Phi \sin mt$ is zero at integer and half-integer values
of
the variable $mt/2\pi$. This allows us to identify
particle
production with time intervals when $\phi(t)$ is very
small.

During each oscillation of the inflaton field $\phi$, the field $\chi$
oscillates many times. Indeed, the effective mass $m_\chi(t) = g\phi(t)$
is
much greater than the inflaton mass $m$  for the main part of the period of oscillation of the field $\phi$
in the broad resonance regime with $ q^{1/2} = {g \Phi \over 2 m } \gg 1$.
As
a result, the typical frequency of oscillation 
 $\omega (t) = \sqrt{k^2 +g^2\phi^2(t)}$
 of the field $\chi$ is much
higher than that of the field $\phi$. Within one period of oscillation of
the
field $\phi$ the field $\chi$ makes $O(q^{1/2})$ oscillations. That is why
during the most of this interval it is possible to talk about
an adiabatically changing effective mass $m_\chi(t)$.
Therefore, in the broad resonance regime
the amplitude of $\chi_k$
is minimal at the points where the frequency is maximal,
$\vert \chi_k \vert \propto \omega(t)^{-{\textstyle {1 \over 2}}}$, i.e.
 at $\phi(t) = \Phi$, and it increases
substantially near the points at which $\phi(t) = 0$.

For very small $\phi(t)$ the change in the
frequency of oscillations $\omega (t)$ ceases to be
adiabatic. The
standard condition necessary for particle production is the absence of
adiabaticity in the change of $\omega (t)$:
\begin{equation}\label{adiab}
{d\omega\over d t} \gtrsim \omega^2 \
{}.
\end{equation}
One should note that for a narrow resonance this condition is not necessary,
because even a small variation of $\omega(t)$
 may be exponentially accumulated in the course of time.
However, for a broad resonance one should expect a considerable effect during
each oscillation, which implies that the condition (\ref{adiab}) should be
satisfied.
To find the time interval $\Delta t_*$ and the typical momenta $k_*$
 when and where it may happen let us remember that for small $\phi$ one
has
$\dot \phi \approx m\Phi$. Therefore our condition (\ref{adiab}) implies that
\begin{equation}\label{adiabA}
{k^2 } \lesssim (g^2\phi m
\Phi)^{2/3} - g^2\phi^2 \ .
\end{equation}
Let us consider those momenta $k^2$ which satisfy   condition
(\ref{adiabA})
as a function
of $\phi(t)$.
This condition becomes satisfied for small $k$ when the field $\phi(t)$
becomes smaller than
 $\sqrt{m\Phi\over g}$.
 The maximal range of momenta for
which particle production occurs corresponds to $\phi(t) = \phi_*$, where
\begin{equation}\label{adiab4}
\phi_* \approx {1 \over 2} \sqrt {m\Phi\over g} \approx {1\over 3} {\Phi }
q^{-1/4} \ .
\end{equation}
 The maximal value of momentum for particles produced at that epoch can be
estimated by $k_{\rm max} = \sqrt{g m \Phi \over 2}$.
 In the main part of the interval $|\phi| \lesssim 2\phi_*$
the range of momentum remains
 smaller but the same order of magnitude
 as $k_{\rm max}$. Thus one may estimate
 a typical value of momentum of particles produced at that stage as
 $ k_*/2$, where
\begin{equation}\label{adiab5}
k_* = \sqrt{g m \Phi } = \sqrt 2 \ m q^{1/4} \ .
\end{equation}
This simple estimate practically coincides with the
result of a more detailed and rigorous investigation which will be performed
in
Sec.
\ref{ANALYTIC}. This is a very important result \cite{KLS}, which we are
going
to use
throughout the paper.\footnote{In this paper we will use both physical momenta
and comoving momenta. Our definition of $k_*$ refers to physical momentum.}
This result implies, in particular, that in the broad
resonance regime $m \ll k_* \ll g\Phi$.

Each time the field $\phi$ approaches the point $\phi = 0$, it spends
time
\begin{equation}\label{adiab7}
\Delta t_* \sim {2 \phi_*\over \dot \phi} \sim {1\over \sqrt {gm\Phi}} \sim
k_*^{-1} 
\end{equation}
 in the domain $|\phi| {\
\lower-1.2pt\vbox{\hbox{\rlap{$<$}\lower5pt\vbox{\hbox{$\sim$}}}}\ } \phi_*$.
During that time $k_* \sim m_\chi=g\phi_*$, so that
 $\omega \sim k_*$. This estimate of $\Delta t_*$
tells us that particle production in the broad resonance regime occurs within
a
time of   order of the period of one oscillation
of the field $\chi$, $\Delta t_* \sim \omega^{-1}$,
in agreement with the uncertainty principle.
 One can easily identify these short intervals in Fig.
\ref{fig3}.

 In the semiclassical regime when the frequency $\omega_k(t)$
is changing adiabatically, $n_k$ is a constant which coincides
with an adiabatic invariant. To appreciate the usefulness of the
introduction
of the adiabatic invariant
$n_k$, one should compare the evolution of the modes $\chi_k$ with the
evolution of the occupation numbers corresponding to each of these modes
shown
in Figs. \ref{fig2} and \ref{fig3}. As we see, in the narrow resonance regime
$\chi_k$ vigorously oscillates, whereas $\ln n_k$ grows like a straight line.
In the broad resonance regime the field amplification occurs near the points
$\phi(t) = 0$ where the process is not adiabatic. The occupation number
$n_k$,
being an adiabatic invariant, changes only during these short time intervals,
when
the number of particles is not well defined.

Analytical
investigation of the broad resonance regime in the context of the theory of
reheating was first reported in \cite{KLS}, see also \cite{Yosh}. Now we are
going to
 perform a much more detailed investigation of this
regime.

\section{\label{STOCHASTIC} Stochastic resonance in an expanding universe}
To understand why the broad resonance regime is so important for the theory
of
reheating in an expanding universe, let us remember that  
resonance in an expanding universe appears only for $q^2 m {\
\lower-1.2pt\vbox{\hbox{\rlap{$>$}\lower5pt\vbox{\hbox{$\sim$}}}}\ } H$,
which
in our case reads
\begin{equation}\label{uslovie}
g\Phi \gtrsim
{2m} \left({H\over m}\right)^{1/4} \ .
\end{equation}

In the simplest inflationary models including the model which we consider now
the value of the Hubble constant at the end of inflation is of the same order
as the inflaton
 mass $m$, but somewhat smaller. Indeed, as we already mentioned, during the
first oscillation the amplitude of the field $\Phi$ is of   order  
$M_p/20$, which gives the Hubble constant $H \sim \sqrt{2\pi \over
3}{m\Phi\over M_p} \sim 0.1 m$. Since dependence of the resonance condition
on $H$ is very weak ($H^{1/4}$), one may conclude that the regime of
explosive
reheating after inflation may occur only if the amplitude of oscillation
satisfies the condition $\Phi > m/g$. Thus explosive decay ends at $\Phi
\lesssim m/g$,
i.e. at $q \lesssim1/4 $.

This means that preheating in this model
cannot begin for $\Phi < m/g$, which would correspond to the narrow resonance
regime. Narrow resonance may be important at the late stages of preheating
\cite{KLS}, but at that stage one should take into account backreaction of
the
particles produced at the previous stage of   broad parametric resonance,
so
the theory of the narrow resonance at the end of preheating is much more
complicated than the one contained in the previous subsection.

In fact, efficient preheating often requires extremely large initial values
of
$q$. Indeed, the amplitude of the scalar field decreases during the expansion
of
the universe much faster than $H^{1/4}$,
so for not very large initial values of $q$ the condition (\ref{uslovie})
becomes violated before the resonance has enough time to transfer the energy
of
the
oscillating field $\phi$ into the energy of $\chi$-particles. As we will
show
in
Sec. \ref{FIRST}, in the model under consideration preheating is
efficient only if the initial
value of $q$ at the end of inflation is very large, $q_0 \gtrsim 10^3$.

In the models
with extremely large $q$ the expansion of the universe makes preheating very
peculiar: instead of a regular resonance process we encounter a rather
unusual
effect which we call {\it stochastic resonance}.

Let us first look at the results of the numerical study of the
development of  broad resonance in an expanding universe, and try to
interpret them. Note that at this stage we do not consider the effects of
backreaction and rescattering of particles; we will discuss these effects
later. Our main strategy here is to study a general picture step by step, and
then correct it later, because otherwise the physical interpretation of the
processes which occur during preheating will remain obscure.

First of all, let us consider Eq. (\ref{38}) for the mode $\chi_k$ in
an expanding universe with $ m^2_{\chi}= 0$, $\xi = 0$ in the asymptotic regime
when $a = ({t\over
t_0})^{2/3}$, and $\Phi(t) = {M_p\over \sqrt {3\pi}mt}$. Strictly speaking,
the
last two conditions are satisfied only for sufficiently large $t$. However,
if
we begin counting time from the end of inflation, taking for definiteness
$t_0
= {\pi\over 2m}$ (which formally corresponds to the
time after a quarter of one oscillation of the field $\phi$), then we will
have
an approximation which is sufficiently good for our purposes.
With these definitions, the initial values of the field $\phi$ and   the
parameter $q$ in our
calculations are given by
\begin{equation}\label{Q}
 \phi_0 = {2 M_p\over \pi\sqrt{3\pi}} \sim {M_p\over 5}, ~~~~ q_0^{1/2} = {g
M_p \over \pi\sqrt{3\pi} m }\sim {gM_p\over 10 m} \ .
\end{equation}
On the other hand, if one wants to investigate the situation numerically, one
can simply solve a combined system of equations for $a(t)$, $\Phi(t)$ and
$\chi_k(t)$. We will not do it here because our main goal is to develop an
analytical approach to the theory of preheating.

The investigation of parametric resonance in an expanding universe can be
simplified if instead of $\chi_k$ one introduces the
function $X_k(t) = a^{3/2}(t)\chi_k(t)$, which is given by ${t\over t_o}\,
\chi_k(t) $ in our case. Then instead of
(\ref{38}) we have a much simpler equation
\begin{equation}
\ddot X_k + \omega_k^2X_k = 0\ ,
\label{39}
\end{equation}
where
\begin{equation}
\omega_k^2= {k^2\over a^2(t)}
+ g^2 \Phi^2\, \sin^2 mt + \Delta ,
\end{equation}
and
$\Delta = m_{\chi}^2 -{3 \over 4}\left({\dot a \over a}\right)^2
-{3 \over 2}{\ddot a \over a}-\xi R$.
 This term is usually very small. Indeed, we
will consider here the case of
light $\chi$ particles, such that $m_\chi \ll k_*$, in which case one can
simply neglect $m_\chi$. Also, soon after the end of inflation one has $H^2 =
\left({\dot a \over a}\right)^2 \sim {\ddot a \over a}
\ll
m^2$. As a result, typically one can neglect the term $\Delta$ altogether.
Eq. (\ref{39}) describes an oscillator with a variable
frequency $\omega_k^2(t)$ due to the time-dependence of the
background field $\phi(t)$ and $a(t)$.
 As an initial condition one should take the
positive-frequency solution
$ X_k(t) \simeq { {e^{-i\omega_kt}}/
\sqrt{2\omega_k}}$.

The series of three figures in this section shows different stages of
development of the fastest growing mode $\chi_k$ in the broad resonance regime in an
expanding universe in the theory ${m^2\phi^2\over 2}$ for an initial value
of
the parameter $q \sim 3\times 10^3$.
Note that during the expansion of the universe the amplitude of   scalar
field
oscillations decreases approximately as $t^{-1}$. Therefore in order to
illustrate the {\it relative} growth of the fluctuations of the field $\chi$
with respect to the amplitude of the oscillating field $\phi$ we show not
the
growing mode $\chi_k$ itself, but its rescaled value $X_k =
\chi_k\, {t\over t_0} $, where $t_0$ corresponds to the beginning of the
calculation. One can construct an  adiabatic invariant for Eq. (\ref{39}), which has an
interpretation of the comoving
occupation number of particles $ n_k$
in the mode $k$ in an expanding universe:
\begin{equation}\label{numberx}
 n_k={\omega_k\over 2} \left( { |\dot X_k|^2 \over \omega_k^2}
+|X_k|^2 \right) - {1 \over 2}\ .
\end{equation}
 Note  that this function does not have any factors inversely proportional to
the volume $a^3$. These factors will appear when we calculate the
number density of particles in physical (not comoving) coordinates.

\begin{figure}[t]
\centering
\leavevmode\epsfysize=10cm \epsfbox{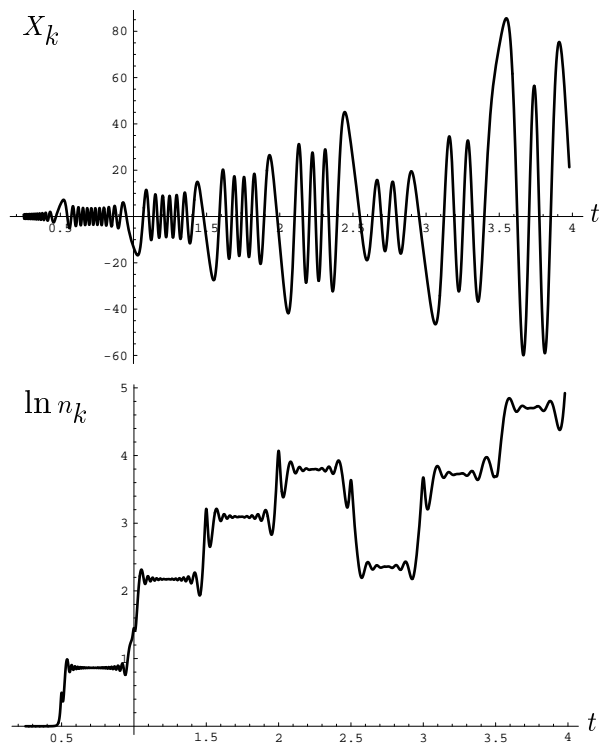}\\
\

\caption[fig4]{\label{fig4} Early stages of parametric resonance in the
theory
${m^2\phi^2\over
2}$ in an expanding universe with scale factor $a \sim t^{2/3}$ for $g =
5\times 10^{-4}$, $m = 10^{-6} M_p$. According to our conventions (\ref{Q}),
initial value of the parameter $q$ in this process is $q_0 \sim 3\times
10^3$.
 Note that the number of particles $n_k$ in this process typically increases,
but it may occasionally decrease as well. This is a distinctive feature of
stochastic resonance in an expanding universe. A decrease in the number of
particles is a purely quantum mechanical effect which would be impossible if
these particles were in a state of thermal equilibrium.}
\end{figure}

In the beginning we have parametric resonance very similar to
the one studied in the previous section, compare Fig. \ref{fig3} and Fig.
\ref{fig4}. As before, one can identify the periods when  
$\chi$-particle production is most efficient with the intervals when the
field
$\phi$ becomes small. An important difference is that because of the gradual
decrease in amplitude of the field $\phi$ the effective mass of the field $\chi$ and,
correspondingly, the frequency of its oscillations decrease in time. As a
result, in the beginning within each half of a period of oscillation of the
field $\phi$ the field $\chi_k$ oscillates many times, but then it
oscillates
more and more slowly.

To understand the rather peculiar behavior of $X_k$ and $n_k$ in this process,
let
us check in which resonance band our process develops.
The number of the band in the theory of the Mathieu equation is given by $n =
\sqrt A$. In our case reheating occurs for $A \sim 2q$, i.e. $n \sim
\sqrt{2q}
\sim {g\Phi\over \sqrt 2 m}$. Suppose we have an inflationary theory with $m
\sim
10^{-6} M_p$, and let us take as an example $g \sim 10^{-1}$. Then after the
first oscillation of the field, according to Eq. (\ref{870}), we have
$\Phi(t)
\sim M_p/20$, which corresponds to $q \sim {10^8\over 16}$. This gives the
band
number $n \sim 3\times 10^3$. After another oscillation the amplitude of the
field drops by a factor of two, and the band number decreases by a factor of
two as well, down
to
 $n \sim 1.5\times 10^3$.

In other words, even during a single oscillation the field does not remain in
the same instability band of the Mathieu equation. Instead of that it jumps over $10^3$
different instability bands! The theory of a broad resonance in
Minkowski space is much less explored than the theory of a narrow resonance,
but the theory of a broad
resonance in an expanding universe proves to be even more complicated. The
standard method
of
investigation of  resonance using the Mathieu equation in a single resonance
band completely fails here.

Still not everything is lost. Indeed, as we have found in the previous
section,
in the broad resonance regime particle production occurs only in a small
vicinity of $\phi = 0$, corresponding to integer and half-integer $N =
mt/2\pi$. Nothing depends on the exact way the field $\phi$ behaves at all
other moments. In this sense the description of the process of particle
production at $\phi = 0$ is very robust with respect to   change in the
shape
of the potential $V(\phi)$ and of the equation describing the field $\chi$, insofar
as it does not alter the behavior of either field at the stage when $\phi(t)$
approaches zero. Therefore some (but not all) of the results related to the
Mathieu equation can be useful for investigation of   broad parametric
resonance in an expanding universe even though the regime we are going to
investigate is fundamentally different.

One of the most important differences between  broad resonance in
Minkowski space and in an expanding universe can be understood by inspecting
the behavior of the {\it phase} of the functions $\chi_k$ near the points
where
$\phi(t) = 0$. Indeed, Fig. \ref{fig3} shows that near all points where $\phi
=
0$ the phases of $\chi_k$ are equal. The physical meaning of this effect is
very simple: In order to open a swinging door by a small force one should
apply
it periodically, ``in resonance'' with the motion of the door.

However, in an expanding universe such a regime is impossible, not only because of the redshift of the momentum ${k\over a}$, but mainly because the
frequency of oscillations of the field $\chi_k$ is proportional to $\Phi$,
which decreases in time.   The frequency of oscillations of the modes
$\chi_k$ changes dramatically with each oscillation of the field $\phi$.
Therefore
for large $q$ the phases of the field $\chi_k$ at   successive moments when
$\phi(t) = 0$ are practically uncorrelated with each other. Using our
analogy,
one may say that the door is vibrating with a large and ever changing
frequency, so it is very difficult to push it at a proper moment of time,
and
successfully repeat it many times in a row. That is why at some moments the
amplitude of the field $\chi_k$ decreases, see Fig. \ref{fig4}.

This could suggest that   broad parametric resonance in an expanding
universe
is simply impossible. Fortunately, this is not the case, for two main
reasons.
First of all, as we are going to show in the next section, even though the
phases of the field $\chi_k$ at the moment when $\phi(t) = 0$ in an expanding
universe with $q \gg 1$ are practically unpredictable, in 75\% of all events
the amplitude of $\chi_k$ grows at that time. Moreover, even if it were not
the
case, and the amplitude would grow only in 50\% of all events, the total
number
of $\chi$-particles would still grow exponentially. Indeed, as we will see,
during each ``creative moment'' $\phi(t) = 0$ in the broad resonance regime
the number of particles at each mode may either decrease by a factor of O(10),
or grow
by a factor of O(10). Thus if we begin with 10 particles in each of the two
modes, after the process we get 1 particle in the first mode and 100
particles in the second. Therefore the total number of particles in this
example grows by more than a factor of 5. The theory of this effect is very
similar to
the theory of self-reproduction of an inflationary universe, where in most
points
the inflaton field rolls down, but those parts of the universe where it jumps
up continue growing exponentially \cite{book}.

As a result, parametric resonance does take place. However, in order to
describe it some new methods of investigation of parametric resonance should
be developed. We will do this in the next section.

\begin{figure}[t]
\centering
\leavevmode\epsfysize=9cm \epsfbox{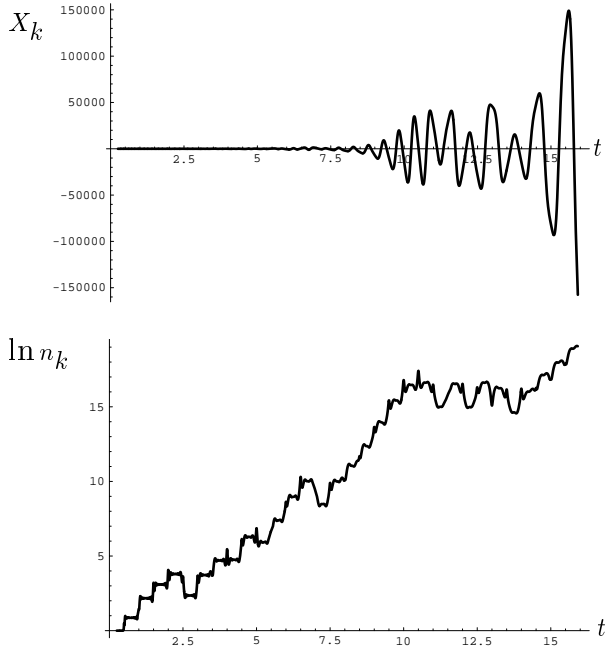}\\

\caption[fig5]{\label{fig5} The same process as in Fig. \ref{fig4} during a
longer period of
time.
The parameter $q = {g^2\Phi^2\over 4 m^2}$ decreases as $t^{-2}$ during this
process, which gradually makes the broad resonance more and more narrow. As
before, we show time $t$ in   units of ${2\pi\over m}$, which corresponds
to
the number of oscillations of the inflaton field. }
\end{figure}

Stochastic resonance occurs only during the first part of the process, when
the
effective parameter $q$ is very large and the resonance is very broad.
Gradually the amplitude of the field $\phi$ decreases, which makes $q$
smaller. Expansion of the universe slows down, 
 the field stays in each resonance band for
a longer time, and eventually the standard methods of investigation based on the Mathieu
equation
become useful again.  As we will show in Sec. \ref{BROADEXPAND}, stochastic resonance ends  and the standard methods become useful after the first  $ {q_0^{1/4}/\sqrt{2\pi}}$ oscillations, which may happen even before the effective parameter $q$ decreases from $q_0 \gg 1$ to  $q \sim 1$, see Eq. (\ref{random}). One of the manifestations of the transition from the stochastic resonance to a regular one is a short
plateau
for $\ln n_k$ which appears in Fig. \ref{fig5} for $10 \lesssim t \lesssim
15$.
This plateau corresponds to the time when the   resonance is no longer stochastic, and
the
mode $X_k$ appears in the region of stability, which divides the second and
the
first instability band of the Mathieu equation, see Fig. \ref{Math}.

\begin{figure}[t]
\centering
\leavevmode\epsfysize=9cm \epsfbox{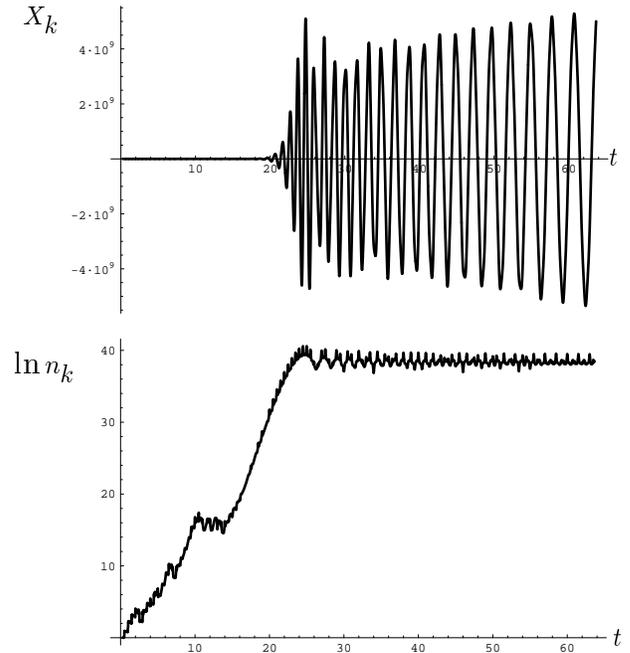}\\
\

\caption[fig6]{\label{fig6} The same process during a longer time, which is
shown in the units ${mt\over 2\pi}$, corresponding to the number of
oscillations $N$. The figures show the growth of the mode $X_k$
for
the momentum $k$ corresponding to the maximal speed of growth of $n_k$. In this
particular case $k \sim 4 m$. Towards
the end of this period, after approximately 25 oscillations of the inflaton
field, the resonance ceases to exist, and the occupation number $n_k$ becomes
constant. }
\end{figure}

\begin{figure}[t]
\centering
\leavevmode\epsfysize=5.6cm \epsfbox{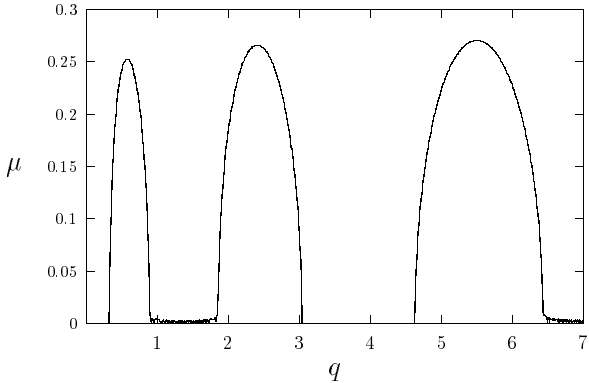}\\
\vskip 0.1cm

\caption[Math]{\label{Math} ~The structure of the resonance bands for the
Mathieu equation along the line $A = 2q$, which correspond to excitations
with
$k = 0$ in our model. The modes with small $k$ are especially interesting
because the momenta of the excitations are redshifted during the expansion of
the universe. A small plateau at $10 \lesssim t \lesssim 15$ on Fig.
\ref{fig5} corresponds to the time where stochastic resonance ceases to
exist,
all modes are redshifted to small $k$, and the system spends some time in the
interval with $1\lesssim q \lesssim 2$, which is outside the instability
zone.
The last stage of the resonance shown in Fig. \ref{fig6} corresponds to the
resonance in the first instability band with $q < 1$.}
\end{figure}

To get a better understanding of this effect one should continue our
calculations for a longer period of time, see Fig. \ref{fig6}. At $t > 15$
the
process does not look  like a broad resonance anymore, but the amplitude still
grows exponentially at a rather high rate
until the amplitude of the field $\Phi$ becomes smaller than $m/g$, which
corresponds to $q \sim 1/3 - 1/4$. Soon after that the resonance ceases to
exist and the amplitude stabilizes at
some constant value.

The time $t_f$ and the number of oscillations $N_f$ at the end of 
parametric resonance in an expanding universe can be estimated by finding the
moment when
$g\Phi \approx {gM_p\over 3 mt}$ is equal to $m$:
\begin{equation}\label{ENDRES}
t_f \approx {gM_p\over 3m^2}, \qquad N_f \approx {gM_p\over 6\pi m}\ .
\end{equation}
As one can check, this estimate for our case ($m = 10^{-6} M_p$, $g = 5\times
10^{-4}$) gives $N_f \sim 26.5$, which is in good agreement with the
results
of our computer calculations shown in Fig. \ref{fig6}. A small
disagreement
(about 10\%) appears because our criterion for the end of the resonance $g\Phi
\sim m$ was not quite precise: the resonance ends somewhat earlier, at $g\Phi
\sim 1.1 m$.

This more exact result can be deduced from Fig. \ref{Math}, which shows that
the first instability band for $k =0$ extends from $q \sim 0.8$ to $q \sim
1/3$. Therefore the growth of all modes with $k \ll m$ terminates not at
$g^2\Phi^2/4m^2 \sim 1/4$, but slightly earlier, at $g^2\Phi^2/4m^2 \sim 1/3$.

At the time $t \sim t_f/2$ one has $q \sim 1$. During the time from
$t_f/2$
to $t_f$ the resonance occurs in the first resonance band, the resonance is
not
very broad and there are no stochastic jumps from one resonance band to
another. At the time just before $t_f/2$ there was no resonance; the field
was
in the stability band between $q = 1$ and $q = 2$, see Fig. \ref{Math}.

 An interesting effect which is shown in Fig. \ref{fig6} is a slow growth of
the
amplitude $X_k$ which continues even after the resonance terminates and $n_k$
becomes constant. This happens because the momentum of each mode gradually
becomes smaller due to the expansion of the universe, and this leads to a
growth
of
$\chi_k$ even though $n_k$ does not change. This is one of the examples which
shows that in order to describe parametric resonance one should use proper
variables such as $n_k$, because otherwise one may get the incorrect idea that
the resonance continues even for $t > 25$.

If one ignores a small island of stability near $t \sim 12$, one may conclude
that during the main part of the process the slope of the curve
$\ln n_k$ remains almost constant. In our case this corresponds to
the exponential growth of the occupation number $n_k$ with an effective
parameter $\mu_k \sim 0.13$. This fact will be very useful for us later, when
we will calculate the number of particles produced during the parametric
resonance. Such a calculation is our main goal. It is also necessary in order
to verify whether one should modify our resonance equations due to the
presence of $\chi$-particles. As we will see, no modifications are needed
for theories with $g \lesssim 3 \times 10^{-4}$. However, for greater values
of
$g$ (and in particular for the case of $g \sim 5\times 10^{-4}$ discussed
above)
the resonance ends in a somewhat different way, see Sec. \ref{SECOND}.

In order to illustrate the stochastic nature of the resonance in this theory,
we will present here at sample of results for the resonance for several
different
values of the coupling constant $g$ in the interval from $0.9 \times 10^{-4}$
to
$10^{-3}$. One might expect   the results to change
monotonically as $g$ changes in this interval. However, this is not the case.
The table contains the results concerning the initial momentum $k$ (in units of
$m$)
corresponding to the fastest growing mode, the total increase of the number
of
particles $\ln n_k$ at the end of the resonance for this mode, the average
value $\mu$ for this mode, and the time $t_f$ (the number of oscillations of
the field $\phi$) at the end of the resonance:

\vskip 0.5cm

\begin{center}
 \begin{tabular}{|c|c|c|c|c|c|}
\hline
 ~~g ~~& ~~$k$~~ & ~~$\mu$~~ & ~~$t_f$~~ &~~ $\ln n_k$~~ \\
\hline\hline
 ~~$0.9 \times 10^{-4}$ & $ 1.5$ & $0.1$ & $5 $&~~ $6$~~ \\
\hline
{}~~$10^{-4}$~~~&~~~$ 2$~~~& ~~~~0.14~~~~&~~$5$~~&~~ $9$~~ \\
\hline
 ~~$1.1\times 10^{-4}$ & $ 0.5$ & $0.17$ & $5.5$&~~ $12$~~ \\
\hline
 ~ $1.2\times 10^{-4}$ & $1.5$ & $~0.12$~ & $6$&~~~$9$~~ \\
\hline
 ~ $1.3\times 10^{-4}$ & $1$ & $~0.13$~ & $6.5$&~~~$11$~~ \\
\hline
 ~ $1.4\times 10^{-4}$ & $2$ & $~0.12$~ & $7$&~~~$11$~~ \\
\hline
 ~ $1.5\times 10^{-4}$ & $0.5$ & $~0.18$~ & $7$&~~~$17$~~ \\
\hline
 ~ $2\times 10^{-4}$ & $3.5$ & $~0.12$~ & $11$&~~~$16$~~ \\
\hline
 ~ $3\times 10^{-4}$ & $0.5$ & $~0.14$~ & $14$&~~~$27$~~ \\
\hline
 ~ $5\times 10^{-4}$ & $4$ & $~0.13$~ & $24$&~~~$40$~~ \\
\hline
 ~ $10^{-3}$ & $6$ & $~0.12$~ & $48$&~~~$75$~~ \\
\hline
\end{tabular}\\

\end{center}
\vskip 0.3cm

Thus we see that the leading mode in this interval of the coupling constant
has initial
momentum comparable to $m$ and slightly smaller than the typical initial width
of the resonance $k_*/2$, which changes from $2 m$ to about $5 m$ for $g$
changing from $10^{-4}$ to $10^{-3}$. The reason why $k$ is usually (though
not
always) somewhat smaller than $k_*/2$ is very simple. The resonance is broad
only during the first half of the time. Narrow parametric resonance which
appears during the second part of preheating typically is more efficient for
smaller $k$. We should note  that for $g \gtrsim 3 \times 10^{-4}$, at the
last stage of preheating one should take into account backreaction of produced
particles, which makes the narrow resonance stage very short, see Sec.
\ref{SECOND}. In such a case the resonance has the width $k_*/2$ in terms of
the value of the momentum $k$ at the beginning of preheating.

Of course, investigation of the leading growing mode is insufficient: One
should integrate over all modes with all possible $k$, which we are going to do
later. However, the number of particles $n_k$ is exponentially sensitive to
$k$. Therefore the main contribution to the integral will be given by the
trajectories close to the leading one. It is similar to what happens, e.g.,
in the theory of tunneling, where one first finds the optimal trajectory
corresponding to the minimum of action, and calculates $e^{-S}$ along this
trajectory. Similarly, one can calculate the rate of growth of the total number
of $\chi$-particles by finding the leading trajectory and calculating the
average value of $\mu$ along the trajectory.

The table clearly demonstrates that the effective values of $\mu$ and
especially the final number of particles $n_k$ produced by the resonance are
extremely sensitive to even very small modifications of $g$, and change in a
rather chaotic way even when $g$ changes by only 10\%. That is why we call
this process ``stochastic resonance.'' We see from the table that for $g\sim
10^{-3}$ the
occupation numbers $n_k$ become incredibly large. It will be shown in Sec.
\ref{FIRST} that for
$g\sim 10^{-4}$ backreaction of created particles is not very important, but
for $g \gtrsim 3\times 10^{-4}$ backreaction
becomes crucial, because it does not allow the resonance to produce
an indefinitely large number of particles. To investigate these issues we
should
first develop the theory of stochastic resonance, and then take into account
backreaction.

\section{\label{ANALYTIC} Analytic Theory of Stochastic
Resonance}

In this section we are going
to develop a new method to study the time
evolution of
the eigenfunctions $\chi_k(t)$ in the most interesting case
of   broad resonance. This method is based on the crucial
 observation made
 in the previous sections: In the broad resonance regime
the evolution of the modes $\chi_k(t)$ is adiabatic and the number of
particles
does not grow in the intervals when $|\phi(t)|> \phi_*$. The number of
particles changes only in the short intervals when $|\phi(t)|\lesssim \phi_*
\ll \Phi$.

The quantum field theory of particle creation in a
 time varying background is naturally formulated
in terms of adiabatic (semiclassical) eigenfunctions.
This formalism is introduced in the next subsection.
Then we will find the change of the particle number density
from a single kick, when $\phi(t)$ crosses zero at some time $t_j$.
For this purpose it is enough to consider the evolution of $\chi_k(t)$
in the interval when $\phi^2(t)$ is very small, so it can be represented by
its quadratic part $\propto (t-t_j)^2$. This process looks like
 wave propagation in a time dependent parabolic potential.
We can combine the action of the subsequent
parabolic potentials to find
the net effect of the particle creation.
Using our formalism, we consider a toy model of   broad resonance
in Minkowski space, and  broad resonance in an expanding universe,
which turns out to have a stochastic nature.

\subsection{\label{WKB}Adiabatic representation of the eigenfunctions}

The semiclassical, or adiabatic evolution of the
eigenfunction $\chi_k(t)$ can be
represented in a specific mathematical form. For
this we adopt a physically transparent method to treat  
Eq. (\ref{39})
 for an arbitrary time dependence of the classical
background field which was originally developed by
 Zeldovich and Starobinsky \cite{ZS} for the problem of particle creation in
a varying gravitational field.

Let us represent solutions of Eq. (\ref{39}) as   products of its
solution  in the adiabatic approximation, $\exp{( \pm i\int dt\
\omega)}$, and some functions $\alpha(t)$ and
$\beta(t)$:
\begin{equation}
a^{3/2}\chi_k(t) \equiv X_k(t) =
{\alpha_k(t)\over \sqrt{2\omega}}\ e^{- i\int^t
\omega dt}
 + {\beta_k(t)\over \sqrt {2\omega}}\ e^{+ i\int^t \omega
dt} \ .
\label{61}
\end{equation}
An additional condition on the functions $\alpha$
and $\beta$ can be imposed by taking the derivative
 of Eq. (\ref{61})
as if $\alpha$ and $\beta$ were time-independent.
Then Eq. (\ref{61}) is a solution of Eq.
(\ref{39}) if the functions
$\alpha_k, \beta_k$ satisfy the equations
\begin{equation}\label{63}
\dot \alpha_k = {\dot \omega \over 2\omega} e^{+ 2 i\int^t \omega
d t}\ \beta_k\ ,~~\ \
\dot \beta_k = {\dot \omega \over 2\omega} e^{- 2 i\int^t
\omega d t}\ \alpha_k \ .
\end{equation}
In terms of classical waves of the $\chi$-field,
quantum effects occur due to  departure from the initial
positive-frequency solution,
therefore the initial conditions at $t \to 0$ are
$\alpha_k = 1$, $\beta_k = 0$.
Normalization gives
 $\vert \alpha_k \vert^2 - \vert \beta_k \vert^2 = 1$.

 The coefficients $\alpha_k(t)$ and
$\beta_k(t)$ in our case coincide with the coefficients of the
Bogoliubov
transformation of the creation and annihilation operators,
 which
diagonalizes the Hamiltonian of the $ \chi$-field
 at each moment of time $t$.
The particle occupation number is $n_k= \vert \beta_k \vert^2$,
see Eq. (\ref{numberx}).
The vacuum expectation value 
 for the particle number density per comoving volume is
\begin{equation}
\langle n_{\chi}\rangle = {1 \over 2\pi^2 a^3}
 \int\limits_0^{\infty}
dk\, k^2 \vert \beta_k \vert^2\
{ }.
\label{number1a}
\end{equation}
In this section we will calculate $\beta_k$, $n_k$ and
$\langle n_{\chi}\rangle$ in the non-perturbative regime of
 broad resonance, where all of these values can be very large.

It is instructive
 to return in the framework of this formalism
 to the simpler perturbative regime which we discussed earlier in Sec.
\ref{OSCILLATIONS}.
 Assuming   $|\beta_k| \ll 1$,
 from Eqs. (\ref{63}) one can
obtain an iterative solution: 
\begin{equation}
\beta_k \simeq
{\textstyle {1 \over 2}} \int\limits_{0}^{t} d t'\,{\dot \omega \over
\omega}\, \exp{\bigl( -
 2i \int\limits_{0}^{t'}d t'' \omega( t'')\bigr)}\ .
\label{BET}
\end{equation}
Using $\omega(t)=\sqrt{({k \over a})^2 + g^2\Phi^2 \sin^2 mt}$, we can
evaluate
Eq. (\ref{BET}) containing an oscillating integrand
by the method of stationary phase \cite{st81}. In the case of the massive
scalar field
decaying via the interaction $g^2 \sigma \phi \chi^2$,
 the dominant contribution is given by the
integration near the moment $t_k$, where $a(t_k)={2k \over m}$.
As we already mentioned, this corresponds to the creation of
a pair of massless $\chi$-particles with momentum $k ={ 1\over 2} a(t_k)m$
 from an inflaton with  mass  (energy)  $m$
 at the instant $t_k$ of the resonance
 between the mode $k$ and the background field.
 The decay rate of the inflaton field calculated with this method
can be described by Eq. (\ref{7}).

For the interaction $\phi \phi \to \chi \chi$, the process in the regime  $|\beta_k| \ll 1$ can be interpreted as  creation of
a pair of $\chi$-particles with momentum $k = a(t_k) m$
 from a pair of massive inflatons with   energy $m$ each.
The decay rate of the massive inflaton field in this case
 rapidly decreases with the expansion of the universe as
${1\over a^4}{d \over dt}(a^4
\rho_{\chi}) \propto a^{-6}.$ Therefore a complete
decay of the massive inflaton field in the theory with the
$\phi^2\chi^2$-interaction
 is impossible. One should have additional terms such as $ g^2\sigma\phi\chi^2$
or $h \bar \psi\psi\phi$. This is a very important conclusion which we already discussed in Sec. \ref{OSCILLATIONS}.

\subsection{Interpretation of parametric resonance in terms of successive
scattering on parabolic potentials}\label{PAR}

We suggest a new analytic method to solve
approximately the basic equations
(\ref{38}) and  (\ref{39}) for the
eigenfunctions $\chi_k$ which correspond to the $\chi$-particles
created by the oscillating inflaton field $\phi(t)$.
This method is rather general; it can be applied to many models
of preheating.
One may also apply it to the idealized case when the universe does not expand
and  backreaction is not taken into account. In the
cases where the equation for the modes $\chi_k$ can be reduced to an
equation with periodic
coefficients (including the Mathieu equation), our
method accurately reproduces the solution of this equation,
and gives us an interesting insight into the physics of  parametric
resonance.
This method is rather powerful; it enables one to investigate some features
of the regime of  broad parametric resonance which, to the best of our
understanding, have not been known before.

In the
realistic situation which we study in this paper, when the
expansion of the universe as well as the
backreaction are taken into account, in some models
(e.g. non-conformal theory) the
 equation for the modes $\chi_k$ cannot be considered as an equation
with
 periodic coefficients,
and the analysis based on standard stability/instability charts
 is not applicable.
This is the situation where our method will be especially useful.

Let us consider the general equation (\ref{39}).
As we noticed, the eigenfunction $X_k(t)$ has
 adiabatic evolution
 between the moments $t_j$, $j=1,2,3, ...$, where
the inflaton field is equal to zero
$\phi(t_j)=0$, (i.e. twice within a period of 
inflaton oscillation). The
 non-adiabatic changes of $X_k(t)$ occur only
 in the vicinity of $t_j$.
Therefore we expect that the semiclassical solution (\ref{61})
of Eq. (\ref{39}) is valid everywhere but around
 $t_j$.
 Let the wave $X_k(t)$
have the form of the adiabatic solution (\ref{61})
before the scattering at the point $t_j$
\begin{equation}
X_k^{j}(t) ={{\alpha_k^{j}} \over \sqrt{2\omega}}\,
 e^{-i\int_0^t \omega dt} +
{{\beta_k^{j}} \over \sqrt{2\omega}}\, e^{+i\int_0^t \omega dt} \ ,
\label{WKB1}
\end{equation}
the coefficients $\alpha_k^{j}$ and $\beta_k^{j}$
are constant for $t_{j-1} < t < t_j$.
Then after the scattering, $X_k(t)$,
 within the interval
 $t_j < t < t_{j+1}$, has the
form
\begin{equation}
X_k^{j+1}(t) ={\alpha_k^{j+1} \over \sqrt{2\omega}}\,
 e^{-i\int_0^t \omega dt} +
{\beta_k^{j+1} \over \sqrt{2\omega}}\,e^{+i\int_0^t \omega dt} \ ,
\label{WKB2}
\end{equation}
and the coefficients $\alpha_k^{j+1}$ and $\beta_k^{j+1}$
are constant for $t_{j} < t < t_{j+1}$.

Eqs. (\ref{WKB1}) and (\ref{WKB2}) are essentially the
 asymptotic expressions
for the incoming waves (for $t < t_j$) and
for the outgoing waves (for $t > t_j$ ), scattered
at the moment
 $t_j$. Therefore the outgoing amplitudes
 $\alpha_k^{j+1} $, $ \beta_k^{j+1} $ can be expressed through
 the incoming
amplitudes
$\alpha _k^{j}$, $ \beta_k^{j}$ with help of the reflection
 $R_k$ and transmission $D_k$
amplitudes of scattering at $t_j$:
\begin{equation}\label{matrix}
\pmatrix{\alpha_k^{j+1} e^{-i \theta_k^{j}}
 \cr \beta_k^{j+1} e^{+i \theta_k^{j}}
 \cr } =
\pmatrix{ {1 \over D_k} & {R^*_k \over D^*_k} \cr
 {R_k \over D_k} & {1 \over D^*_k} \cr}
\pmatrix{\alpha_k^{j}e^{-i \theta_k^{j}} \cr \beta_k^{j}
e^{+i\theta_k^{j}} \cr}
 \ .
\end{equation}
Here $\theta_k^{j}=\int\limits_0^{t_j} dt~ \omega(t)$ is the phase
accumulated by the moment ${t_j}$.

Now we specify the scattering at the moment $t_j$.
The interaction term $g^2\phi^2(t)$ in Eq.  (\ref{39})
 has a parabolic
form around all the points $t_j$:
 $g^2\phi^2(t) \approx g^2\Phi^2m^2 (t-t_j)^2 \equiv k_*^4(t-t_j)^2$,
where the current amplitude of the
fluctuations $\Phi$ is defined in (\ref{870}), and
the characteristic momentum $k_*=\sqrt{g\Phi m}$.
In the general case $k_*$ depends on time via the time dependence of
$\Phi \propto a^{-3/2}$.
Figure 8 illustrates two possible outcomes of the scattering of the wave $X_k(t)$ on the parabolic potential near zeros of
the function $g^2\phi^2(t) $. Depending on the phase of the incoming wave, the corresponding number of particles   may either decrease or grow. 
\vskip -1cm

\begin{figure}[t]
\centering
 \hskip -1.3 cm
\leavevmode\epsfysize= 8cm \epsfbox{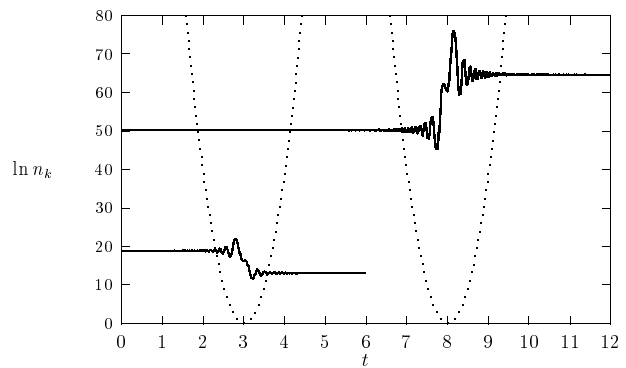}\\
\vskip -1cm
 \caption[fig7]{\label{fig7}
 The change of the comoving particle number $n_k$ due to  
scattering at the parabolic potential,
calculated from Eq.  (\ref{parab1}). The dotted lines show the sequence of the parabolic potentials $g^2\phi^2(t) \approx g^2\Phi^2m^2 (t-t_j)^2$ where scattering occurs.
 Time is given in units of ${2\pi \over \kappa}$.
The number of particles can either increase or decrease
at the scattering, depending on the phase of the
incoming wave }
\end{figure}

First, let us consider the mode equation
around a single parabolic potential.
In the vicinity of $t_j$
 the general equation (\ref{39}) is transformed to the equation
\begin{equation}
{d^2 X_k \over dt^2} + \left( {k^2 \over a^2} +
 g^2\Phi^2m^2 (t-t_j)^2 \right) X_k =0 \ .
\label{parab}
\end{equation}
For simplicity we introduce a new time variable
$\tau=k_* (t-t_j)$ and a scaled momentum $\kappa={k \over {a k_*}}$.
Notice that $\kappa^2= (A_k -2q )/2\sqrt{q}$.
In general, $k_*$ and $\kappa$ depend on $t_j$ through $a(t_j)$,
and should be marked by the index $j$, which we drop for the moment.
Then Eq.  (\ref{parab}) for each $j$ is reduced to the simple
equation
\begin{equation}
{d^2 X_k \over d{\tau}^2} + \left( \kappa^2 +
 \tau^2 \right) X_k =0 \ .
\label{parab1}
\end{equation}
The asymptote of this equation, which corresponds
to the incoming wave, is matches to the form (\ref{WKB1}).
The asymptote corresponding to the outgoing wave 
 matches  the form (\ref{WKB2}).
Therefore   the reflection
 $R_k$ and transmission $D_k$
amplitudes of scattering at $t_j$
are essentially the reflection and transmission
amplitudes of scattering at the parabolic potential.
 Thus the problem is reduced to the well-known problem
of  wave scattering at a (negative) parabolic potential
\cite{LL}, which we consider in the next subsection.

\subsection{\label{ParabB}Particle creation by parabolic potentials}

 A general analytic solution of Eq. (\ref{parab1})
is the linear combination of the
 parabolic cylinder functions \cite{AS}:
 $ W \left(-{\kappa^2 \over 2} ; \pm\sqrt{2} \tau \right)$.
The reflection $R_k$ and transmission $D_k$
amplitudes for  scattering on the parabolic potential
can be found from these analytic solutions:
\begin{equation}
R_k=- { i e^{ i \varphi_k} \over {\sqrt{1+e^{\pi \kappa^2} }} } \ ,
\label{R}
\end{equation}
\begin{equation}
D_k= { e^{-i \varphi_k} \over {\sqrt{1+e^{-\pi \kappa^2}}}} \ ,
\label{D}
\end{equation}
where the angle $\varphi_k$ is
\begin{equation}
\varphi_k= \arg
 \Gamma \left({1+i\kappa^2 \over 2}\right)+{\kappa^2\over 2}\left(1+ \ln{2\over \kappa^2}\right) \ .
\label{angle}
\end{equation}
The angle $\varphi$ depends on the momentum $k$.
Notice the following properties of these coefficients:
$R_k=-iD_ke^{-{{\pi \over 2} \kappa^2}}$,
$|R_k|^2+|D_k|^2=1$.
Substituting (\ref{R}) and (\ref{D}) into (\ref{matrix}), we
can obtain
the evolution of $\alpha _k^{j}$, $ \beta_k^{j}$ amplitudes
from a single parabolic scattering
in terms of the parameters of the parabolic potential
and the phase $\theta_k^{j}$ only.

The mapping of $\alpha _k^{j}$, $ \beta_k^{j}$
into $\alpha _k^{j+1}$, $ \beta_k^{j+1}$ reads as
\begin{eqnarray}\label{matrix1}
&&\pmatrix{\alpha_k^{j+1} \cr \beta_k^{j+1} \cr } = \\
 &&
\pmatrix{ \sqrt{1+e^{-\pi \kappa^2}} e^{i\varphi_k} &
 ie^{-{{\pi \over 2} \kappa^2} +2i\theta_k^{j}} \cr
 -ie^{-{{\pi \over 2} \kappa^2} -2i\theta_k^{j}}
 & \sqrt{1+e^{-\pi \kappa^2}} e^{-i\varphi_k} \cr}
\pmatrix{\alpha_k^{j} \cr \beta_k^{j} \cr}
 \ . \nonumber
\end{eqnarray}

Since the number density of $\chi$-particles with momentum ${\bf k}$
is equal to $n_k=|\beta_k(t)|^2$, from Eq. (\ref{matrix1}) one can calculate
the number
density of outgoing particles $n^{j+1}_k= |\beta^{j+1}_k|^2$ after the
scattering on the parabolic
potential out of $n^{j}_k= |\beta^{j}_k|^2$ incoming particles:
\begin{eqnarray}
n^{j+1}_k &=& e^{-\pi \kappa^2} + \left( 1 +2 e^{-\pi \kappa^2} \right)
n^{j}_k\nonumber\\
 &-&
2e^{-{\pi \over 2} \kappa^2}\sqrt{1+ e^{-\pi \kappa^2}}
\sqrt{ n^{j}_k (1 + n^{j}_k)}
 \sin \theta_{tot}^{j} \ ,
\label{particles}
\end{eqnarray}
where the phase $\theta_{tot}^{j} = 2\theta_k^{j}-\varphi_k + \arg
\beta_k^{j}-\arg \alpha _k^{j} $.

Before we apply the formalism (\ref{matrix}) and (\ref{particles}) to
specific models, we shall analyze these generic equations.
Although we did not specify yet the phase
 $\theta_{tot}^{j}$, we already can learn a lot from
the form (\ref{particles}).
First of all, the number of created particles is a step-like function of
time.
The value of $n^{j}_k$ is a constant between two successive
scatterings at points $t_j$ and $t_{j+1}$. The number of particles is changed
exactly at the instances $t_j$ in a step-like manner, in full agreement
with the exact numerical solution, see Figure 4.
The effect of particle  creation is significant if $\pi
\kappa^2 {\
\lower-1.2pt\vbox{\hbox{\rlap{$<$}\lower5pt\vbox{\hbox{$\sim$}}}}\
} 1$, otherwise the exponential term $e^{-\pi
\kappa^2}$
suppresses the effect of particle  accumulation. This gives us the important
general criterion
for the width of the resonance band \cite{KLS}:
\begin{equation}
\kappa^2= {{A-2q} \over 2\sqrt{q}} \leq \pi^{-1} \ ,
\label{criterion}
\end{equation}
where $A = {k^2\over a^2m^2} + 2q$, $q = {g^2\Phi^2\over 4 m^2}$. Equivalently,
one can write this condition in the following form:
\begin{equation}
{k^2\over a^2} \leq k_*^2/\pi = g m \Phi/\pi . \
\label{criterion1}
\end{equation}
This estimate of the resonance width $k \lesssim k_*/\sqrt\pi$ practically
coincides with the estimate $k \lesssim k_*/2$ (\ref{adiab5})
derived in Sec. \ref{BROAD} by elementary methods.

Next, let us consider the large occupation number limit, $n_k \gg 1$.
 From Eq. (\ref{particles}) we derive
\begin{equation}
n^{j+1}_k \approx \left(
1 +2 e^{-\pi \kappa^2} -
2\sin \theta_{tot}^{j}~
e^{-{\pi \over2} \kappa^2}~\sqrt{1+ e^{-\pi \kappa^2}}
\right)n^{j}_k \ .
\label{particles2}
\end{equation}

The factor in the r.h.s. of this equation depends
on the coupling constant $g$ through $\kappa^2 \propto g^{-1}$.
This dependence has the structure $ \exp {(-1/g)} $, which is
 a non-analytic function of $g $ at $g= 0$.
Therefore the number of particles generated in the broad resonance regime
cannot be derived using a perturbative series with respect to
coupling parameter $g$.
Thus formula (\ref{particles2})
 clearly manifests the
non-perturbative nature of the resonance effects.

The growth index $\mu_k$ is defined by the formula
\begin{equation}
n^{j+1}_k =n^{j}_k \exp( 2\pi \mu_k^{j}) \ .
\label{index}
\end{equation}
Comparing (\ref{index}) and (\ref{particles2}) we find
\begin{equation}
\mu_k^{j}={1 \over 2\pi} \ln \left(
1 +2 e^{-\pi \kappa^2} -
2\sin \theta_{tot}^{j}~
e^{-{{\pi \over2} \kappa^2}}~\sqrt{1+ e^{-\pi \kappa^2}}
\right) \ .
\label{mu}
\end{equation}

The first two terms in Eq. (\ref{particles2}) correspond to the effect of
spontaneous particle creation, which always increases the number of
particles.
The last term corresponds to   induced particle creation, which can
either increase or decrease the number of particles.
At first glance it looks paradoxical that
the number of particles $n^{j}_k$ created from the
time-varying external field can not only increase but
sometimes decrease, i.e. the growth index $\mu_k$ can be not only
positive but sometimes negative.
 Indeed, it is well known that if the
$|in\rangle$-state of the quantum field $ \chi$ corresponds to $n$ particles,
then the number of particles in the $|out\rangle$-state due to the
interaction with the external field will always be greater than $n$.
This is how to resolve the paradox:
the particles created
from the vacuum by the time-varying external field
 are not in the $n$-particle $|in\rangle$-state but are
in the squeezed $|in\rangle$-state. In this case the interference
of the wave functions can lead to a decrease of the
particle number.

The whole effect of the particle production
crucially depends on the interference of the wave functions, i.e.
the phase correlation/anticorrelation between successive scatterings
at the parabolic potentials.
The maximal value of $\mu$ is reached for  positive interference when $\sin \theta_{tot} =-1$ and is equal
to
 $\mu={1 \over \pi} \ln \left(1 +\sqrt{2}\right) \approx 0.28$,
see also \cite{KLS}, \cite{Yosh}.
The typical value of $\mu$ corresponds to $\sin \theta_{tot} =0$ and is
equal to
$\mu={1 \over 2\pi} \ln 3 \approx 0.175$.
The value of $\mu$ is negative for  negative
 interference when $\sin \theta_{tot} =1$.
Therefore the behavior of the resonance essentially depends on the
behavior of the phase $\theta_k^{j}$ as a function of
$k$ for different time intervals $j$, see Fig. 8.
In the case of a fixed amplitude of the background field
 $\Phi(t)=const$ and $a(t)=const$, the phases $\theta_k^{j}$
do not depend on time but only on $k$. In this case we expect
the existence of separate stability and instability $k$-bands.
However, this separation is washed out as soon as
the phases $\theta_k^{j}$ are significantly varying with time due to
 changes in the parameters of the background field,
for instance, in $\Phi(t)$ and $a(t)$.

Now we estimate the net effect of particle  creation after
a number of oscillation of the inflaton field.
Eqs. (\ref{matrix}) and (\ref{particles}) are recurrence
relations for the
$\alpha^{j}_k$ and $\beta^{j}_k$ coefficients and for the number of
particles $n^{j}_k $ after successive actions of the parabolic
potentials centered at
$t_1, \ t_2, \ ... $.
 To find the number of particles created up to the
moment\, $t_{j}$,
one has to repeat the formulas $j$-times for the initial values
$\alpha^0_k=1$,
$\beta^{j}_k=0$, $n^0_k=0 $ and a random initial phase
$\theta^0_k$.

After a number of inflaton oscillations,
the occupation number of $\chi$-particles is
\begin{equation}
n_k(t)={ 1 \over 2}~ e^{ 2\pi\sum_j \mu_k^{j}} \approx
{ 1 \over 2}~ e^{ 2m\int^t dt \mu_k(t)} , \
\label{index2}
\end{equation}
where we convert the sum over $\mu_k^{j}$
 to an integral over $ \mu_k(t)$.
In some cases
the index $\mu_k(t)$ does not depend on time. In a more general case
one can replace $\mu_k(t)$ by an effective index $\mu_{k}^{\rm eff}$
 defined by the
relation
 $ \int^t dt \mu_k(t) = \mu_{k}^{\rm eff} t$, which, for brevity, we will
write simply as $\mu_k t $.
Then
the total number density of created
particles is given by
\begin{equation}
n_\chi(t)={1 \over (2\pi a)^3}\int d^3k~n_k(t) =
{1 \over 4 \pi^2 a^3}~ \int dk~ k^2
 e^{ 2m \mu_{k} t} \ .
\label{total}
\end{equation}
The function $ \mu_{k}$
has a maximum $\mu_{\rm max} \equiv \mu$ at some $k=k_m$. The integral
(\ref{total})
can be evaluated by the steepest descent method:
\begin{equation}
n_\chi(t) \simeq {1 \over 4\pi^2 a^3}~ {{ k_m^2 ~ e^{2 \mu m t}}
 \over
{ \sqrt{ 2\pi\,mt\,\mu_{k}^{''} } }} \simeq
{1 \over 8\pi^2 a^3}~
 { { \Delta k ~ k_m^2 ~ e^{2\mu m t}}
\over { \sqrt{\pi \mu m t} }} \ .
\label{totalx}
\end{equation}
where $\mu_{k}^{''}$ is the second derivative of the function
$\mu_{k}$ at $k=k_m$ which we estimated as
 $\mu_{k}^{''} \sim 2\mu /{\Delta k^2}$,
${\Delta k}$ being the width of the resonance band.
Thus the effect of particle creation is defined
by the leading value of the growth index $\mu$, by
the leading momentum $k_m$ and by the width of the resonance band
${\Delta k}$. In practice typically $k_m \sim {\Delta k} \sim k_*/2$,
so we can use an estimate
\begin{equation}\label{ESTN}
n_\chi(t) \sim
{k_*^3 \over 64\pi^2 a^3 \sqrt{\pi \mu m t} } ~
 e^{2\mu mt} \ .
\end{equation}
In order to calculate $n(t)$ one should find the values of the parameters
$\mu
$ and $k_*$.

In what follows in this section we will apply the
 general formalism of successive parabolic potentials
first to the toy model without the expansion of the universe,
where $a(t)=const$ and $\Phi(t)=const$, in the case of broad
resonance, $q \gg 1$. We will find the resonance zones and
the number of particles which would be created in such a model.
Then we consider a realistic case with the expansion of the universe
taken into account. It turns out that the resonance in an expanding universe
is very different from that without expansion.

\subsection{\label{BROADMINK} Broad parametric resonance without expansion
 of the universe}

Let us apply the general formalism of the previous subsection
to the toy model neglecting the expansion of the universe. This is equivalent
to taking $a(t) = 1$. Thus, we will study the evolution of the eigenfunctions
in the case with  fixed values of the background parameters
and without backreaction of created particles.
In this case   Eq. (\ref{38}) is reduced to the
standard Mathieu equation
(\ref{M1ux}) with $A_k= {k^2 \over m^2 }+2q$, $q = {g^2\phi^2\over
4m^2} $, $z= mt$.

As we saw in Sec. \ref{STOCHASTIC},
for  the realistic situation with the expansion of the
universe the Mathieu equation is applicable
only at   the last stages of the resonance   when $q \leq 1$.
 For   $q \gg 1$
this equation has only a
  heuristic meaning for our problem.

For the Mathieu equation with a large value of $q$ (which is a
constant in this subsection) we have the broad resonance regime.
In this case  the parameters $\kappa^2$ and $\varphi_k$
of   matrix (\ref{matrix1}) are time-independent, i.e.
 they are the same for different $j$.
The phase $\theta_k^j$ is simple: $\theta_k^j=\theta_k \cdot j$.
Here $\theta_k=\int_{t_{j-1}}^{t_j} dt~ \omega_k $ is the phase
 accumulating between two successive zeros of $\phi(t)$,
i.e. within one half of a period of the inflaton oscillations,
$\pi/m$, so that $\theta_k=\int_0^{\pi \over m} dt~ \omega_k $.
To find $\alpha_k^{j}$ and $ \beta_k^{j}$
 we have to apply  the same matrix (\ref{matrix1}) $j$ times.
We are mainly interested in the regime with a large number of
created particles, $n_k^j= \vert \beta_k^{j}\vert^2 \gg 1$.
In this regime $ \vert \alpha_k^{j}\vert \approx \vert \beta_k^{j}\vert$,
so $\alpha_k^{j}$ and $ \beta_k^{j}$ are
distinguished by their phases only.
In this case there is a simple solution of the
matrix Eq. (\ref{matrix1}) for an arbitrary $j$:
\begin{equation}\label{a1}
\alpha_k^{j}= { 1 \over \sqrt{2}}\cdot e^{(\pi \mu_k + i\theta_k)\cdot j }
 \ ,
\end{equation}
\begin{equation}\label{a2}
\beta_k^{j}= { 1 \over \sqrt{2}} e^{i\vartheta}
 \cdot e^{(\pi \mu_k - i\theta_k)\cdot j} \ ,
\end{equation}
where  ${\vartheta}$ is a  constant phase.
In principle, it is possible to construct not only the
 asymptotic solution (\ref{a1}), (\ref{a2}), but the general
solution which starts with $ \beta_k^{0}=0$. However,
the general solution very quickly converges to the simple solution
(\ref{a1}), (\ref{a2}), which contains all the physically relevant
information.
 From (\ref{a2}) the number of particles created by the time
 $t \approx {\pi j \over m}$ is
\begin{equation}\label{asymp}
n_k={ 1 \over 2}~e^{2\pi \mu_k \cdot j} ={ 1 \over 2}e^{2\mu_k m t} \ ,
\end{equation}
where $ \mu_k $ from (\ref{a1}), (\ref{a2}) is indeed
the growth index.
Substituting the solution (\ref{a1}), (\ref{a2}) into
Eq. (\ref{matrix1}), we get  a  complex
 equation for the parameters $\mu_k$ and $\theta_k$
\begin{equation}\label{cond1}
 e^{(\pi \mu_k + i\theta_k)}=
 \sqrt{1+e^{-\pi \kappa^2}}e^{-i\varphi_k}
+ie^{-{\pi\over 2} \kappa^2-i {\vartheta}} \ .
\end{equation}
Alongside  the solution (\ref{a1}), (\ref{a2}),
there is another asymptotic solution of the matrix equation
 (\ref{matrix1})
\begin{equation}\label{b1}
\alpha_k^{j}= {1 \over \sqrt{2}}
\cdot e^{(\pi \mu_k + i\theta_k +i \pi)\cdot j } \ ,
\end{equation}
\begin{equation}\label{b2}
\beta_k^{j}= {1 \over \sqrt{2}} e^{i\vartheta} \cdot
 e^{( \pi \mu_k - i\theta_k- i \pi )\cdot j} \ ,
\end{equation}
with the condition
\begin{equation}\label{cond2}
- e^{( \pi \mu_k + i\theta_k)}=
 \sqrt{1+e^{-\pi \kappa^2}}e^{+i\varphi}
+ie^{-{{\pi \over 2} \kappa^2}-i {\vartheta}} \ .
\end{equation}
 Excluding the phase ${\vartheta}$  from
 the complex equations (\ref{cond1}) and (\ref{cond2}),
it is easy to find a single equation for the growth index $\mu_k$
valid for  both solutions:\footnote{Notice that the number of particles
calculated with
Eq. (\ref{cond4}) is in an agreement with the general
formula (\ref{particles2}).
 From the definition of $\theta_{k,tot}$ and the solutions
$\alpha_k^{j}$ and  $\beta_k^{j}$  we have
  $\theta_{k,tot}=\varphi_k-\vartheta$. Therefore
from the complex equations (\ref{cond1}) and (\ref{cond2})
 we have additionally that $\cos {(\theta_{k,tot})}=
\sqrt{1+e^{\pi \kappa^2}}\sin {(\theta_k-\varphi_k)}$.}

\begin{eqnarray}\label{cond4}
 e^{\pi \mu_k}&=&| \cos (\theta_k- \varphi_k)|\sqrt{1+e^{-\pi
\kappa^2}}\nonumber\\
&+&\sqrt{(1+e^{-\pi \kappa^2}) \cos^2 (\theta_k- \varphi_k) -1 } \ .
\end{eqnarray}
In the instability bands, the parameter $\mu_k$
in  Eq.  (\ref{cond4}) should be real.
Therefore, the
 condition for the momentum $k$ to be in the resonance band is
$\cos (\theta_k- \varphi_k) \geq 1/\sqrt{(1+e^{-\pi \kappa^2})}$, or
\begin{equation}\label{cond3}
| \tan (\theta_k- \varphi_k)| \leq e^{-{{\pi \over 2} \kappa^2}} \ .
\end{equation}

To further analyze the constraints on the width  (\ref{cond3})
 and strength (\ref{cond4}) of the resonance,
we should find how the phases $\theta_k$ and $\varphi_k$
depend on the momentum $k$.
 The angle $\varphi_k$ as a function of $k$
is defined by   Eq. (\ref{angle}).
For the phase $\theta_k$ we have
\begin{eqnarray}\label{phase}
\theta_k&=&\int_{0}^{\pi \over m} dt \sqrt{ {k^2 }+g^2 \phi^2(t)}
\nonumber\\
&\approx&
{{2g \Phi} \over m}+{\kappa^2\over 2} \left( \ln {{g\Phi} \over {m\kappa^2}}
+4\ln2 + 1 \right)\\
 &= &
4\sqrt{q}
+  {k^2\over 4\sqrt{q} m^2}  
 \left( \ln {\Bigl[ 4q \left( {{ m} \over k}\right)^2 \Bigr] }
+4\ln2 + 1 \right) \ .\nonumber
\end{eqnarray}
To obtain  these estimates we used the condition that
 $\kappa^2 \ll {{g \Phi} \over m}$
for the resonant modes.
In Eq. (\ref{phase}) we presented $\theta_k$ in two equivalent forms: first
in terms of the
physical parameters $g$, $\Phi$, and  $\kappa$, and second in
 terms of the parameters $q$ and  $k$.
Combining Eqs. (\ref{phase}) and (\ref{angle})
 for the phases $\theta_k$ and $\varphi_k$,
we can find how $ \theta_k - \varphi_k$ depends on $k$.
The leading term in $ \theta_k - \varphi_k$
 for large values of $q$ is the term
${{2 g\Phi }\over m}=4\sqrt{q}$ which does not depend on $k$.
Substituting $ \theta_k - \varphi_k$
into Eq. (\ref{cond3})
 we get the equation  for the
width of the resonance explicitly in terms
of $k$ for a given parameter $q$.
Eq. (\ref{cond3}) transparently
shows the presence of a sequence of stability/instability bands
as a function of $k$.
Typical half-width of a resonance band is $k^2 \sim 0.1 k_*^2$.
 Substituting $ \theta_k - \varphi_k$
 into Eq.  (\ref{cond4}),
we find the strength of the resonance as a function of $k$.
The effect of amplification is not a monotonic function of
$q$. The strongest amplification
is realized for   discreet values of the parameter $q$:~
$q=\left({ n \pi \over 4}\right)^2 $, where $n$ is an integer.
For this case $\mu_k$ has a maximum at $k=0$.
We can illustrate our results graphically for this case, since
the function $\arg \Gamma ({1+i\kappa^2\over 2})$ involved in the expression for $\varphi_k$ (53) has a particularly simple form for
 $\kappa^2 \ll 1$:
\begin{equation}\label{angle2}
 \arg
 \Gamma \left({1+i\kappa^2 \over 2}\right)\approx -0.982~ \kappa^2 \ .
\end{equation}
Then we have
\begin{equation}\label{phase3}
 \theta_k - \varphi_k \approx
4\sqrt{q}
+ {k^2\over 8\sqrt{q} m^2}  
 \bigl(  \ln  q  
+9.474 \bigr) \ .
\end{equation}
The function $\mu_k$ derived with the formulas (\ref{cond4})
 and (\ref{phase3})
is plotted in Fig. 9 for $q=\left({32 \pi}\right)^2$.
We also plot  $\mu_k$ derived numerically from the
Mathieu equation (\ref{Mu}). We conclude that the predictions of
the analytic theory
developed here for the Mathieu equation with large $q$
are rather accurate.

 \begin{figure}[t]
\centering
\leavevmode\epsfysize=5.9cm \epsfbox{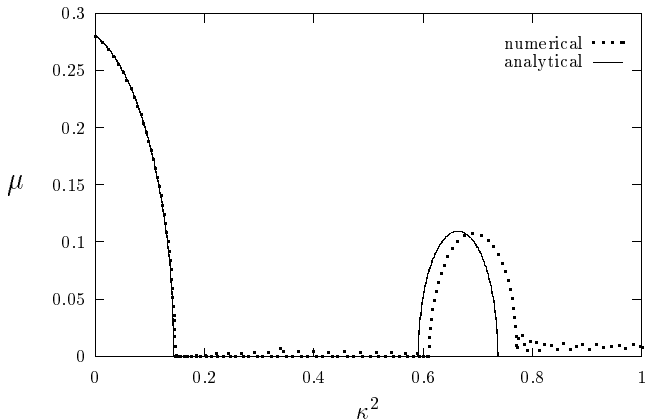}\\
\

\caption[fig8]{\label{fig8}
The characteristic exponent $\mu_k$ of the
Mathieu equation (\ref{Mu}) as a function
of $\kappa^2\equiv {k^2\over k_*^2}$ 
for   $q=\left({32 \pi }\right)^2$.
The dotted curve is obtained from a numerical solution. Two instability
bands are shown. The solid curve for these instability bands
was derived analytically
with Eqs. (\ref{cond4}) and (\ref{phase3})
  where the simple
approximation (\ref{angle2}) was used. The numerical and analytical results  are  in a perfect agreement for the first band where the approximation (75) is accurate. By improving expansion (75), one can reach similar agreement for the higher bands as well.}
\end{figure}

\subsection{\label{BROADEXPAND} Stochastic resonance in an expanding
universe }

Let us consider the creation of  $\chi$-particles by  harmonic
oscillations of the inflaton  field in an expanding universe.
Due to the expansion of the universe,  there are few complications
in  Eq.  (\ref{38}) for the modes $\chi_k$ in an expanding universe
in comparison
with the Mathieu equation. The effect of the term $3H \dot \chi$ can be
eliminated
by using $X_k=a^{3/2}\chi_k$, see Eq.  (\ref{39}).
The redshift of momenta $k \to {k \over a(t)}$ should be taken into account,
especially at the latest stages.
The most important change is the time-dependence of the
  parameter $q={{ g^2 \Phi^2} \over 4 m^2}$:
 $q \propto t^{-2} \propto N^{-2} $. For the broad resonance case where
$q \gg 1$,
this parameter significantly varies within
a few inflaton oscillations; hence, the concept of the static
 stability/instability chart
of the Mathieu equation cannot be utilized
in this important case.

Surprisingly, the most interesting case when the parameter $q$
is large and time-varying can also
 be treated analytically
by the method of  successive parabolic scatterings.
Indeed, the matrix mapping for the
$\alpha_k^j$ and $\beta_k^j$ developed in 
 subsections B and C is also valid
in the case of an expanding universe.
Let us consider the phase accumulating between two successive
zeros of the inflaton field:
\begin{eqnarray}\label{random1}
\theta_k^j&=&\int_{t_j}^{t_{j+1}} dt \sqrt{ {k^2 \over a }+g^2 \phi^2(t)}
\nonumber\\
&\approx&
{{2g \Phi} \over m}+{\kappa^2\over 2} \left( \ln {{g\Phi} \over {m\kappa^2}}
+4\ln2 + 1 \right)\\
 & \approx &
{{g M_p} \over {5 m j}}+ O( \kappa^2) \ ,\nonumber
\end{eqnarray}
where we used Eq. (\ref{870}) for the amplitude of oscillations, $\Phi$,
as a function of the number of oscillations, $j \approx 2N$. If the initial
value    ${{g M_p} \over {10 m }} \equiv \sqrt{ q_0}$   is
large, then variation of the phase
$\delta \theta_k^j$  between successive scatterings due to the
$j$-dependence is $\delta \theta_k^j \simeq {{g M_p} \over {5 m j^2}}$,
or in terms of the number of oscillations
\begin{equation}\label{var}
\delta \theta_k \simeq {{g M_p} \over {20 m N^2}}={{ \sqrt{q_0}} \over 2N^2} \ .
\end{equation}
The crucial observation is the following:
for large initial values of $q$, the phase variation
$\delta \theta_k$ is much larger than $\pi$
 for all relevant
$k$. Therefore, all the phases $\theta^{j}$ in Eqs.
(\ref{matrix})
and
(\ref{particles}) in this case can be considered to be random numbers.
For given $q_0$, the phases are random for the first
\begin{equation}\label{random}
N_{\rm stoch} \simeq {q_0^{1/4}\over \sqrt{2\pi}}
\end{equation}
oscillations. For example, for $q_0=10^6$ the phases are random
for the first dozen oscillations, and for $q_0=10^9$, neglecting backreaction
effects, the phases would be
random for the first hundred oscillations.  During this time each mode experiences chaotic behavior in the standard terms of the theory of chaotic systems \cite{CHAOS}: a small change in the values of parameters and/or initial conditions can lead  to large changes in the final results. 

We will show in Sec.
\ref{SECOND}  that  the  backreaction of created particles leads to an
exponentially  rapid decrease of $q$ down to $q \sim 1/4$ at the last moments
of preheating. This means that the
parameter $q$ in this regime remains very large and phases remain random until the very last
stages of preheating.

The stochastic character of the phases, $\theta_k^j$,
significantly simplifies the analysis of the matrix equation
 (\ref{matrix}).
Indeed, since there is no memory of the phases,
 each mapping   can be considered as
independent  of the previous ones.

As we see  in  Eq. (\ref{particles2}), the number of
created particles depends on the
phase $\theta_{tot}=\varphi_k +2 \theta_k^j +\arg \beta_k-
\arg \alpha_k$. In principle, from the matrix equation
 (\ref{matrix}) one can derive a series of equations which allow one
to express the phases $\arg \beta_k$, $ \arg \alpha_k$, and eventually
 $\theta_{tot}$ through the random phase $\theta_k^j$.

For qualitative analysis we simply assume that
$\theta_{tot}$ is a random phase.
As a result the number of particles $n_k^{j+1}$
 obeys the  recursion equation
\begin{equation}
n^{j+1}_k \approx \left(
1+ 2 e^{-\pi \kappa_j^2} -
2\sin \hat \theta~
e^{-{\pi \over2} \kappa_j^2}~\sqrt{1+ e^{-\pi \kappa_j^2}}
\right)n^{j}_k \ ,
\label{particles3}
\end{equation}
where $\hat \theta$ is a random phase in the interval $(0, 2\pi)$,
and $\kappa^2_j$ is slowly changing with $j$ as
 $\kappa^2_j={ k^2 \over a^2  g m \Phi}  \propto j^{-1/3}$.

Eq. (\ref{particles3}) defines the number of
particles at an arbitrary moment as a function of the random phase.
Therefore, $n^{j}_k$ is a random variable  which
 can either increase or decrease
depending on the realization of the phase.
Qualitatively, each mapping corresponds to one of  the  two
possibilities depicted in Fig. 8.
Therefore, the whole process of particle  creation is the
superposition of   elementary processes where $n_k$
jumps up or down. This explains the random behavior
of $n_k$ in Fig. 5.
On average the number of particles
is amplified with time, i.e. $n_k$ increases more often than it
decreases.

\begin{figure}[t]
\centering
\leavevmode\epsfysize=6cm \epsfbox{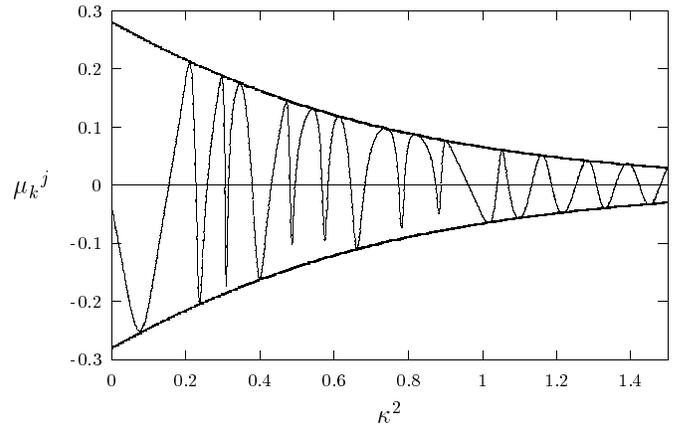}\\

\

\caption[muboth]{\label{muboth}
The characteristic exponent $\mu_k$ of the
mode Eq. (\ref{39}) in an expanding universe
 as a function
of $\kappa^2$ 
for the initial value of the parameter
 $q=\left({{32 \pi}  }\right)^2 \approx 10^4$,
 obtained from a numerical solution.
The curve is obtained at the time after the first 5   oscillations,
which corresponds to $\mu_k^j$ with $j=10$.
 The envelope of the curve is obtained from Eq. (\ref{mu1}) by taking there $\sin \hat \theta = \pm 1$. We see that there is a complete agreement between the analytical prediction of the amplitude of $\mu_k$ (\ref{mu1})  and the results of the numerical investigation.
Contrary to the static case of   Fig. \ref{fig8},
the resonance is much broader, there are no
distinguished stability/instability bands, and for certain values of momenta the function $\mu_k^j$ is negative. During the stochastic resonance regime, this function changes dramatically with every half   period of
 the inflaton oscillations.  Comparison of  Figs. \ref{fig8} and   \ref{muboth} shows that it is incorrect to use the structure of the resonance bands of the static Mathieu equation for investigation of the stage of stochastic resonance, unless one is only looking for a very rough estimate of $\mu$. }
\end{figure}

Stochastic resonance is different in many aspects from the
usual broad parametric resonance of the Mathieu equation,
considered in the previous subsection.
Let us investigate the basic features of the stochastic resonance.
First, the structure of Eq. (\ref{particles3})
does not imply the existence of separate  stability or instability bands.
Indeed, the loss of the phase interference appears for  any $k$ within the broad interval $k \leq k_*$, where
the coefficients of the mapping (\ref{particles3})
are not exponentially suppressed. Therefore, as one can see by comparison of   Figs. \ref{fig8} and \ref{muboth},  the  stochastic resonance is significantly broader (almost by an order of magnitude) than
each of the stability zones of the Mathieu equation, $\Delta k \sim k_*$.
 It makes stochastic resonance  more stable with respect to
possible mechanisms which, in principle, could terminate parametric
resonance. For instance, the conclusion that the
$g^2\phi^2 \chi^2$ interaction can terminate broad parametric
resonance in   Minkowski space-time  \cite{Prokopec}
cannot be easily generalized to the case of an expanding universe, where the
broad
resonance is stochastic and much wider.

Second, the   exponent $\mu_k$ is also a random variable:
\begin{equation}
\mu_k^{j}={1 \over 2\pi} \ln \left(
1 +2 e^{-\pi \kappa_j^2} -
2\sin \hat \theta~
e^{-{{\pi \over2} \kappa_j^2}}~\sqrt{1+ e^{-\pi \kappa_j^2}}
\right) \ .
\label{mu1}
\end{equation}
The functional form of $\mu_k$ for stochastic resonance is
different from that for broad parametric resonance.
It  changes with every half   period of
 the inflaton oscillations.
An example of $\mu_k$ calculated at  intermediate stage of 
stochastic resonance (for $j = 10$) with  the initial value of the parameter  $q \approx10^4$ is plotted in Fig. \ref{muboth}.

\

Equation  (\ref{mu1}) implies that for $\pi\kappa^2 \ll 1$ the value of $\mu_k^{j}$ is positive (i.e. the number of particles {\it grows}) for $ {\pi\over 4}< \hat\theta < {3\pi\over 4}$. This occurs for one quarter of all possible values of $\hat\theta$, in the range of $-\pi < \hat\theta < \pi$. Therefore, positive and negative occurrences
of $\mu_k$ for $\kappa \ll 1/\sqrt\pi$ appear   in the proportion $3$:$1$,
so that the probability for the  number of particles to increase is three times higher than the probability of its decreasing; see Sec. \ref{STOCHASTIC}.
Computer simulations of this process confirm this result.
However, there will be also a ``natural selection effect'': among all modes $\chi_k$ there will be some modes for which positive occurrences
of $\mu_k$ appear more often than in the proportion $3$:$1$, and these modes will give the dominant contribution to the total number of produced particles.
The typical mean value of the characteristic
 exponent is $\mu_k \sim 0.13$, but the actual number is very sensitive to even a very small change of parameters; see the table in Sec. \ref{STOCHASTIC}.
Based on the central limit theorem, we expect that the statistics
of the random variable $n_k$ obey the log-normal distribution
in the regime of the stochastic resonance.

From Eq. (\ref{mu1}) one could expect that the suppression of particle production occurs not at $\kappa^2 > \pi^{-1}$, but at $ \kappa^2 > 2\pi^{-1}$. However, the situation is more complicated. As soon as the second term under the logarithm becomes small, the probability for the  number of particles to increase becomes equal to  the probability of its decreasing, so the process of particle production becomes much less efficient.

The stochastic resonance occurs for $N_{\rm stoch}$ oscillations of the
inflaton field defined by Eq. (\ref{random}).
When  the parameter $q$ decreases because of the expansion of the universe  and
becomes smaller than $O(1)$, which happens for
$N > N_{\rm stoch}$, the resonance becomes very similar to the usual
parametric resonance  with $q \lesssim 1$.
However, at some stage it may become necessary to correct this description by
taking into account backreaction of the created particles.

 \section{\label{BACKREACTION} Resonance, Backreaction and Rescattering}

Until now we have treated  the field $\chi$
as a test field in the presence of the background fields $\phi(t)$ and $a(t)$
which have  independent dynamics. We found the effect of the
resonant amplification of $\chi_k(t)$, which  corresponds
to the exponentially fast creation of $n_{\chi}$ particles.
As we have seen, the resonance   in an expanding universe in the beginning
may be very broad, then it becomes narrow, and then eventually
disappears.

Because of the exponential instability of the $\chi$ field,
we expect its backreaction
 on the background dynamics to   gradually accumulate  until it  
affects the process of resonance itself.
Therefore the development of  resonance is divided into two stages.
At the first stage of the process, the backreaction of the created particles
 can be neglected.
 As we will see, this stage is in fact
rather long, and if the initial value of $q$ was small enough ($q_0 \lesssim
 10^3$) preheating may end
 before the backreaction becomes important (see also \cite{KhTk2}).
However, if   $q_0$ is greater than about $10^3$, then at some moment  the
description of the parametric resonance
changes. We enter the second stage of preheating where the backreaction
should be taken into account. In what follows we will treat the first and  
second stages of preheating separately.

There are several ways in which backreaction can alter the process. First of
all, interaction with particles created by parametric resonance may change
the effective masses of all particles and the frequency of oscillation of the
inflaton field. Also, scattering of the particles off each
other
and   their interaction with the oscillating field $\phi(t)$ (we will
vaguely
call both processes ``rescattering'') may lead to  additional particle
production and to the removal of   previously produced particles from the
resonance.

In our model there will be two especially important effects. First,
$\chi$-particles  may change the frequency $m$ of oscillations of the field
$\phi(t)$. This may increase the value of $m$ in
 the mode equation, which can make the resonance narrow
and eventually shut it down.

The second effect is the production of  $\phi$-particles, which occurs due to
interaction of $\chi$-particles with the oscillating field $\phi(t)$. One can
visualize this process as scattering of $\chi$-particles on the oscillating
field $\phi(t)$. In each act of interaction, each $\chi$-particle takes one
$\phi$-particle away from the homogeneous oscillating field $\phi(t)$. When
many $\phi$-particles are produced, they may change the effective mass of the
field $\chi$, making $\chi$-particles so heavy that they no longer can be
produced. Also, scattering, when it occurs for a sufficiently long time, can
destroy the oscillating field $\phi(t)$ by decomposing it into separate
$\phi$-particles.

In this section we will derive the general set of  equations
which describe  the self-consistent dynamics of the
classical homogeneous inflaton field $\phi(t)$ , as well as the fluctuations of
the fields $\chi$ and  $\phi$.
We will then discuss different feedbacks of the amplified
fluctuations. In particular, we will check the energy balance between the
background
homogeneous inflaton field $\phi(t)$, the fluctuations $\chi(t, \bf x)$,
and the fluctuations $\phi(t, \bf x)$.

\subsection{\label{self}Self-consistent evolution of $\phi$ and $\chi$ fields}

We can describe all of these effects within a full set
of self-consistent equations.
The Friedmann equation for a universe containing classical field $\phi(t)$
and particles $\chi$ and $\phi$ with densities $ \rho_{\chi} $ and
 $ \rho_{\phi} $ is
\begin{equation}
3H^2={{8\pi} \over M_p^2} \left(
{\textstyle {1 \over 2}} \dot \phi^2 + {\textstyle {1 \over 2}}m^2 \phi^2 +
\rho_{\chi}
+ \rho_{ \phi}
 \right) \ ,
\label{en}
\end{equation}
 where $\rho_{\chi}$
and $\rho_{ \phi}$ are the energy densities of $\chi$-particles and
$\phi$-particles respectively.

 The mode Eq. (\ref{39}) for $X_k(t) = a^{3/2}(t)\chi_k(t)$
now should include a term describing the coupling between $ \chi$ and
$\phi$ fluctuations:
\begin{eqnarray}\label{qmode1}
& &\ddot X_k(t) +
 \left( {k^2\over a^2}+ g^2 \Phi^2\, \sin^2 mt \right) X_k(t) =
\nonumber \\
& &
-\int dt' X_k(t') \,\Pi_{\chi} (t, t'; \bf k) \ ,
\end{eqnarray}
where the polarization operator for the field $\chi_k=a^{-3/2}X_k$ is
$\Pi_{\chi} (t, t'; {\bf k}) \equiv \int d^3 x\, e^{i\bf k( \bf x-\bf x')}
 \Pi_{\chi} (t, t'; \bf x-\bf x')$.

\begin{figure}[t]
\centering
\leavevmode\epsfysize=5cm \epsfbox{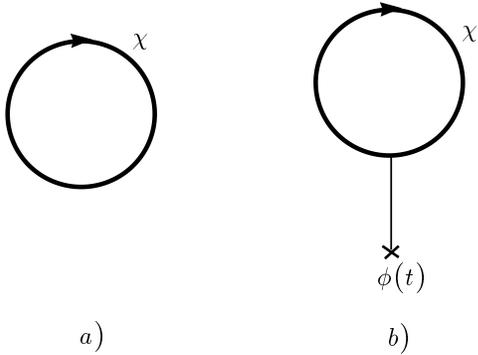}\\
\

\caption[energy1]{\label{energy1} The one-loop diagrams for the contribution
of $\chi$-particles  to the effective action
 of the field $\phi$ (Fig \ref{energy1}a) and to its derivative
$\Gamma_\phi(t)$ (Fig \ref{energy1}b). The thick line corresponds to the Green
function of the $\chi$-particles in the
external field $\phi(t)$.
}
\end{figure}

We will also
consider
quantum fluctuations of the inflaton field $ {\delta \phi}(t, \bf x)
=\phi(t, \bf x) - \phi(t) $
which can exist on  top of the homogeneous inflaton condensate $\phi(t)$.
The mode equation for $ \varphi_k(t) \equiv a^{3/2} \delta \phi_k(t)$ is
\begin{equation}\label{qmode2}
 \ddot \varphi_k(t) + \left( {k^2\over a^2} + m^2 \right) \varphi_k(t) =
-\int dt' \varphi_k(t') \Pi_{\phi} (t, t'; \bf k) \ ,
\end{equation}
where $ \Pi_{\phi} (t, t'; \bf k)$ is a corresponding polarization operator
for the field $\delta \phi_k(t) \equiv a^{-3/2} \varphi_k(t)$.
The equation for the homogeneous condensate $\phi(t)$ is
\begin{equation}\label{qcond}
 \ddot \phi (t) + 3H\dot\phi(t) + m^2 \phi(t) =
-\Gamma_\phi(t) = -\Pi^1_\phi(t) \, \phi(t) \ .
\end{equation}
Here $\Gamma_\phi(t)$ is the tadpole diagram, representing the derivative of the
effective action of the field $\phi$ (not  the decay rate!). The one-loop
diagram representing $\Gamma_\phi(t)$ is shown in Fig.
\ref{energy1}. The thick line corresponds to the exact solution of
the classical equation of motion of the field $\chi$ in the external field
$\phi $.

\begin{figure}[t]
\centering
\leavevmode\epsfysize= 9.5cm \epsfbox{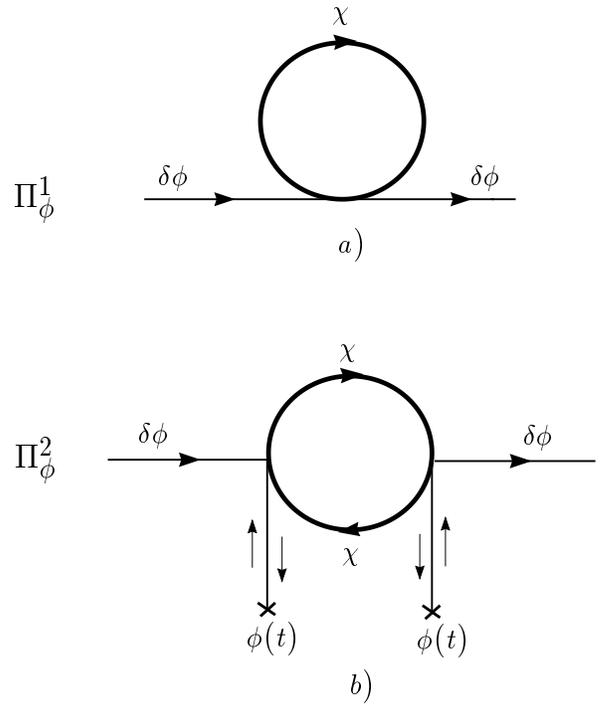}\\
\

\caption[polariz]{\label{polariz} Two diagrams for the polarization operator
of
the field $\phi$. Thin and thick lines represent the fields $\phi$ and $\chi$
respectively. The first diagram corresponds to the Hartree approximation
which
takes into account the contribution of $\langle\chi^2\rangle$.  The
contributions of these two diagrams to the effective mass of  $\phi$-particles
can be comparable
to each other.}
\end{figure}

To get an expression for the polarization operator of the field
$\phi$,
one should differentiate the effective action twice with respect to the scalar
field
$\delta\phi$. The result can be represented as a sum of two polarization operators
shown in Fig. \ref{polariz}.
$\Pi^1_\phi$   can be identified with the contribution of the fluctuations of
the field $\chi$ to the mass squared of the field $\phi$:~ $\Delta m^2 =
g^2\langle\chi^2\rangle$. Note that it is directly related to $\Gamma_\phi$:~
$\Gamma_\phi = -\Pi^1 \phi $. The polarization operator $\Pi^2_\phi$ has a more
complicated structure; it
contains an external scalar field $\phi(t)$ in each of its vertices due to
the
interaction $g^2\delta\phi\, \phi(t)\chi^2$.

The self-consistent dynamics described by Eqs. (\ref{en}) --
(\ref{qcond})
is rather complicated and not very well investigated.
There are several different approximations which can be used to solve these
equations in the context of preheating.
We will describe them in this section.

\subsection{\label{HARTREE}Hartree approximation}

The simplest way to take into account the backreaction of the amplified
quantum fluctuations $\chi$ is to use the Hartree approximation,
\begin{equation}
\ddot \phi + 3H \dot \phi + m^2 \phi +
 g^2 \langle \chi^2\rangle \phi = 0\ ,
\label{35}
\end{equation}
where the vacuum expectation value  for $\chi^2$ is
\begin{equation}
\langle \chi^2 \rangle
 = {1 \over 2\pi^2 a^3} \int\limits_0^{\infty} {dk\, k^2}
 \vert X_k(t) \vert^2 .
\label{fluc}
\end{equation}
Quantum effects contribute to the effective mass $m_\phi$ of the
 inflaton field as follows:
$m^2_\phi=m^2 + g^2 \langle \chi^2\rangle$.
The  Hartree approximation corresponds to the
first of the two diagrams of Fig. \ref{polariz}.

Initially, we have no
fluctuations $\varphi_k(t)$,
and we can use Eq. 
 (\ref{39}) for the modes $X_k$. One can express $\langle \chi^2\rangle$
in terms of the $\alpha_k(t)$ and $\beta_k(t)$
coefficients describing the resonance:
\begin{equation}
\langle \chi^2 \rangle
 = {1 \over 2\pi^2 a^3} \int\limits_0^{\infty} {dk\, k^2\over
\omega} \left( \vert \beta_k \vert^2 + \mbox{Re} \,
\Bigl(\alpha_k\, \beta^*_k \, e^{-2i\int_0^t
\omega d t}\Bigr) \right) \ .
\label{65}
\end{equation}

This formal expression may need to be renormalized. The WKB expansion of the
solution of equations
 (\ref{63})
provides a natural scheme of regularization \cite{ZS}.
However, in our case the coefficients $\alpha_k$ and $\beta_k$ of the
Bogoliubov transformation
appear due to  particle production (as opposed to  vacuum polarization), so the
integral in Eq. (\ref{65}) is finite and does not
require further regularization.

Let us estimate $\langle \chi^2 \rangle$ from Eq. (\ref{65})
using the results of the previous section.
For the resonant creation of $ \chi$ particles we have
$\vert \beta_k \vert^2 \equiv n_k \approx {\textstyle {1 \over 2}}
e^{2\mu_k\,m\, t}$,
$\mbox{Re} \,(\alpha_k\, \beta^*_k \, e^{-2i\int
\omega d t}) \approx
 \vert \beta_k \vert^2
\cos( 2{\int_0^t\omega d t} - \arg \alpha_k +\arg \beta_k)$.
For $\omega \approx g\phi(t) = g\Phi \sin mt$ the phase in this expression is
equal to ${{2g \Phi} \over m}\cos mt$
plus a small correction $O(\kappa^2)$. Due to this small correction,  the term
${{2g \Phi} \over m}\cos mt$ acquires a numerical factor
$C<1$ after the   integration $\int d^3 k$:
\begin{equation}
\langle \chi^2 \rangle
\approx {{ 1 + C\,\cos {{2g \Phi \cos mt} \over m} }
 \over {2\pi^2 a^3}}~ \int\limits_0^{\infty} {{dk\, k^2}\over
\omega} n_k \ .
\label{100}
\end{equation}
In the broad resonance case when $\phi > \phi_*$ (i.e. for most  of
the time), one has ${k \over a } \ll g\phi $, $\omega \approx g
 \vert \phi(t) \vert$,
 and therefore,
\begin{equation}
\langle \chi^2 \rangle
\approx \Bigl({ 1 + C\,\cos {{2g \Phi \cos mt} \over m} }\Bigr)
 {n_\chi\over g \vert \phi(t) \vert}\ .
\label{100a}
\end{equation}

This means in particular that in the broad resonance regime
the effective mass squared of the background field $\phi(t)$ in the Hartree
approximation
\begin{equation}\label{hartreemass}
m^2_\phi = m^2 +
  \left({ 1 + C\,\cos \Bigl( {{2g \Phi} \over m}\cos mt\Bigl) }\right)
      {gn_\chi\over \vert \phi(t)\vert  } \ ,
\end{equation}
oscillates with two frequencies. One is the frequency of
oscillation of  $\vert \phi(t)\vert$, which is equal to $2m$.
In addition,  when  $\phi(t) \not = \Phi$, the effective mass squared  $m^2_\phi$
oscillates with a very
  high frequency $\sim 2g\Phi \gg m$.  The amplitudes of both
oscillations are
 as large as the maximal value of $g^2\langle\chi^2\rangle $.
One can easily identify both types of oscillations   of   $\langle \chi^2
\rangle $
 in the numerical simulations of
 Khlebnikov and Tkachev \cite{Khleb}.

The resulting equation for the field $\phi(t)$ looks as follows:
\begin{eqnarray}\label{35u}
\ddot \phi &+& 3H \dot \phi + m^2 \phi \nonumber\\ &+&
 g {n_\chi} \Bigl({ 1 + C\,\cos {{2g \Phi \cos mt} \over m}
}\Bigr){\phi\over|\phi|} = 0\ .
\end{eqnarray}
The last term in this equation oscillates with a frequency $\sim 2g\Phi$.  In
the broad resonance regime with $g\Phi \gg m$ the high-frequency oscillation of
this term does not much affect  the evolution of the field $\phi(t)$ because
the
overall sign of the term ~$C\,\cos {{2g \Phi \cos mt} \over m}$ changes many
times during each oscillation of the field $\phi$. One may wonder, however,
whether these high-frequency oscillations may lead to a copious production of
$\phi$-particles. A preliminary investigation of this issue shows that the
quasi-periodic change of the last term in Eq. (\ref{35u})
does not lead to  parametric resonance, but a non-resonant particle production
is possible because the effective mass changes in a very nonadiabatic way:
${dm\over dt } \sim gm\Phi \sim k_*^2 \gg m^2$.

In the first approximation one may   neglect this effect and write Eq.
(\ref{35u}) as follows:
\begin{equation}\label{35uu}
\ddot \phi  +  3H \dot \phi + m^2 \phi + g {n_\chi} {\phi\over|\phi|} = 0\ .
\end{equation}

Even in this simplified  form the last term of this equation looks rather
unusual. It is not proportional to $\phi$, which would be the case if
$\chi$-particles  gave a $\phi$-independent   contribution to the
effective mass of the field $\phi$. In our case this contribution is inversely
proportional to $|\phi|$.  As a result, the   field $\phi$ behaves  as if it were oscillating in the effective potential $g {n_\chi} {|\phi|}$.

To estimate the change in the frequency of oscillations of the field
$\phi$ due to the
term $g {n_\chi} {\phi\over|\phi|}$ in Eq. (\ref{35uu}), one can neglect the
term $3H\dot \phi$ in the equation
for
the homogeneous
field $\phi$, because $H \ll m$ at the end of the first stage of preheating,
when  the term $g {n_\chi} {\phi\over|\phi|}$ becomes important. Let us find
when
the frequency increase due to the interaction with $\chi$-particles becomes
greater than the initial frequency ${m }$. In order to do this one
should solve the equation $\ddot\phi = - g n_\chi$ in the interval $0 < \phi
<\Phi$. The time during which the field $\phi$ falls down from
$\Phi$ to $0$   is $\Delta t = \sqrt{2\Phi \over
g
n_\chi}$. This time corresponds to one quarter of a period of an oscillation.
 This gives the following expression for the frequency of oscillations of the
field $\phi$ in the regime when it is much greater than its bare mass
squared $m^2$:
\begin{equation}\label{EFFMASS}
\omega_\phi = {\pi \over 2\sqrt 2} m_\phi  \approx m_\phi\ .
\end{equation}
Here $m_\phi$ is the value of the effective mass of the field $\phi$ at the
moment when    $\phi(t) = \Phi$ (the oscillations of $\langle \chi^2
\rangle$ being ignored). Therefore to estimate the
change of the frequency of oscillations of the scalar field $\phi$ one can
use the standard expression $m^2_\phi = m^2 + g^2\langle \chi^2
\rangle$
for the effective mass squared of the field $\phi$, where by $\langle \chi^2
\rangle$ one should understand its {\it smallest} value per period, which
appears for $\phi(t) = \Phi$. This implies that the
frequency of oscillations of the inflaton field does not change until the
number of $\chi$-particles  grows  to
\begin{equation}\label{stage1}
n_\chi \simeq {m^2\Phi \over g} = {2m^3 \over g^2}\, q^{1/2} \ .
\end{equation}
This is a very important criterion  which defines the duration of the first
stage of preheating where the
backreaction of the created particles can be neglected.

For future reference we include here   expressions for the energy density and
pressure of the nonrelativistic $\chi$-particles. The contribution of
$\chi$-particles to the energy
density $\rho_\chi(\phi)$ of the oscillating field
$\phi $ in terms of $\alpha_k(t)$ and $\beta_k(t)$ is given by
\begin{equation}
 \rho_{\chi}(\phi) = {1 \over 2\pi^2 a^3}
 \int\limits_0^{\infty}
dk\, k^2 \omega\ \vert \beta_k \vert^2 \ ,
\label{energy}
\end{equation}
where   $|\beta_k|^2 = n_k$. This expression does not have  any high-frequency
modulations which we have found for the Hartree term ${g^2\over 2}
\langle\chi^2\rangle \phi^2$.
 During the main part   of each oscillation
 of the field $\phi$, the field $\chi$ has mass much greater than
the range of the integration $\sim k_*$,
 which
means that $\omega \approx g|\phi(t)|$, and
\begin{equation}\label{EFFPOT'}
\rho_\chi(\phi) = {g|\phi| \over 2\pi^2 a^3 } \int\limits_0^{\infty}
dk\, k^2 \, n_k
 = g|\phi| n_\chi \ .
\end{equation}

The contribution of $\chi$-particles to  pressure in terms of $\alpha_k(t)$ and
$\beta_k(t)$    is given by
\begin{eqnarray}\label{PRESS}
p_\chi(\phi) &=&  -{ 1\over 2\pi^2 a^3} \int\limits_0^{\infty} {dk\, k^2\,
\omega }\ \Bigl[ \mbox{Re} \,
\bigl(\alpha_k\, \beta^*_k \, e^{-2i\int_0^t
\omega d t}\bigr)  +
  \nonumber\\
&+&  {k^2\over 3\omega^2} \vert\beta_k \vert^2 \Bigr]=
 -g|\phi| n_\chi \,  C \cos {{2g \Phi \cos mt} \over m}
  \ .
\end{eqnarray}
The last equality holds in the nonrelativistic limit, for $\phi \gg \phi_*$.
Average   pressure in this regime is equal to zero, as it should be  for   nonrelativistic particles.

\subsection{Is the Hartree approximation sufficient for the calculation of particle
masses?}

In the previous subsection we investigated the change of frequency of
oscillations of the classical background field $\phi(t)$ due to its interaction
with $\chi$-particles, see Eqs. (\ref{35uu}) and (\ref{EFFMASS}).  What 
about the spectra of  perturbations $\delta\phi$?  In order to answer this
question one should calculate both diagrams shown in  Fig. \ref{polariz}. The
first of these diagrams,   Fig. \ref{polariz}a, gives the same contribution
$\Pi_\phi^1 = g^2\langle \chi^2\rangle$ as the one which we already calculated
when we studied oscillations of the field $\phi(t)$.  As we have seen, in the
situation where fluctuations $\chi_k(t)$ are amplified by   resonance, even
the calculation of this simple diagram is rather nontrivial and leads to an
unusual result (\ref{35u}). The calculation of the polarization operator
$\Pi_\phi^2$,\, Fig. \ref{polariz}b, is much more involved. Similar diagrams have been ignored in all previous papers on preheating. Let us
try to understand, however, whether $\Pi_\phi^2$ can be   neglected as compared
with $\Pi_\phi^1$. A positive answer to this question would imply that the
Hartree approximation is sufficient not only for the investigation of the
oscillations of the field $\phi(t)$, but also for finding the spectrum of
perturbations of the field $\phi$.

Usually when one calculates similar diagrams at high temperature, the
polarization operator $\Pi_\phi^1$ in the high-temperature limit is
proportional to $T^2$,
whereas $\Pi_\phi^2$ is less divergent at large momenta and therefore grows
only as $T$. Therefore in the high-temperature approximation, the first
diagram, which corresponds to the Hartree approximation, gives the leading
contribution.   In our case this issue should be reconsidered because the
leading contribution to the diagrams is given by particles with large
occupation numbers and relatively small momenta.

The backreaction of created particles becomes essential only at later stages of
reheating, when, as we will see shortly, $H \ll m$. Therefore at that stage one
can neglect the expansion of the universe when calculating polarization
operators,
and it is more convenient to perform all calculations in terms of the usual,
physical (rather than comoving) momenta $k$ and the modes $\chi_k(t)$.
Therefore throughout the rest of the paper we will use  physical momenta,
$k$, $p$, etc. During the last stages of reheating they remain almost
constant, but in order to relate them to the original physical momenta for each
mode $\chi_k$ one should remember that physical momenta are redshifted as
$a^{-1}(t)$.

 To calculate $\Pi_\phi^2$ one needs to know the  Green function of the field
$\chi$ in an external field $\phi(t)$, which is given by
\begin{equation}\label{G}
G_{\chi}(x, x') = \int d^3k\  T \bigl[\chi_k(t)\chi_k(t') \bigr] e^{i \bf k(\bf
x -\bf x')},
\end{equation}
where $T$ stands for time-ordering.
The calculation of the diagram for $\Pi_\phi^2$, Fig. \ref{polariz}b,
using this Green function for the internal lines of the field $\chi$ is rather
tedious. Therefore, we will make certain simplifications.
Consider the broad resonance regime $q \gg 1$ at a time when $\phi(t) \gg
\phi_*$. At this stage there is no particle production,
and the adiabatic form (\ref{61}) can be used for the eigenfunction
$\chi_k(t)$. Consider a time interval $\Delta t < m^{-1}$
near the time when the inflaton field
$\phi(t)$ reaches its maximum, $\Phi$. During this short interval, one
can neglect the expansion of the
universe and the change of the field $\phi(t)$, i.e. one may take $\phi(t)
\approx
\Phi$. The Green function in the space-time representation
consists of two parts. The first part is similar to the
standard Green function in   Minkowski space in the fixed background field
$\phi$. The second part contains the high frequency modulation
$e^{i\omega (t+t')}$. Both terms are of the same order.
One can show that in this
regime the first term in the expression for the Green function (\ref{G})
has a simple form in the momentum representation:
\begin{equation}\label{G1}
G(k) = {i\over k^2-m^2_\chi} + 2\pi n_k \delta (k^2 - m^2_\chi)
 \ .
\end{equation}
Here $m_\chi = g\phi(t)$, and $k$ is  a physical momentum.
 The first
term in this equation
 is the standard Green function for quantum fluctuations in the vacuum. The
second term is proportional to the occupation
number $n_k = \vert \beta_k \vert^2$ of the
$\chi$-particles.

The second part of the full Green function
 containing the modulation $e^{i\omega (t+t')}$
 does not have a simple interpretation in the
momentum representation.
 Omitting this part does not
affect the order-of-magnitude estimate of the polarization operator. This can
be most easily  seen for the diagram Fig. \ref{polariz}a,
where the
calculations are much simpler.
 Indeed, with the complete Green
function (\ref{G}) one can immediately reproduce the
result (\ref{35u}) for the diagram in Fig. \ref{polariz}a.
Meanwhile, if one uses Eq.  (\ref{G1}), then in the large $n_k$ limit one gets
the first, nonoscillating term in the
brackets of (\ref{hartreemass}):
\begin{equation}\label{P1}
\Pi^1_\phi \simeq {g^2\over (2\pi)^4 }\int d^4p\, 2\pi \delta (p^2 - m^2_\chi)
n_p
= {g n_\chi \over |\phi(t)|} \ .
\end{equation}
The part of the Green function
containing the modulation $e^{i\omega (t+t')}$
in this case gives us
the second (rapidly oscillating) term in Eq. (\ref{hartreemass}).

Thus, whereas in the first approximation one can interpret
the growing modes of the field $\chi$ during   parametric resonance as
normal
particles on the mass shell with the standard Green function (\ref{G1}), this
interpretation in general is not quite adequate and
may lead to the loss of some  terms such as the oscillating term
discussed above. Still we correctly reproduced the most important part of the
polarization operator $\Pi_\phi^1$.

 Let us  try to estimate the
polarization operator $\Pi_\phi^2$ using the simple Green function (\ref{G1})
for $|\phi(t)| \approx \Phi$. The
general structure of the
polarization operator is given by
\begin{equation}\label{G2}
 \Pi^2_\phi(k) \sim - i\ {g^4 \Phi^2 \over (2\pi)^4} \int d^4 p\, G(p)
G(p-k \pm q ) \ .
\end{equation}
 The sign of $ q$ depends on whether
the external field $\phi(t)$ brings the momentum $q_0 = m, {\bf q} = 0$
to the two vertices of the polarization operator  or takes this momentum away.

It is not our purpose now to perform a complete calculation of $\Pi^2_\phi$ in
this paper because we do not need to know the exact spectrum of perturbations
$\delta\phi$. Our main goal here is to find out whether or not $\Pi^2_\phi$ may
contain terms comparable to the Hartree operator $\Pi^1_\phi$. And indeed, if
one
calculates, for example, the diagram where the external field $\phi(t)$ brings
a momentum $q_0 = m, {\bf q} = 0$ to the first vertex and takes it away from
the second vertex, one finds (ignoring   factors $O(1)$) that this
contribution to
the real part of $\Pi^2_\phi$ for $k_0 = m,\, {\bf k } = 0$
 in the limit $n_p \gg 1$ has the same structure as $\Pi^1_\phi$:
\begin{equation}\label{G3}
 {\rm Re}\, \Pi^2_\phi \sim
 -{g^4 \Phi^2\over (2\pi)^3 } \int { n_p d^3p\over p_0~\bigl({ p_0}^2 - m^2
\bigr)}
\sim
-{g n_\chi \over \Phi } \ .
\end{equation}
Here $p_0\equiv \omega = \sqrt{{\bf p}^2 +
g^2\Phi^2 } \approx g \Phi$ for a typical resonant mode with $g^2\Phi^2 \gg
{\bf p}^2\sim {g m\Phi } \gg m^2 $.
Thus, for $|\phi(t) | \approx \Phi$ the second polarization operator of Fig.
\ref{polariz}
contains   terms of the same order of magnitude as the value of the
polarization operator in
the Hartree approximation.
This result indicates that one may need to go beyond the Hartree approximation
used in many   papers on preheating.

This result looks paradoxical. In particular, one could argue that
the Hartree approximation is closely related to the $1/N$ approximation, which
is expected to give exact results in the limit $N \to \infty$. Indeed,
instead of a single $\chi$-field  one can take
 $N$ fields $\chi_i$ with the interaction
${g^2\over 2N}\phi^2\chi_i^2$. The Hartree diagram is proportional to $g^2$,
i.e. it survives in the limit $N\to \infty$ , whereas
 the expression for the polarization operator $\Pi^2_\phi$ is proportional to
${g^4\over N}$.
That is why usually at large $N$ one can neglect contributions like
$\Pi^2_\phi$
as compared with $\Pi^1_\phi$.
 Indeed, this would be true in our
case as well if the field $\chi$ had  a large $\phi$-independent mass.
But in the theory we are discussing now its mass squared is ${g^2 \over N}
\phi^2$. As we have seen, when one calculates $\Pi^2_\phi$ this mass squared
appears
in the
denominator. As a result, the factor ${g^4\over N}$ in front of the diagram
becomes $g^2$, so that this diagram
also survives in the limit $N\to \infty$ and has
the same order of magnitude as the Hartree diagram
in the $1/N$ approximation.
 This means, in particular, that without a complete calculation of $\Pi_\phi^2$
one cannot be sure that the $1/N$
approximation gives a correct spectrum of particles in the
limit $N \to \infty$
when applied to the theory of
preheating.

To avoid misunderstandings we should reiterate that this problem appears in the
calculations of the effective masses of the $\phi$-particles but not in the
calculation of corrections to the equation of motion of the background field
$\phi(t)$, which was our main goal in Sec. \ref{BACKREACTION}.

\subsection{\label{CLASS}Classical approximation to the self-consistent
dynamics}

Fluctuations of bose fields generated from vacuum by an external field
in the large occupation number limit can be considered
as classical waves with  gaussian statistics,
see e.g. \cite{ps}. Therefore in the first approximation  all  fields $\chi$,
$\delta \phi$
can be  treated as interacting classical waves.
 This makes it possible to study  preheating by investigating a system of
nonlinear classical equations   or by lattice numerical
simulations of the interacting classical scalar fields
\cite{KhTk,KhTk2,Prokopec,Khleb}.

The Fourier decomposition of the Klein-Gordon equations of the
interacting fields can be reduced to   mode equations.
The mode equation for $X_k=a^{3/2}\chi_k$ is
\begin{eqnarray}\label{mode1}
& & \ddot X_k +
 \left( {k^2\over a^2}+ g^2 \phi^2(t) \right) X_k \nonumber\\
&&= -
{g^2 \phi(t) \over {(2\pi)^3 a^{3/2}} } \int d^3 k' X_{\bf k -\bf k'}
\varphi_{\bf k'} \nonumber\\
& &
-{g^2 \over {(2\pi a)^3 } } \int d^3 k' d^3 k'' X_{\bf k -\bf k' +\bf k'' }
\varphi_{\bf k'} \varphi_{\bf k''} .
\end{eqnarray}

The mode equation for $\delta \phi_k(t) \equiv a^{-3/2} \varphi_k(t)$ is
\begin{eqnarray}\label{mode2}
& & \ddot \varphi_{ k} + \left( {k^2\over a^2} + m^2 \right) \varphi_k =
-{g^2 \phi(t) \over {(2\pi)^3 a^{3/2}} } \int d^3 k' X_{\bf k -\bf k'}
X_{\bf k'}
\nonumber \\
& &
-{g^2 \over {(2\pi a)^3 } }
 \int d^3 k' d^3 k''~ \varphi_{\bf k -\bf k' +\bf k'' }~
 X_{\bf k'}~X_{\bf k''} .
\end{eqnarray}
The first term in the r.h.s. of this equation describes rescattering of
$\chi$-particles on the classical field $\phi(t)$, which leads to  
$\phi$-particle production. The second term describes scattering of
$\phi$-particles and $\chi$-particles. Corrections to the effective mass of the
modes $\phi_k$ appear as a result of the  iterative solution of the system of
equations which we now present.

The equation for the oscillating background field $\phi(t)$  looks as follows:
\begin{eqnarray}\label{mode2a}
& & \ddot \phi + 3H \dot \phi + m^2 \phi =
-{g^2 \phi \over {(2\pi)^3 a^3} } \int d^3 k'   X_{\bf k'}^2
\nonumber \\
& &
-{g^2 \over (2\pi)^3 a^{9/2 } }
 \int d^3 k' d^3 k''~ \varphi_{\bf k'' -\bf k'}~
 X_{\bf k'}~X_{\bf k''}.
\end{eqnarray}

The first term on the r.h.s. of this equation is proportional to the
polarization operator $\Pi^1_\phi$, which is shown in Fig. \ref{polariz}a. The
second
term describes rescattering, which is related to the imaginary part of the
polarization operator $\Pi^2_\phi$, Fig. \ref{polariz}b. Neglecting this term,
one
reproduces Eq. (\ref{35}) with
the term containing $ \int d^3 k \vert X_{\bf k} \vert^2$
playing the role of the induced mass. Thus the classical
approximation reproduces the Hartree approximation, but it also takes into
account effects related to rescattering.

In the beginning one can neglect $\varphi_k(t)$ and the corresponding
integral
terms in Eq. (\ref{mode1}).
Later, the fluctuations $X_k(t)$ are amplified by the resonance
and give rise to   $\varphi_k(t)$ fluctuations via the
integral terms in Eq. (\ref{mode2}). When the amplitude of fluctuations
 $\varphi_k(t)$ grows   significantly,
they begin to contribute to the integral terms of Eq.  (\ref{mode1}).
We will show (see Sec. \ref{VALIDITY})
 that the amplitude $\varphi_k(t)$ grows
with time as $e^{2\mu m t}$.
 Therefore the number of particles corresponding to
$\delta \phi$ fluctuations grows as $e^{4\mu mt}$ , i.e. much faster
than $n_{\chi}$.
 The interaction terms in Eqs. (\ref{mode1}) and  (\ref{mode2}) can be interpreted
as
scattering of $\chi$ particles
on the inflaton field. Because of the very fast generation of $\delta \phi$
fluctuations, $\vert \delta \phi\vert^2
\propto e^{4\mu t}$,
the process of rescattering can be very important.
However, it is not so easy to evaluate its full
  significance for the efficiency of the resonance.
For example, if the particles $\phi$ produced during rescattering have small
momenta $k$, they cannot be
distinguished from the homogeneous oscillating scalar field,
 and therefore
they
do not make any difference to the development of the
resonance, see the discussion of this issue in Sec. \ref{RESCATTERING}.
Therefore we need to know not only
how many
 $\delta \phi$ particles
are produced, but also whether they are ``hard'' particles with large momenta
or ``soft'' particles with small momenta.   We will return to this question
in Sec. \ref{RESCATTERING}.

\section{\label{FIRST} Two stages of preheating, rescattering being neglected
}

Previously, we were mainly following the evolution of each particular mode
$\chi_k$.
Now we will study their integral effect in an expanding universe.

As we have found in the previous section, the development of   broad
parametric resonance can be divided into two stages.
In the first stage  $n_\chi < {m^2\Phi \over g}$, backreaction of the
particles
$\chi$ can be neglected, and the frequency of oscillations of the field
$\phi$
is determined by its mass $m$. (We will argue later that at this stage their
scattering also does not lead to any important effects.) In the second stage
$n_\chi > {m^2\Phi\over g}$, and the frequency of oscillations of the field
$\phi$ becomes determined not by its bare mass, but by its interaction with
$\chi$-particles.
Now we will study the first and second
 stage of   broad parametric resonance.

 We begin with the first stage when the backreaction of created particles
can be neglected.
Then we consider the second stage where backreaction is important assuming
a certain hierarchy of the feedback effects:
effective mass of the inflaton is changed first, and rescattering
may become important afterwards. In this section we will neglect rescattering.
In the next section we will discuss rescattering and the validity of the
assumption mentioned above.

\subsection{\label{FIRST1} The first stage of preheating: no backreaction and
no rescattering}

In the first stage of preheating
one can ignore the backreaction of created particles on the frequency of
oscillations of the field $\phi(t)$.
 As we have found in
Sec. \ref{FIRST1}, this stage ends at the moment $t_1$ when
\begin{equation}\label{stage1a}
n_\chi(t_1) \simeq {m^2\Phi(t_1) \over g} \ .
\end{equation}
In the next section we will show that the effects related to rescattering
  also do not
alter the development of the resonance during this
stage.
 In this section we will estimate the duration of the first stage
    $t_1$, the number of  inflaton oscillations
$N_1$ at the time $t = t_1$, the number of created particles $n_\chi(t_1)$, the
 energy density of these particles $\rho_\chi(t_1)$
and the value of $\langle \chi^2(t_1) \rangle$. We will use symbols
$\Phi$, $q$ and $k_*$
without any  indices for the running (time-dependent) values of the amplitude
of the field $\phi(t)$, of
the $q$-factor, and of $\sqrt{gm\Phi(t)}$, whereas, for example, $q_0$ will
correspond to the value of $q$ at the beginning of preheating, and $q_1$ will
correspond to its value in the end of the first stage of preheating.

 One can use Eq.  (\ref{ESTN}) to estimate $n_\chi$.
First one should
determine which fluctuations $\chi_k$ are amplified during the entire period of
the
resonance. The fluctuations amplified by the broad resonance have
physical momenta
 $k \lesssim k_*/2 \sim
\sqrt {g m \Phi}/2 $, see Eq.  (\ref{adiab5}). (More precisely, one may expect $k \lesssim
k_*/\sqrt\pi$, see Eq. (\ref{criterion}).) Then the amplitude $\Phi$ in this expression
decreases as about $M_p/3mt$. Therefore, the resonance width
decreases as $k \sim t^{-1/2}$, whereas
redshift of the momenta of
previously produced particles occurs
as $a^{-1} \sim t^{-2/3}$, i.e. somewhat faster.
 (In terms of comoving momenta $k$, the resonance width
grows as $k \simeq a(t) \sqrt{g m \Phi }/2 \propto t^{1/6}$.)
This means that those
modes
which have been amplified at the first stages of the process continue to be
amplified later on. There are modes which
were outside of the resonance band in the very beginning, but
entered the resonance band later.
 However,
 after a time $\sim (2\mu m)^{-1}$ the
fluctuations which have been amplified from the very beginning will
be exponentially
larger than the ``newcomers''.
Therefore the modes which do not enter the resonance
band  from the beginning typically give a subdominant
contribution to the net effect.

Thus, with   reasonably good accuracy,
during the first stage of preheating one may consider only those fluctuations
which
have
been amplified from the very beginning.\footnote{The
total duration of the first stage of preheating in our model typically is about
$10^2
m^{-1}$. If one compares the redshifted value $k_*(t_0)/a(t)$ of the physical
momentum which was equal to $k_*(t_0) \sim {\sqrt{gm\Phi_0}} $ at the
beginning of preheating, and the running value of $k_*(t) \sim {\sqrt{gm\Phi
(t)}} $, one finds out that in the beginning these two quantities coincide,
whereas at the end of the first stage of preheating the running value of
$k_*$ is greater than
the redshifted one  by only a factor $\sim 2$. Therefore at the end of the
first stage and at the beginning of the second stage of preheating instead of
calculating the redshifted value of $k_*(t_0)/2$ one may simply use
the condition $k \lesssim k_*(t)/4$ for the growing modes.} This is important
because it
means that in all integrals one should consider only
momenta which initially, when $a(t_0) = 1$, $\Phi(t) = \Phi_0$, were in the
interval
\begin{equation}\label{KMAX}
k(t_0) \leq {k_*}(t_0)/2 \simeq {\sqrt{gm\Phi_0}}/2 \simeq {m\,
q_0^{1/4}}/\sqrt 2 .
\end{equation}
where
 $q_0 = {g^2\Phi_0^2\over 4 m^2}$.

The most important element of our calculations is the exponentially growing
occupation number of particles with $k < k_m$: $n(t) \propto e^{2\mu m t}$.
Here $\mu $ is an effective index
 which describes an average rate of growth for modes with $k \lesssim k_*$, see
Sec. \ref{ParabB}.
In our model $\mu$
depends on $g$, but not very strongly, see the table in Sec.
\ref{STOCHASTIC}. Typically it is in the range  $0.1 - 0.2$. For definiteness,
in our estimates we will use $\mu = 0.13$
which we have found numerically for a certain range of values of the coupling
constant $g$,
see the table in Sec. \ref{STOCHASTIC}.
As we will see, in the context of our approach
 an error in our estimate of $\mu$, say of 10\%,
 does not create an exponentially large error in
the final result (contrary to the remark of \cite{Boyan1});
 it only leads to an error of 10\% in the calculation
 of the duration of the first stage of preheating. Our final results will be
even less sensitive to the value of the
subexponential
factor in Eq.  (\ref{ESTN}).

Substituting Eq. (\ref{KMAX}) into
Eq.  (\ref{ESTN}), we find
\begin{equation}\label{ESTN2}
n_\chi(t) \simeq
{(gm\Phi_0)^{3/2} \over 64\pi^2 a^3 \sqrt{\pi \mu m (t-t_0) } } ~
 e^{2\mu m(t-t_0)} \ ,
\end{equation}
where $t_0$ is the beginning of the inflaton oscillation.
The convention we used in Sec. \ref{STOCHASTIC}
is that $t_0 = \pi/2m$, which gives
 $\Phi_0 \simeq M_p/5$ and $q_0 = 10^{10} g^2 $.
Our choice is also very close to the convention of ref.
\cite{Khleb}.
(This particular choice is not going to be important
because the total duration of the process is much greater than $t_0$.) With
this choice of $t_0$ we have $a(t)
 = \bigl({2mt\over \pi}\bigr)^{2/3} $.
For $t \gg t_0$ one has\footnote{Equation (\ref{number2}) is a starting point for our further estimates.  To derive this equation we used the theory of successive parabolic scatterings.  However, the general structure of Eq. (\ref{number2})  can be easily understood even without any use of this theory. As we already mentioned, the value of $\mu$ can be obtained by solving the Mathieu equation numerically in an expanding universe, see Sec. VI. One can make a simple estimate  of $\mu$  even without  using a computer. Indeed, we know that the parameter $\mu$ along the line $A = 2q$ changes from $0$ to $0.28$ \cite{KLS}. An average of these two numbers, $0.14$, provides an excellent approximation to the true value of $\mu$. }
\begin{equation}\label{number2}
n_\chi(t) \simeq 10^{-4}\ { \left({gm M_p}\right)^{3/2}\over (m
t)^{5/2} \mu^{1/2} }\
e^{2\mu m t} \ .
\end{equation}

Now we have to substitute Eq. (\ref{number2}) and $\Phi(t) \simeq M_p/3 mt$
into Eq. (\ref{stage1a}).
 The result can be transformed into an equation for $t_1$:
\begin{equation}\label{number4}
t_1 \simeq {1\over 4\mu m}\ln {10^{6}\, m\, (mt_1)^3 \over g^5
M_p } \ .
\end{equation}

An approximate solution of Eq. (\ref{number4}) for
 $\mu \simeq 0.13$ is
$t_1 \simeq {1\over 4\mu m}\ln { 10^{12} m \over g^5 M_p }$ \cite{KLS}. As we
will see soon, this is
a good estimate not only for the duration of the first stage of preheating, but
for the duration of the whole process, because the second stage of preheating
typically is rather short.

For a realistic value $m \simeq 10^{-6}M_p$ in chaotic
inflation in the theory $m^2\phi^2/2$, our estimate gives \cite{KLSSR}
\begin{equation}\label{number6}
t_1 \simeq {5\over 4\mu m}\ln {15 \over g }
\ .
\end{equation}
For instance, for $\mu = 0.13$ and $g = 0.1$ one has
$t_1 \simeq 50 m^{-1}$;
for $g = 10^{-2}$ one has $t_1 \simeq 70 m^{-1}$; for $g = 10^{-3}$ one has
$t_1
\simeq 90 m^{-1}$, etc.

The value of the field $\Phi_1 \equiv \Phi(t_1)$ at the end
 of this first stage is given by
\begin{equation}\label{number9}
\Phi_1 \simeq {M_p\over 3mt_1 }=
 {4\mu M_p\over 3}\ln^{-1} {10^{12} m \over g^5 M_p }.
\end{equation}
Another important quantity is the value of the parameter $q = {g^2\Phi^2\over
4
m^2}$ at the end of the first stage:
\begin{equation}\label{number9a}
q_1^{1/2} = {g\Phi_1\over 2 m} = {2 g\mu M_p\over 3m}\ln^{-1} {10^{12} m
\over
g^5 M_p } \ .
\end{equation}

To find the typical occupation numbers at the end of the first stage of
reheating, let us remember that $ n_\chi = {1\over 2\pi^2} \int k^2 dk\,
n_k$,
and that integration typically goes from $0$ to the physical momentum $\sim
k_*/2$. This gives an estimate
\begin{equation}\label{np}
n_k \simeq {48\pi^2 n_\chi \over k_*^3} \ .
\end{equation}

The occupation numbers of $\chi$-particles $n_k(t_1)$
by the end of
 that stage
 can be estimated as $n_k(t_1) \simeq 3\times10^2 g^{-2} q_1^{-1/4}$, see
Eqs. (\ref{index2}) and (\ref{number4}).

Using the results of this section,
 for different values of the coupling constant $g$
one can estimate the initial value $q_0$ of the parameter $q$,
its value $q_1$ at the end of the first stage of preheating,
 the value $\Phi_1$, and the number of oscillations
$N_1$
which the field $\phi$ makes from the end of inflation to the end of the
first stage. In the table below we give
 somewhat rounded numbers:

\vskip 0.5cm

\begin{center}
 \begin{tabular}{|c|c|c|c|c|c|}
\hline
 g & $q_0$ & $q_1$ & $\Phi_1/M_p$ & $N_1$ \\
\hline\hline
{}~~~$10^{-3}$~~~&~~~$ 10^4$~~~& ~~~~3~~~~&~~$3.5\times 10^{-3}$~~&~~ $15$~~ \\
\hline
 ~$10^{-2}$ & $ 10^6$ & $550$ & $5\times 10^{-3}$&~~ $11$~~ \\
\hline
 ~ $10^{-1}$ & $10^8$ & $~ 10^5$~ & $7\times 10^{-3}$&~~~$8$~~ \\
\hline
\end{tabular}\\
\end{center}
\vskip 0.3cm

The energy density at the end of the first stage is given by
\begin{equation}\label{number12}
{m^2\Phi_1^2\over 2} \simeq {8\mu^2 m^2 M_p^2\over 9}\ln^{-2} {10^{12} m
\over
g^5
M_p }.
\end{equation}
It is worth comparing the frequency of the inflaton
 oscillations $m$ with
the Hubble parameter at
that time:
\begin{equation}\label{number13}
H(t_1) \approx m \sqrt{\pi\over 3} {2 \Phi_1 \over M_p} \simeq m {8\mu \over
3
}\ln^{-1}
{10^{12} m \over g^5 M_p }.
\end{equation}
For instance, for $\mu = 0.13$, $g = 10^{-2}$, $m
= 10^{-6} M_p$ one has
\begin{equation}\label{number14}
H(t_1) \sim 10^{-2} m.
\end{equation}
Thus, at the last stages of preheating
 (though not at the beginning) one can, in the first approximation,
neglect the expansion of the universe.

 At that time, when $g^2\langle\chi^2\rangle \simeq m^2$, the total energy
density (on the r.h.s. of Eq. (\ref{en}))
 becomes approximately equally
distributed
between the interaction energy $V_\chi(\phi) = g\Phi_1 n_\chi = { m^2\Phi^2_1
} $
 and the potential energy density $m^2 \Phi_1 ^2/2$ of the
field
$\phi$.
 The kinetic
energy of $\chi$-particles  can be estimated as
$ \langle ( \nabla \chi)^2\rangle \simeq
k_*^2 \langle\chi^2\rangle \simeq g\Phi_1 m
\langle\chi^2\rangle
 \simeq m^2\Phi_1^2 \ {m\over g\Phi_1} \simeq m^2\Phi_1^2
\ q_1^{-1/2}$.

If preheating does not end with the end of the
first stage, i.e. if $q_1 \gg 1/4$, then
 the kinetic energy remains small:
$ \langle ( \nabla \chi)^2\rangle \simeq m^2\Phi_1^2\ q_1^{-1/2}
 \ll g^2\Phi_1^2 \langle\chi^2\rangle \simeq m^2 \Phi_1^2$.
However, if at the end of the first
stage $q_1 \sim 1$, then at that time a considerable fraction of the energy
of
the inflaton field will have been transformed into the kinetic energy of the
$\chi$-particles:
$ \langle ( \nabla \chi)^2\rangle
 \simeq m^2\Phi_1^2\ q_1^{-1/2}
\simeq m^2 \Phi_1^2$.

Let us find the range of values of the coupling constant $g$ for
which preheating ends during the first stage and for which investigation of
backreaction
is not necessary. Without taking account   of the
backreaction preheating ends at the time $t_f$ when $g\Phi(t)$ drops down to
$m$, which gives
$t_f \approx {gM_p\over 3m^2}$ (see Sec. \ref{STOCHASTIC}).
Therefore, preheating ends in the first
 stage if $t_f \lesssim t_1$, i.e. if
\begin{equation}\label{ENDRES2}
g \lesssim {4m\over \mu M_p} \ln {15\over g} \ .
\end{equation}
For our values of parameters this gives the condition $g \lesssim 3\times
10^{-4}$. In our convention, this corresponds to
an initial value $q_0 \lesssim 10^3$.

In this regime the total number density of $\chi$-particles created during
preheating is given by
\begin{equation}\label{number1c}
n_\chi \simeq {m^4 \over g M_p} \exp{ 2g\mu M_p\over 3m}
 \ ,
\end{equation}
and the $\chi$-fluctuations  at the end of this stage are
given by
\begin{equation}\label{number1k}
\langle\chi^2\rangle \simeq \ {m^3 \over {g M_p} } \exp{ 2g\mu
M_p\over 3m} ~.
\end{equation}
Eq. (\ref{ENDRES2}) implies that for $g\approx 3 \times 10^{-4}$ this
quantity should coincide with the value of $\langle\chi^2\rangle$ at the end
of the first stage of preheating, $\langle\chi^2\rangle = {m^2\over g^2}$.
Thus, for
 $g \approx 3 \times 10^{-4}$ one has
\begin{equation}\label{maxfluct}
\sqrt {\langle\chi^2\rangle} \simeq 3\times10^{16}~ {\rm GeV} \ .
\end{equation}

 The possibility to obtain enormously large fluctuations of the field $\chi$
is one of
the most remarkable features of preheating. For comparison, if the field $\chi$
were
in a state of thermal equilibrium, the dispersion of its fluctuations would be
given by $\sqrt {\langle\chi^2\rangle} = T/2\sqrt 3$ \cite{Kirzhnits}.
Therefore in order to obtain $\sqrt {\langle\chi^2\rangle} \sim
3\times10^{16}~ {\rm GeV} $ one would need to have $T\gtrsim 10^{17}$ GeV,
which is practically impossible in the context of inflationary
cosmology.
Here such fluctuations can be
generated prior to thermalization  due to the resonance at the
stage of preheating.
 Fluctuations
(\ref{maxfluct}) change the effective masses of particles interacting with the
field $\chi$.
 The simplest way to study this possibility is to add to our model another
scalar field $\eta$ with a potential describing symmetry breaking, for
example, $V(\eta,\chi) = \lambda [(\eta^2-\eta_0^2)^2 + \eta^2\chi^2]$. For
sufficiently small $\lambda$ this addition does not affect preheating and does
not change any of our results concerning ${\langle\chi^2\rangle}$. It is
obvious that the generation of perturbations ${\langle\chi^2\rangle}$ leads to
symmetry restoration in this model for $\eta_0 \lesssim \sqrt
{\langle\chi^2\rangle}$ on a scale up to $\eta_0 \sim 10^{16}$ GeV
\cite{KLSSR,tkachev}.
Such effects may have important cosmological implications
\cite{PhaseTr}.

Thus, we can distinguish between different scenarios
depending on the coupling constant $g$.

For $g \ll 3 \times 10^{-4}$ the broad resonance ends during the
first stage. In this case parametric resonance is not efficient enough to
transfer a significant part of the energy of the inflaton field to the energy
of $\chi$-particles. The
most important part of the process of preheating in such theories is described
by the elementary theory of reheating \cite{DL,AFW,PERT}.

 For $g \sim 3 \times 10^{-4}$, at the end of the first stage $q_1 \sim 1/4$,
and the energy becomes approximately equally distributed between the energy
of
the
oscillating scalar field $\phi$ and the energy of $\chi$-particles produced
by
its oscillations.

For $g > 3 \times 10^{-4}$ the broad resonance
continues after the end of the
first stage. To investigate the further development of the
resonance
one should study quantum effects which could
be produced by the $\chi$-fluctuations interacting with the oscillating field
$\phi(t)$.

Before doing so, let us remember that the presence of the interaction
$g^2\phi^2\chi^2$ typically leads to radiative corrections to the effective
potential of the type   ${g^4\phi^4\over 32\pi^2}\ln\phi$. For $g \gtrsim
10^{-3}$ this term becomes greater than the term $m^2\phi^2\over 2$ for $\phi
\sim
4 M_p$, when  the density perturbations determining the structure of our part
of the universe were produced.
Thus one may argue that in  models of the type considered above $g $
should
be smaller than $10^{-3}$. If $g$ belongs to the narrow interval between
$3\times 10^{-4}$ and $10^{-3}$, reheating ends soon after the end of the
first
stage, and the effects of backreaction are only marginally important. For the
description of preheating in such theories it is sufficient to use the
simple
estimates obtained in this section.

However, in
supersymmetric theories radiative corrections from bosons and fermions have a
tendency to cancel each other. In such theories the coupling constant $g$ can
be much greater than $10^{-3}$. Therefore, we will continue to
consider  all possible values of the coupling constant $g$ without assuming
that $g < 10^{-3}$.

\subsection{\label{SECOND}The second stage of preheating, neglecting
rescattering}

We defined the second stage of
preheating, $t > t_1$,
as the stage when the frequency
of inflaton oscillations due to the feedback
of amplified $\chi$ fluctuations is no longer $m$
but is determined   by the backreaction of $\chi$-particles.
In this section we will investigate preheating   neglecting
rescattering. In the next section we
will discuss the validity of this assumption.
The frequency of the inflaton oscillations
during this stage was derived in Section \ref{HARTREE} and
 given by Eq. (\ref{EFFMASS}).
Since this frequency is much greater than the bare mass $m$,
 the second stage
 is much shorter than the first one. Indeed, at this stage each oscillation
takes a
time which is much shorter  than $2\pi
m^{-1}$, whereas the number of particles, as before,  grows as $e^{4\pi\mu N}$,
where N is the number of oscillations.   Therefore the number of particles can
grow exponentially within a   time which is much shorter than $H^{-1}$. This
implies that  one can neglect expansion of the universe and the corresponding
decrease of the total energy density of matter during the second stage of
preheating.

Let us consider the inflaton oscillations $\phi(t)$
during the second stage. From Eq. (\ref{35uu}) in the limit $H \ll m_\phi$ we have
\begin{equation}\label{KG1}
\ddot \phi + m^2 \phi+ g  n_{\chi} {\rm sgn} \phi = 0 \ ,
\end{equation}
where $ {\rm sgn} \phi$ is $\pm 1$ depending on the sign of the value $ \phi $,
 $ n_{\chi}(t)$ is a function of time,
the expansion of the universe is neglected, and $ m^2 \phi \ll g^2 n_{\chi}
{\rm sgn} \phi$.
The solution of this equation $\phi(t)$ consists of a sequence of segments of
parabolas with
opposite orientation that are symmetric relative to the
 $t$-axis and match at $\phi=0$.
The   equation for the modes $\chi_k(t)$ will
contain the square of $g \phi (t)$  instead of $g^2\Phi^2\sin^2mt$.
 Thus, the behavior
of $\chi_k(t)$ for   $\phi(t)$
determined by  Eq. (\ref{KG1})
is somewhat different from the behavior of $\chi_k$ as described by the Mathieu
equation.
Nevertheless, this is not a real problem here.

Indeed, if one does not take backreaction into account, then, according to our
investigation in Sec. \ref{STOCHASTIC}, the system spends half of the time in
the broad resonance regime, and another half of the time in the regime with $q
\sim 1$, so this regime is very important. However, let us consider the effects
of backreaction. The parameter $q = g^2\Phi^2/4m_{\phi}^2$ at the second stage
can be estimated using the ``effective mass'' (or, more exactly, the frequency
of oscillations of the
field $\phi$)\, $m_\phi^2 \sim g n_\chi/\Phi \sim g^2\langle\chi^2\rangle$
(\ref{EFFMASS}). This gives
$q \sim g\Phi^3/4n_{\chi} $. The end of the resonance, as before,
occurs at $q \sim 1/4$, see below. The number of $\chi$ particles grows
exponentially, so
during the previous oscillation one had $q \sim e^{4\pi\mu}/4 \sim 1$,
and during the previous oscillation $q$ was much greater than $1$. Therefore
during
all the time except the last one or two oscillations the parameter $q$ was very
large, the resonance was very
broad, and it could be described by the theory of stochastic resonance. This
theory is very robust; it depends only on the speed of the field $\phi$ near
$\phi = 0$. Thus, the difference between the Mathieu
equation and the equation for the modes $\chi_k$ in the field $\phi(t)$ satisfying
Eq. (\ref{KG1}) in this context becomes unimportant.

On the other hand, at the time when $q$ decreases, the structure of the first
resonance band becomes important. We investigated this issue by solving
equations for the modes $\chi_k$ numerically. We found that if the
field $\phi(t)$ obeys Eq. (\ref{KG1}), the structure of the first resonance
band for $\chi_k$ at small momenta is very similar to that of the Mathieu
equation. Therefore, the second stage of preheating in this case
ends when
\begin{equation}\label{}
q = {g^2\Phi^2\over 4m_{\phi}^2} \simeq {g^2\Phi^2\over 4
g^2\langle\chi^2\rangle} \simeq {g\Phi^3\over 4n_{\chi}} \sim
1/4,
\end{equation}
just as before. This happens at some moment $t_2$ when
\begin{equation}\label{number16}
g\Phi_2 \simeq m_\phi(t_2),~~ \Phi_2 \simeq \sqrt {\langle\chi^2\rangle_2},~~
n_\chi(t_2) \simeq {g |\Phi_2|^3\over 4 }.
\end{equation}

At this time the total energy density becomes approximately equally
distributed
between the kinetic energy of $\chi$-particles $\sim {gm_\phi \Phi\over 8}
\langle\chi^2\rangle$ and the energy $  \sim g\Phi n_\chi$  of
their
interaction
with the field $\phi$   (which includes the potential energy of the field
$\phi$). This energy should be equal to the total energy of the system at the
time $t_1$, which is given by ${3 m^2\Phi^2_1\over 2}$. The final value of the
inflaton field at the end of  resonance is
\begin{equation}\label{number19}
\Phi_2 \simeq \Phi_1 q_1^{-1/4} \ .
\end{equation}
Thus, $\Phi_2$ is somewhat smaller than $\Phi_1$ for $q_1>1 $:
\begin{equation}\label{number20}
\Phi_2\simeq \sqrt{\langle\chi^2\rangle}_2 \simeq
\left({8\mu m M_p \over 3 g} \ln^{-1} {10^{12} m \over g^5
M_p}\right)^{1/2} .
\end{equation}

To find the typical occupation numbers of the modes with $k \sim k_*$ at the
end of the second stage of
reheating, let us remember that $n_k \simeq {48\pi^2 n_\chi \over k_*^3} $.
This corresponds to enormously large occupation numbers \cite{KLS}
\begin{equation}\label{number15aa}
n_k(t_2) \simeq   10^2~ g^{-2 } \ .
\end{equation}

This result indicates potential problems with the perturbative
investigation of preheating at the end of its second stage. Adding extra
internal lines of the diagrams may introduce enormous factors $n_k  \simeq
10^2~ g^{-2 } $, which may cancel extra degrees of $g^2$ which appear in the
higher order corrections.

In order to calculate the duration of the second stage let us note that
 $n_\chi(t_2) \approx n_\chi(t_1) e^{4\pi\mu N_2}$.
 One can show that
${n_\chi(t_2)\over n_\chi(t_1)} \simeq 4 q_1^{1/4}$.
Therefore, the duration of the second stage is
\begin{equation}\label{NNN}
N_2 \simeq {1\over 4\pi\mu} \ln 4 q_1^{1/4} \ .
\end{equation}
Using the table of values of $q_1$ given in the previous subsection, one can
conclude that the second stage may take from 2 oscillations (for $g =
10^{-3}$) to about 10 oscillations (for $g= 10^{-1}$). Because of the growth of the effective mass of the inflaton field, each oscillation takes much smaller time than ${2\pi\over m}$, so Eq. (\ref{number4}) for the duration of the first stage of preheating gives a good estimate for the total duration of the stage of broad resonance \cite{KLS}.

Numerical estimates of $\Phi_2\sim \sqrt{\langle\chi^2\rangle}_2$ show that
it can be in the range of $10^{15}$ to
$10^{16}$ GeV. As an example, for $g = 10^{-2}$, which corresponds to $q_0
\simeq 10^6$, one has $\Phi_2 \sim
\sqrt{\langle\chi^2\rangle}_2 \simeq 10^{16}$ GeV. An interesting
feature of Eq. (\ref{number20}) is the
inverse dependence of $\sqrt{\langle\chi^2\rangle}_2$ on the value of the
coupling constant.

Note that in addition to the high-frequency oscillations with frequency
$\sim
g\Phi$ discussed in Sec. \ref{HARTREE}, the amplitude of fluctuations
$\sqrt{\langle\chi^2\rangle}$ experiences
oscillations with a frequency $2m$. At the end of the second stage these two
frequencies coincide. In all our estimates we calculated the {\it minimal}
value of $\sqrt{\langle\chi^2\rangle}$ which occurs when
$|\phi(t)|\simeq \Phi$. It was important for us  because this is the time which
determines the frequency
of oscillations of the field $\phi(t)$. Near $\phi (t) = 0$ the amplitude of
fluctuations $\sqrt{\langle\chi^2\rangle}$ is greater than at $|\phi(t)|\simeq
\Phi$, but close to the end of the second stage of preheating this difference
becomes less significant.

 The results of numerical calculations of $\sqrt{\langle\chi^2\rangle}$
performed in \cite{Khleb} are in agreement with our estimates for $g
\lesssim 3\times 10^{-4}$ but give a few times greater value of
$\sqrt{\langle\chi^2\rangle}_2$ for larger $g$. The difference
 can be interpreted as a result of rescattering of
$\chi$-particles during the second stage of preheating.

\section{\label{RESCATTERING} Rescattering}

Theoretical considerations contained in \cite{KhTk,Son,rt,Kofman,Khleb} and
numerical
lattice simulations of preheating   \cite{KhTk,KhTk2,Prokopec,Khleb}
indicate
that there is another
effect which should be incorporated into the preheating scenario. In the
context
of the model investigated in this paper, one should consider the generation of
inflaton fluctuations $\delta \phi$ due to the interaction of $\chi$ particles
with the oscillating inflaton
field $\phi(t)$, and subsequent interaction between $\chi$ and $\delta \phi$
fluctuations.
We already discussed in   Sec. \ref{CLASS} the possibility to describe this
process by
equations for classical waves. One may also  represent the classical scalar
field as a  condensate of $\phi$-particles with zero momentum, and interpret
$\phi$-particle production as a result of  rescattering of $\chi$-particles and
the
$\phi$-particles in the condensate
\cite{KhTk,Son,Prokopec,Khleb}. This ``particle-like'' interpretation of the
interaction allows one to use the concept of cross-section of the
interacting particles, and the Boltzmann equation for the occupation numbers.

The theory of this process is rather complicated, and its interpretation in
terms of the rescattering of elementary particles is not universally valid, see
Sec. \ref{VALIDITY}. Still we can formulate the following apparently general
results. First, there is a significant generation of rapidly growing
fluctuations $\delta \phi \propto e^{2\mu m_\phi t}$ due to the interaction
between $\chi$-particles and the oscillating field $\phi(t)$. Second, the
generation of   large fluctuations of $\delta \phi$ can terminate the
resonant creation of $\chi$ particles only at the end of the second stage of
reheating. In this section we will try to justify these statements.

\subsection{\label{GENERATION}Generation of $\phi$-particles by
rescattering}

To evaluate the effects of rescattering we will use here the
``particle-like'' interpretation of rescattering. First, one should make an
estimate of the cross-section $\sigma$ for the scattering of $\chi$
particles with an effective mass $g\phi(t)$ and a typical physical momentum
$\simeq k_*/2 = \sqrt{gm\Phi}/2$ on $\phi$
particles of mass $m$ with zero initial momentum  which constitute the
oscillating field $\phi(t)$. The effective mass of the field $\chi$ is
time-dependent.
This makes investigation of their scattering rather complicated. However, in
the broad resonance regime during the main part of the oscillation (for $|\phi|
> \phi_* \simeq {1\over 3} {\Phi } q^{-1/4}$ (\ref{adiab4})), the field
$\chi$ changes adiabatically. During this time, the effective mass of the field
$\chi$ also changes adiabatically, so one may consider $\chi$-particles as
ordinary particles with an effective mass $g\phi(t)$.
 We will also consider the oscillating scalar field $\phi(t)$ as a collection
of particles $\phi$ with an effective mass $m_\phi$ and   number density
$n_\phi = m_\phi\Phi^2/2$.

We will suppose now that in such situation one can use the standard result for
the cross-section for elementary particles $\phi$ and $\chi$ with constant
masses:
\begin{equation}\label{r1}
\Bigl({d\sigma\over d\Omega}\Bigr)_{CM} = {|p_\phi| {\cal M}^2\over
64\pi^2E_{\phi} E_\chi (E_\phi+E_\chi) |v_\phi - v_\chi|} \ .
\end{equation}
Here all energies $E_{\phi}, E_\chi$ and velocities $v_\phi, v_\chi$ are
given in the center-of-mass (CM) frame and refer to the initial state, except
for $p_\phi$ which refers to the final state. ${\cal M}^2$ is the square of
the matrix element, which is given by $g^4$ \cite{Peskin}.

During most of an oscillation one has $|\phi| > \phi_* \simeq
{1\over 3} {\Phi } q^{-1/4}$, and $m_\chi = g\phi \gg k_* \sim
\sqrt{gm_\phi\Phi}$.
 In this case both the $\phi$-particles and $\chi$-particles are
nonrelativistic. If one goes to the CM frame one finds that the $\phi$
particles have a small speed $v_\phi \approx {\textstyle {1\over
2}}\sqrt{m_\phi\over g\phi} \gg v_\chi$.
Thus
$E_\phi = m_\phi$, $E_\chi \approx g\phi$. For $g\phi \gg m_\phi$ the absolute
value of
the momentum of the $\phi$-particles does not change after scattering,
$|p_\phi|
\approx {m_\phi\over 2} \sqrt{m_\phi\over g\phi} \ll m_\phi$. This gives, after
the
integration of Eq. 
(\ref{r1}) over $d\Omega$, a single particle cross-section
$ \sigma_1 \sim { g^4\over 16\pi E^2_\chi} = { g^2\over 16\pi \phi^2}$.

Now one should take into account that the actual cross-section will be much
greater because the scattering occurs not in a vacuum, but in a state which
already
contains many bosons $\phi$ and $\chi$. There are many $\chi$-particles from
the resonance and many inflaton particles $\phi$. Naively one would expect that
the cross-section should be proportional to the product of the occupation numbers
$n_p^\phi $ and $n_k^\chi$ in the final state.
 However, the corresponding terms
disappear in the collision integral
in the Boltzmann equation,
 which takes into account all the channels of scattering.
 Therefore in the investigation of enhancement of  the cross-section due to the large occupation numbers of particles in the final state, 
one should consider terms proportional either to $n_p^\chi$ or  $n_k^\phi$,
but not to $n_p^\phi n_k^\chi$. In the beginning of the process $n_\chi \gg
n_{\phi}$, and the cross-section $ \sigma_1$ should be multiplied by
$n_p^\chi \simeq {48\pi^2 n_{\chi}(t) \over k_*^3}$. This gives, for $\phi(t)
\simeq \Phi$, $ \sigma \sim { 3\pi g^2n_\chi\over k_*^3\Phi^2 }$.

 Using this result, one can estimate the time for each
$\chi$-particle to experience one scattering with a $\phi$-particle belonging
to
the oscillating field $\phi(t)$: $\tau = {1\over
\sigma n_\phi v_\phi} \simeq 0.5 { \Phi^2 \over
n_\chi}$. In particular, at the end of the first stage, $n_\chi \simeq
m_\phi^ 2\Phi_1/g $, which yields
\begin{equation}\label{r41}
\tau \simeq m_\phi^ {-1} q_1^{1/2} \ .
\end{equation}
For $g \sim 10^{-3}$ this time is of the same order as the time of one
oscillation of the field $\phi$, see the table in Sec. \ref{FIRST}. However,
just one oscillation before the end of the first stage the density of particles
was much smaller and rescattering was inefficient.
 For $g \gtrsim 10^{-2}$ this time is much greater than the
time of one oscillation, which means that rescattering occurs only during the
second stage of preheating.

 In the ``particle-like'' picture the number of $\chi$ particles does not
change in each act of interaction (apart from its growth due to the
resonance), but each
collision releases one $\phi$-particle from the homogeneously oscillating
field $\phi(t)$ Since the scattering time for each $\chi$-particle $\tau
\propto n_\chi(t)$, one may conclude that the number
of free $\phi$-particles grows with time
as $ n_\phi \propto 5n_\chi^2/\Phi^2m_\phi \propto e^{4\mu m_\phi t}$. However,
the true dependence is more complicated because during each interaction
the $\chi$-particles will slow down. This affects their subsequent
interactions.

\subsection{\label{VALIDITY}On the validity of the ``particle-like''
interpretation
of rescattering}

In the previous subsection we considered rescattering of 
particles during time intervals when
 $\phi(t) > \phi_*$. At that stage $\chi$ particles are
nonrelativistic. In contrast, during the short time intervals $\Delta t_*
\simeq k_*^{-1}$,
 when $|\phi(t)|< \phi_*$, $\chi$-particles are
ultrarelativistic, and their effective mass $g\phi$
is very small comparing to their typical momenta
$\sim k_*/2 $.
If one uncritically repeats the calculation of the
rescattering for the case of ultrarelativistic
$\chi$ particles in the time interval $\Delta t_*$,
one obtains a much higher cross-section
and a much shorter rescattering time
$\tau \simeq { m_\phi^ 2 \over 3\pi^2 g^2 n_{\chi}}$
than that of the non-relativistic case of the previous subsection.

However, within the
 very short time interval $\Delta t_* \simeq q^{-1/4}m_\phi^ {-1}$,
one cannot use the standard methods of calculation  \cite{Peskin}
developed for the investigation of  processes which begin  at $t = -\infty$ and
end at $t = +\infty$. The uncertainty principle tells us that during the time
$\Delta
t_*$ one cannot specify the energy of particles with an accuracy better than
$k_*$. Therefore during the short interval $\Delta t_*$ one cannot tell the
difference between a $\phi$-particle with momentum $k = 0$, belonging to the
classical field
$\phi(t)$, and a free $\phi$-particle with
momentum $k< k_*$, i.e. one cannot tell whether scattering occurred or not.
This question can be answered only by observing the system for a longer time,
comparable to $m_\phi^ {-1}$, but during the main part of such intervals the
effective mass of each $\phi$-particle is large, and cross-section is much
smaller than the cross-section which one would obtain by naive application of
the S-matrix approach during a small interval $\Delta t_*$. In other words, we
cannot use the standard formalism of particle scattering to describe
  scattering around zeros of the inflaton field. 
 Another
 element missing in this formalism is that the
 field $\chi$ is not in an $n_{\chi}$-particle quantum state, but is in the
squeezed state. (We have discussed already one of the nontrivial consequences
of this fact, namely  the high-frequency modulation of $\langle
\chi^2\rangle$.) Thus
one may wonder whether one can trust the results of our calculations for the
more safe situation when $\phi > \phi_*$, and what we can say about
the contribution of the intervals with $\phi < \phi_*$
to the net rescattering effect?

Here we will outline   a possible way to answer this question.
Let us consider the self-consistent set of
 equations (\ref{mode1}) and (\ref{mode2})
 for the interacting fields in the classical
approximation.
Eq. (\ref{mode2}) describes the
evolution of the
$\delta \phi_k(t)$ fluctuation.
Let us concentrate on the first integral term in Eq. (\ref{mode2}),
assuming for the moment that the second term is subdominant until
$\delta \phi_k(t)$ increases sufficiently.
What we obtain is
 the equation for the forced oscillations of
$\delta \phi_k(t)$.
The homogeneous part of
this inhomogeneous linear differential equation
has a simple Green function $\propto \sin \Omega_{\bf k} (t-t')$, where
$\Omega_{\bf k}^2= {\bf k}^2+m_\phi^ 2$.
Then the solution of   Eq. (\ref{mode2}) with only
the first integral term is
\begin{eqnarray}
\delta\phi_k(t)&=&-{{g^2 } \over {(2\pi)^3 \Omega_{\bf k} }}
\int_0^t dt' \sin \Omega_{\bf k} (t-t') \phi(t') \nonumber \\
&\times&
 \int d^3 k' \chi_{\bf k -\bf k'}(t')\chi_{\bf k'}(t') + h.c. \ .
\label{grow}
\end{eqnarray}
Here, as before, $k$ is a physical momentum. This solution expresses the
function   $\delta \phi_k(t)$
via the known functions $\phi(t)$ describing the inflaton oscillations, see Eq. (\ref{870}), and the
functions $\chi_k(t)$, see Sec. \ref{ANALYTIC}.
Eq. (\ref{grow}) takes into account all the
regimes of $\phi(t)$, as well as the resonant amplification
of $\chi_{ k}$.
In particular, from this it follows that the amplitude $\varphi_k(t)$
 grows
with time as $e^{2\mu m_\phi t}$, because the amplitude $\chi_{  k}$ grows as
 $e^{\mu m_\phi t}$. Therefore the number of particles corresponding to
$\delta \phi$ fluctuations
is proportional
 to $ n_{\chi}^2(t) \propto e^{4\mu m_\phi  t}$ , i.e. grows much faster
than $n_{\chi}$.
Another specific prediction which follows from Eq. (\ref{grow})
is that the random field $\delta\phi(t, \bf x)$
at the early stages of its generation will have non-gaussian
statistics in contrast to the random gaussian field $\chi_{t, \bf x}$.

Let us further investigate the solution (\ref{grow}).
The inner integral $\int d^3k' \chi_{\bf k -\bf k'}(t')\chi_{\bf k'}(t')$
 is time-dependent.
It is convenient to
change the order  of integration of  $\int dt'$ and $\int d^3 k'$.
Then the r.h.s. of the solution (\ref{grow}) will
contain   terms like
\begin{eqnarray}\label{collision}
&& {{g^2 } \Phi e^{ i \Omega_{\bf k} t} \over {(2\pi)^3 \Omega_{\bf k}
}}
 \int d^3k' \int_0^t dt'
{ \beta_{\bf k'} \beta^*_{\bf k- \bf k'} \over
\sqrt{ \omega_{\bf k -\bf k'}(t') \omega_{\bf k'}(t')}} \nonumber\\
 &\times & e^{- i\Omega_{\bf k} t'+ im_\phi t'
 - i \int^{t'} dt''\omega_{\bf k -\bf k'}(t'')
+ i \int^{t'} dt''\omega_{\bf k'}(t'')} \ ,
\end{eqnarray}
where $\omega_{\bf k}^2(t)= {\bf k } ^2+g^2\phi^2(t) $.
 During each  half of the
oscillation $\beta_{\bf k} $
is constant, see Section \ref{WKB}. (Note that
 the coefficients $\beta_{\bf k}$ corresponding to the
classical waves will be dimensionless if one uses
discrete modes ${\bf k}$.)

It is easy to see that different choices of signs
 in Eq. (\ref{collision}) correspond
to different channels of scattering between
$\delta \phi$, $\chi$ and $\phi$ waves.
The terms (\ref{collision}) correspond to
the generation of $\delta \phi$ fluctuations due to
the scattering of $\chi$ and $\phi$ waves.
Obviously, one can leave in the inner integral
$\int_0^t$ only the segment $\int_{t_j}^t$
 (where $t < t_j+{\pi \over m_\phi }$)
 from the
most recent cycle of the inflaton oscillation, when
 $\beta_{\bf k'}$ is the largest. During this interval $\beta_{\bf k'}$ is
constant.
Therefore to further investigate the inner integral
 $\int_{t_j}^t$, we
shall consider the variation of the phase of the exponent
in Eq. (\ref{collision})
$\theta \simeq - \Omega_{\bf k} t'+ m_\phi t '
 - \int^{t'} dt''\omega_{\bf k -\bf k'}(t'')
+ \int^{t'} dt''\omega_{\bf k'}(t'')$ within this time interval
$t- t_j < {\pi \over m_\phi }$,
which describes the interference of the
four interacting waves $\phi(t)$, $\delta \phi_{\bf k}$,
$\chi_{\bf k'}$ and $\chi_{\bf k -\bf k'}$.
 Earlier we estimated the
integral  $\int^{t'} dt''\omega_{\bf k'}(t'') \approx {{2 g \Phi}
\over m_\phi}\cos m_\phi t ' + O(\kappa^2)$, see (\ref{random1}).
The crucial observation is that
for the process
 $\chi_{\bf k'} \phi_0 \to \delta \phi_{\bf k} \chi_{\bf k' - \bf k} $
 the large terms
 $ {{2 g \Phi }\over m_\phi}\cos m_\phi t $ in the expression for $\theta$ are
cancelled
and
the phase $\theta$ does not
 oscillate within each half of the period, $t- t_j < {\pi \over m_\phi }$.
As a result, the integral $\int dt$ cannot be
reduced to the usual delta-function
 $\delta \left(- \Omega_{\bf k} +m_\phi -\omega_{\bf k -\bf k'}
+ \omega_{\bf k'} \right)$, as one would expect in
the ``particle-like'' picture. Instead, in the wave picture
we will have nonvanishing contributions from the
bunches of modes ${\bf k}$ and ${\bf k'}$ for which
 the phase $\theta \simeq \pi$, which
corresponds to the interaction of   packets of $\chi$ and
$\delta \phi$ waves.
In contrast  to the process of rescattering, the annihilation process
 $\chi_{\bf k} \chi_{\bf k'} \to \delta \phi_{\bf k''}
\delta \phi_{\bf k + \bf k'- \bf k''} $ and the inverse process
will be suppressed because the corresponding time integrals
have very rapidly oscillating
 exponents $e^{\pm i {{4 g \Phi }\over m_\phi}\cos m_\phi t }$.

The analysis of Eq. (\ref{collision}) shows
the hard component $\delta \phi$
 with $k \simeq k_*$ can be generated only
during the very short time intervals $\Delta t_* \simeq k_*^{-1}$
around zeros of the inflaton field.
The soft component with momenta $k \ll k_*$ is generated all the time. Soft
particles produced at $|\phi| >\phi_*$ have very small momenta
 in the range of $0 < k < m$.
It makes sense to talk about such particles as free $\phi$-particles removed
from the coherently oscillating field $\phi(t)$ only at  time intervals
$\tau \gg m_\phi^ {-1}$.
An estimate of the soft component from (\ref{collision})
at the beginning of the process is
$\langle \phi^2 \rangle_{\rm soft} \simeq g^2 n_{\chi}^2/m_\phi^ 4$,
whereas for the hard component one has
$\langle \phi^2 \rangle_{\rm hard} \simeq \langle \phi^2 \rangle_{\rm soft}
/\sqrt{q}$.
Since $\delta \phi$ grows very fast, one has to be
careful with the range of validity of the solution (\ref{grow}).
Indeed, Eq. (\ref{grow}) is only the first term in the iterative solution of
Eq. (\ref{mode2}). As soon as $\delta \phi$ grows,
we have to consider the   iterative
solutions of   both Eqs. (\ref{mode1}) and (\ref{mode2}). We have to
take into account the corrections to $X_k$ due to the
 $X$ and $\varphi$ coupling on the r.h.s. of Eq. (\ref{mode1})
as well as the second bilinear term on r.h.s. of Eq. (\ref{mode2}).
Due to the exponential growth in the number of particles, these corrections
to the  simple solution (\ref{grow}) very quickly become important, which makes
 further investigation rather complicated.

   One should note that in addition to rescattering, there may exist other mechanisms of $\phi$-particle production. For example, let us consider fluctuations $\delta\phi$ with    effective
mass squared  $g^2\langle\chi^2\rangle$. As we already emphasized, this term is
time-dependent. First of all, it experiences quasiperiodic high-frequency
modulation, which, as we already noted in Sec. \ref{HARTREE}, may serve as an
additional source of $\phi$-particles. In addition, the term
$g^2\langle\chi^2\rangle$ oscillates with   period ${\pi\over m_\phi}$.
During each oscillation it changes from its minimal value $g\Phi n_\chi$ (for
$|\phi(t) |= \Phi$) to a much greater value $\sim  3g\Phi  n_\chi q^{1/4}$ (for
$|\phi(t) |= \phi_*$). This leads to a significant periodic change in the
properties of $\phi$-particles, which is especially pronounced when $|\phi(t)|
\lesssim \phi_*$. A preliminary investigation of this issue indicates the
possibility of a parametric resonance with $\phi$-particle production.

Our main purpose here was not to give the final analysis of this issue but
rather to outline  different approaches to the problem of rescattering and
$\phi$-particle production, which should provide a proper framework for  
future investigation.

\subsection{\label{ENDREHEAT} Rescattering and the end of preheating}

Can rescattering kill the resonance?
In Sec. \ref{GENERATION} we found that rescattering can be rather efficient at
the second stage of preheating.
What can we say about the influence of rescattering on the development of
  parametric resonance?

The simplest idea would be to estimate the effective mass of the
$\chi$-particles induced by the fluctuations $\langle\phi^2\rangle$:~ $\Delta
m_\chi^2 \sim g^2 \langle\phi^2\rangle$. However, this would not be quite
correct.
Indeed, the whole process of $\chi$-particle production occurs in the interval
$|\phi| \lesssim \phi_*$ during the time $t_* \sim (gm_\phi \Phi)^{-1/2} =
k_*^{-1}$, see Eq.
(\ref{adiab7}).   If oscillations of the modes $\delta \phi$ occur during a
longer time, then from the point of view of the creation of $\chi$-particles
they
cannot be distinguished from the oscillations of the field $\phi(t)$, and
therefore they do not   harm  the development of stochastic resonance.
We called such modes ``soft,'' and the modes with $k \gtrsim k_*/4$ ``hard.''

Fluctuations of the scalar field $\phi$ can be
harmful to the development of the resonance if they can considerably
alter the motion of the field $\phi$ in the interval $|\phi| \lesssim \phi_*$.
The only fluctuations which can change the direction of their motion during the
short time $t_* \sim k_*^{-1}$ are the modes with $k \gtrsim 2\pi k_* \gg k_*$.
 This effect does not seem to be very important. At the time when the
homogeneous mode $\phi(t)$ enters the interval
$|\phi| > \phi_*$, it has a kinetic energy $ \dot\phi ^2/2 \sim
m_\phi^ 2\Phi^2/2$.
In order to alter the motion of the field $\phi$ the ``hard'' fluctuations
$\delta\phi$ should (occasionally) have comparable (and opposite) speed, and
therefore they should have a kinetic energy   comparable to $m_\phi^
2\Phi^2/2$.
Thus, the resonance disappears only after the kinetic energy of
$\phi$-particles with momenta $k \gg k_*$ becomes comparable to the total
energy of the oscillating
field $\phi(t)$. This could happen only at the very end of preheating.

However, there is another mechanism which may harm the resonance. Each mode
$\chi_k$ ``probes'' space on a length scale $\Delta l \sim 2\pi k^{-1}$. If the
field $\delta\phi$ is homogeneous on this scale, it acts as a homogeneous
background for the mode $\chi_k$. On the other hand, if $\delta\phi$ is
inhomogeneous on this scale, then the field $\chi_k$ has an integrated
interaction with all inhomogeneities of the field $\delta\phi$ on the scale
$\Delta l \sim 2\pi k^{-1}$, i.e. it interacts with the contribution to
$\langle\phi^2\rangle$ from the modes with momenta greater than $k$. This
corresponds to the appearance of an ``effective mass squared'' $\Delta
m_\chi^2 \sim g^2 \langle\phi^2\rangle$, but only the modes with momenta
greater than $k$ should be taken into account in this calculation. Thus, from
the point of view of the development of parametric resonance, one can introduce
a new notion of an effective mass squared $\Delta m_\chi^2(k) \sim g^2
\langle\phi^2\rangle_{k}$, where the index $k$ means that we take into account
only the modes with momenta greater than $k$.

If the effective mass squared $\Delta m_\chi^2(k) $ becomes greater than
$k^2$, the equation of motion for such modes $\chi_k$ changes considerably.
This
effect kills the resonance for the mode $\chi_k$ if $\Delta m^ 2_\chi (k) $
becomes greater than the width of the resonance.
The resonance for the leading modes with $k \sim k_*/4$ ends when $\Delta
m^ 2_\chi (k_*) \sim g^2{\langle\phi^2\rangle_{\rm hard}}$ becomes greater than
$k_*/4$.

The difference between the total value of $\langle\phi^2\rangle$ and
$\langle\phi^2\rangle_{\rm hard}\equiv \langle\phi^2\rangle_{k_*/ 4}$ can be
quite significant. The number of $\phi$-particles produced in
each scattering
 is equal to the number of $\chi$-particles, each $\phi$-particle taking away
some portion of the momentum $k $ of the corresponding
$\chi$-particle. If this portion is small,
 $\delta\phi$ fluctuations corresponding to these particles have momenta
much smaller than $k_*/4$. Therefore, they do not give any contribution to the
effective mass $ \Delta m_\chi^2(k \sim k_*/4)$, so they do not hurt the
resonance at such momenta.
 If in the first collision a $\chi$-particle  with   momentum
 $k \sim k_*/4 $
gives a significant portion
 of its energy to a  $\phi$-particle, then it loses
its energy, and in subsequent collisions it
 will produce only harmless $\delta\phi$ fluctuations with $k \ll k_*/4$.

 Thus, one may argue that if rescattering is efficient,
 the number of ``hard" $\phi$-particles  produced by
$\chi$-particles should be similar to the initial number
of
$\chi$-particles with momenta $\sim k_*/4$, i.e.
$n_\phi^{\rm hard} \lesssim n_\chi$, whereas the total number of
$\phi$-particles produced by rescattering
 may be much greater. At the second stage of reheating, when
$g^2\langle\chi^2\rangle \gg m_\phi^ 2$, one can use an estimate
\begin{equation}\label{phisqr}
\langle\delta\phi^2\rangle\simeq {1\over 2\pi^2}\int {k^2 dk\ n_k^\phi\over
\sqrt{k^2 + g^2\langle\chi^2\rangle}} \ .
\end{equation}
If $\langle\delta\phi^2\rangle$ is dominated by  soft fluctuations with $k^2
\ll g^2 \langle\chi^2\rangle$,
then at the second stage of the resonance one should expect a strong
anticorrelation
 between oscillations of $ \langle\chi^2\rangle$ and
$ \langle\delta\phi^2\rangle$.
This prediction is in agreement with the numerical
results of \cite{Khleb}.

Now let us concentrate on the ``hard''
 fluctuations with typical momenta $\sim k_*/4$. They can
 hamper the resonance if they make the
field $\chi$ massive, with an
induced effective mass squared $\Delta m^2
\sim g^2\langle\delta\phi^2\rangle_{\rm hard}$
 comparable to the square of the typical
momentum of $\chi$-particles $k \sim k_*/4$:
\begin{equation}\label{f}
g^2\langle \delta\phi ^2 \rangle_{\rm hard} \gtrsim {g m_\phi\Phi/16 } \ .
\end{equation}
Suppose that a fraction $\gamma $ of all energy
$ {m_\phi^2\Phi^2\over 2}$ is transferred to the kinetic energy $ {k^2\over
2}\langle \delta\phi ^2
\rangle_{\rm hard} $ of ``hard'' fluctuations,
\begin{equation}\label{ff}
{gm\Phi\over 32} \langle \delta\phi ^2 \rangle_{\rm hard} \simeq \gamma {
m_\phi^2
\Phi^2\over 2} \ .
\end{equation}
This gives
\begin{equation}\label{fff}
g^2 \langle \delta\phi^2 \rangle_{\rm hard} \simeq 16\gamma g m_\phi \Phi
 \ .
\end{equation}

Comparison of Eqs. (\ref{fff}) and (\ref{f}) shows
 that $\gamma \gtrsim 1/256$, i.e. the
 resonance may slow down and eventually terminate only
 when the oscillating field $\phi$
transfers at least $\sim 1/256$ of its energy to the
 ``hard'' fluctuations $\phi$. The
total energy of all $\phi$-particles
will be somewhat greater than that. These particles get
their kinetic energy from the kinetic energy of $\chi$-particles $\sim
{g m_\phi\Phi\over 8}
\langle \chi ^2 \rangle$, so one may expect that the resonance terminates only
after
${g m_\phi\Phi\over 32} \langle \chi ^2 \rangle$ becomes greater than $
{1\over 256}
{ m_\phi^2 \Phi^2\over
2}$.
This can only occur  close to the end of preheating.
Let $ \sqrt{\langle \chi ^2 \rangle}_r$ and
$ \Phi_r$ be the values of $\chi$-fluctuations
and amplitude of the background field at the
moment $t_r$ when the parametric resonance is terminated by
rescattering.
Taking into account that at the second stage of preheating
 $m^2_{\phi} \simeq g^2
\langle \chi ^2 \rangle$ one finds that at the end of preheating
\begin{equation}\label{ffu}
 \sqrt{\langle \chi ^2 \rangle_r} \gtrsim { \Phi_r }/16 \ ,
\end{equation}
Note also that $\sqrt{\langle \chi ^2 \rangle_r} \lesssim { \Phi_r }$,
because this would correspond to the result which we obtained in Sec.
\ref{SECOND} neglecting rescattering. In our subsequent calculations we will
use the estimate $ \sqrt{\langle \chi ^2 \rangle_r} \sim 10^{-1} { \Phi_r
}$. This value is somewhat smaller than $\sqrt{\langle \chi ^2
\rangle}_2 \simeq { \Phi_2 }$ which we obtained in Sec. \ref{SECOND}
neglecting rescattering. However, the difference between these two values is
in fact not very large because $\Phi_r > \Phi_2$.

We are going to find $\sqrt{\langle \chi ^2 \rangle}_r$ and $\Phi_r$, which
should
replace   our previous estimates for $\sqrt{\langle \chi ^2 \rangle}_2$ and
$\Phi_2$ at the end of the second stage neglecting rescattering.
Again we will use   energy conservation.
 At the end of the first stage the energy
density was equal to the potential energy density $ m^2\Phi^2_1/2$ of the
inflaton field plus the energy of its interaction $g \Phi_1 n_\chi \sim
 m^2\Phi^2_1 $, where $m$ is the bare inflaton mass.
 At the end of the resonance (at the second stage),
 with an
account taken of rescattering, the kinetic energy of the $\chi$-particles
remains small, so the whole energy $\sim 3 m_\phi^ 2\Phi^2_1/2$ transforms to
the
energy density of interaction between $\chi$-particles and the field $\phi$,
$\rho_\chi = g\Phi_r n_\chi \sim g^2\langle \chi ^2 \rangle_r \Phi_r ^2 \sim
10^{-2} g^2\Phi_r^4$. Note that $\rho_\chi$ includes the energy of the
oscillating scalar field $\phi(t)$. Energy conservation implies that
$\Phi_r \sim 3.5 \sqrt {m\Phi_1/g} \sim 2.5 \Phi_1 q_1^{-1/4}$. However,
$\Phi_r$ obviously cannot be greater than $\Phi_1$. This means that
rescattering can terminate the resonance  either if $\sqrt{\langle \chi
^2 \rangle_r}\gg 10^{-1} { \Phi_r } $, in which case we essentially recover
the previous results of Sec. \ref{SECOND}, or if $q_1 \gtrsim 10^2$. In the
last case one has $\sqrt{\langle \chi ^2 \rangle_r} \sim 0.35 \sqrt {m
\Phi_1\over g}$, which yields
\begin{equation}\label{number90}
\sqrt{\langle \chi ^2 \rangle_r} \sim \left({\mu m  M_p\over 6 g}\ln^{-1}
{10^{12} m \over
g^5 M_p
}\right)^{1/2}.
\end{equation}
This estimate should replace Eq. (\ref{number20}) derived without account
taken of
rescattering.
In particular, for $g= 10^{-2}$, which corresponds to $q_0 = 10^6$, and $q_1
\sim 550$, we
get $\sqrt{\langle \chi ^2 \rangle_r} \approx 2.5\times 10^{15}$ GeV. To
compare this result to the result of \cite{Khleb} one should note that the
definition of $q_0$ in \cite{Khleb} differs slightly  from ours, so it is
better
to compare our results for a given $g$ rather than for a given $q_0$. In
particular, one should compare their results for $q_0 = 10^6$ with our results
for $g= 10^{-2}$: $\sqrt{\langle \chi ^2 \rangle_r} \approx 3\times 10^{15}$
GeV. This result agrees, to within a factor of $2$, with the results of the
lattice simulation of \cite{Khleb}.

One should not overemphasize the significance of this agreement. The theory of
the
last stages of preheating is extremely complicated, and there are many points
in which our rough estimates could be improved. One should remember also that
we are discussing stochastic
resonance, which is extremely sensitive to even minor changes of parameters,
see the table in Sec. \ref{STOCHASTIC} \cite{FOOT}. From this perspective it
is even somewhat surprising that one can describe many features of this
process by analytical methods with   rather good accuracy.

Strictly speaking, the condition which we derived does not imply that the
resonance is completely terminated. The leading modes, which have been
amplified from the very beginning, stop growing when the effective mass of the
field $\chi$ becomes greater than $k \sim k_*/4$. However, the sub-leading
modes still continue their growth until the effective mass becomes greater than
$k_*/2$. This process is very inefficient, but $\langle \chi^2\rangle$
continues slowly growing for a while. Moreover, $\langle \chi ^2 \rangle$ may
grow a little even when the resonance is completely terminated and new
particles are no longer produced. Indeed, due to the decay of the field
$\phi(t)$, the effective mass of the $\chi$ particles becomes smaller, and
therefore $\langle \chi ^2 \rangle$ may become greater even if $n_\chi$
remains constant. These effects are not very significant, but they make it
difficult to clearly recognize the end of parametric resonance by looking at
the behavior of $\langle \chi ^2 \rangle$. That is why throughout this paper,
 alongside   the dispersion of the fluctuations which is
studied in most papers on preheating,
 we use the number density of particles to investigate the resonance.

An estimate of the density of $\chi$-particles at the end of the resonance can
be obtained by
multiplying $\langle \chi ^2 \rangle_r$ by $g\Phi_r \sim 16g \sqrt{\langle
\chi ^2 \rangle_r}$. It is given by
\begin{equation}\label{BBBB}
n_\chi \sim 0.4 g^{-1/2} \left({\mu m M_p}\ln^{-1} {10^{12} m \over
g^5 M_p}\right)^{3/2}.
\end{equation}

It is useful to compare this number with the number of $\phi$-particles
$n_\phi$ in the oscillating field $\phi(t)$ which survive the rescattering. To
distinguish the particles $\phi$ in the oscillating field and the free
$\phi$-particles created by rescattering, we will denote the number of
particles in the classical field as $n_\phi^c$. At the end of the resonance it
is given by $m_{\phi}\Phi_r^2/2$, where $m_{\phi}$ is the effective mass
$ g\sqrt {\langle \chi ^2 \rangle_r}\simeq 0.1 g\Phi_r$. Meanwhile
$n_\chi \sim g\Phi_r \langle \chi ^2 \rangle_r \sim 10^{-2} g\Phi_r^3$.
Therefore,
\begin{equation}\label{number91}
{n_\chi } \sim 10^{-1} n_\phi^c \ .
\end{equation}

Eq. (\ref{number91}) says that at the end of the resonance
$\chi$-particles need to rescatter only 10 times
 to destroy the coherent
oscillations of the classical field, i.e. to decompose it into separate
$\phi$-particles.
Therefore one may expect that at the end of the resonance or very
soon after it $\chi$-particles may destroy the classical field $\phi(t)$
completely, in agreement with \cite{Khleb}.
 This means that the final stage of decay of the homogeneously oscillating
classical scalar field in our model is determined not
by   resonance but by rescattering.

The decay of the   classical scalar field $\phi(t)$ is not the end of the
story, but
rather the beginning of a new stage of
reheating. As we pointed out in \cite{KLS}, it does not make much sense to
calculate the reheating temperature at this stage of the process. Indeed,
from
the point of view of the energy stored in the $\phi$-particles, it is not very
important whether it is in the form of
$ \phi$ fluctuations or in the form of a coherently
oscillating field $\phi$. According to our estimates, the kinetic energy of
$\chi$-particles may constitute only about $10^{-2}$ of the total energy at the
end of   parametric resonance. This estimate may be too pessimistic, but even
if the true energy is much higher, the main fraction of energy after the end of
the resonance remains stored in the energy of $\phi$-particles, and the energy
of their interaction with $\chi$-particles.
 The total energy of $\chi$-fluctuations
 at large $t$
decreases as $a^{-4}$, whereas the energy of $\phi$-fluctuations
 as well as the
energy of the oscillating field $\phi(t)$ at large $t$ decreases as $a^{-3}$.
Even if the total energy of the oscillating field $\phi(t)$ and of
$\phi$-particles were very small after preheating,
eventually it would again dominate the energy density of the universe. Eq.
(\ref{number91}) gives
us additional information: the number of $\phi$-particles after preheating is
at least ten times greater than the number of $\chi$-particles. If these
particles do not decay, they will always dominate the energy density of the
universe, which is unacceptable.
Therefore when preheating ends one should apply the elementary (perturbative)
theory of reheating \cite{DL,AFW} to describe the decay of the remnants of the
classical oscillating field $\phi(t)$ as well as the decay of the large amount
of $\phi$-particles created by rescattering. We will return to the theory of
this process in a subsequent publication \cite{PERT}.

\section{\label{SUPERMASSIVE}Production of Superheavy Particles during
Preheating}
One of the most interesting effects which may become possible during
preheating is the copious production of particles which have a mass greater
than
the inflaton mass $m$. This question is especially interesting in the context
of the theory of GUT baryogenesis, which may occur in a rather unusual way if
superheavy particles with mass $M$ a few times heavier than $m$ can be produced
\cite{Kolb}. Such processes are impossible in perturbation theory and in the
theory of narrow parametric resonance. However, we are going to show that
superheavy $\chi$-particles with mass $M \gg m$ can be produced in the regime
of a broad parametric resonance.

In order to study this regime let us return to Sec. \ref{BROAD}, where we
made a simple derivation of the width of the resonance band, see Eq.
(\ref{adiabA}). The only modification which should be made to this equation in
the
case where the field $\chi$ has a $\phi$-independent mass $m_\chi(0) \equiv M$
is
to add it to $k^2$ on the l.h.s. of the equation:
\begin{equation}\label{adiabAAA}
{k^2 + M^2} \lesssim (g^2\phi m_\phi
\Phi)^{2/3} - g^2\phi^2 \ .
\end{equation}
As before, the maximal range of momenta for
which particle production occurs corresponds to $\phi(t) = \phi_*$, where
$\phi_* \approx {\textstyle {1 \over 2}} \sqrt {m_\phi\Phi\over g}$. The
maximal value of momentum for particles produced at that epoch can be
estimated by $k^2_{\rm max} + M^2 = {g m_\phi \Phi \over 2}$.  The resonance becomes efficient for
\begin{equation}\label{MASSIVE}
 g m_\phi \Phi \gtrsim 4 M^2 \ .
\end{equation}
Thus, the inflaton oscillations may lead to a copious production of
superheavy particles with $M \gg m$ if the amplitude of the field $\Phi$ is
large enough, $g\Phi \gtrsim 4M^2/m$.

However, in an expanding universe $\Phi$ and $m_\phi$ are time-dependent. One
should not only have a very large field at the very beginning of the process;
one should continue to have $gm\Phi \gtrsim 4M^2 $ until the end
of preheating.

During the second stage of preheating both $m_\phi$ and $\Phi$ change very
rapidly, but their product remains almost constant because the energy density
of the field $\phi$, which is proportional to $m^2_\phi \Phi^2/2$, practically
does not change until the very end of preheating. Therefore it is sufficient to
check that $gm\Phi \gtrsim 4M^2 $ at the end of the first stage of preheating.
One can represent this criterion in a simple form:
\begin{equation}\label{QQQ}
M \lesssim {m\over \sqrt 2}\, q_1^{1/4} \approx  {m } \left({ g\mu M_p\over 3m}\ln^{-1} {10^{12} m
\over
g^5 M_p }\right)^{1/2} .
\end{equation}
For example, one may take $M = 2m$ and $g \approx 0.007$, which corresponds to
$q_0 = 10^6$ in the normalization of Ref. \cite{Khleb}. In this our condition (\ref{QQQ}) is satisfied, and an investigation with an account taken of rescattering shows a relatively insignificant suppression of $\langle \chi^2 \rangle$, approximately  by a factor of $  3$. Our investigation suggests that for $ g \gg 10^{-2}$ this process should not be suppressed at all. Eq. (\ref{QQQ}) shows that for sufficiently large $g$ one can produce superheavy particles with $M \gg m$. For example,  production of $\chi$-particles with $M = 10 m$ is possible for   $g \gtrsim 0.065$. 

In fact, suppression of superheavy particle production may be even less significant. Indeed,   the resonance becomes strongly suppressed if it occurs only for $k^2 \ll {k_*^2\over 4} \sim {g m_\phi \Phi \over 4}$. As a result, the condition for the efficient preheating (\ref{MASSIVE}) can be slightly relaxed:
$ g m_\phi \Phi \gtrsim 2 M^2$. This small modification implies that heavy particle production is not strongly suppressed for $M \lesssim {m }\, q_1^{1/4} \approx  {m } \left({2 g\mu M_p\over 3m}\ln^{-1} {10^{12} m
\over
g^5 M_p }\right)^{1/2}$. For $M = 10 m$ this leads to a rather mild condition $g \gtrsim 0.036$.

 We conclude that at least in our simple model,
the production of superheavy particles is possible. However, with an increase of $g$ the total
number of produced particles becomes smaller, see Eq. (\ref{BBBB}). It would be most interesting
to investigate this issue in realistic models of elementary particles and to
apply the results to the theory of baryogenesis.

\section{\label{DISCUSSION} Discussion}

In this paper we discussed the theory of preheating for the simple model of
a massive inflaton field $\phi$ interacting with another scalar field $\chi$.
As we have seen, the theory of preheating is very complicated even in such
a simple model. Our main purpose was not to answer all questions related to the
theory of preheating, but to develop  an adequate framework in which these
questions should be investigated.

In the beginning   particle production occurs in the
regime of a
 broad parametric resonance, which gradually becomes narrow and then
terminates. If the resonance is narrow from the very beginning, or even if it
is
 not broad enough, it remains inefficient. We have found that   broad
 resonance in an
expanding universe is actually a stochastic process.
 The theory of this process,
which can be called stochastic resonance, or stochastic amplification, is
dramatically different from the theory of parametric resonance in Minkowski
space. Therefore one cannot simply apply the standard methods of investigation of
  parametric resonance
 in Minkowski space; it was necessary to develop
new analytical methods for the  investigation of stochastic resonance in an
expanding
universe. We have found the typical width of the resonance $\sim k_*/2$ and the
typical rate of the exponential growth of the number of produced particles in
this regime. An important feature of our formalism of investigation of the broad resonance regime is its robustness with
respect to modification of the form of the effective potential.   Our methods
should apply not only to   theories with the potential $m^2\phi^2/2$, but to
any potential $V(\phi)$   when the resonance is broad.

 One should note,   that the main reason why   broad resonance has a stochastic nature is the expansion of the universe. In the conformally invariant theories such as the theory ${\lambda\over 4}\phi^4 + {g^2\over 2}\phi^2\chi^2$ with $g^2\gg \lambda$ the resonance is broad but not stochastic because expansion of the universe does not interfere with its development \cite{GKLS}. In realistic theories where the inflaton field $\phi$ has mass $m$ the conformal invariance is broken and one could expect that the broad resonance  becomes stochastic as soon as the amplitude of the oscillations of the field $\phi$ becomes smaller than $m/\sqrt\lambda$. Indeed, for $\Phi \lesssim m/\sqrt \lambda$ the resonance is described by the model of a massive inflaton field considered in this paper. A more detailed investigation of this question shows that   in models with $g^2 \gg \lambda$   the resonance becomes stochastic even earlier, at $\Phi \lesssim {g\over\sqrt\lambda}\, { \pi^2 m^2  \over 3\lambda  M_p}$ \cite{GKLS}.

In our investigation of preheating we took into account the interaction of the
oscillating inflaton field $\phi$ with the particles produced during
preheating.
  We have found, in  particular, that the correction  to the effective mass
squared of the oscillating field $\phi$ is proportional to ${gn_\chi \over
|\phi|}$, and
the equation of motion of the field $\phi$ acquires a term $\sim
gn_\chi{\phi\over |\phi|}$.  This term  experiences  quasiperiodic
oscillations with a very
high  frequency $\sim 2g\Phi$, which do not much affect  the motion of the
field
$\phi(t)$ but may serve as an additional source of $\phi$-particles.

 We have found that if the coupling constant $g^2$ in the interaction term ${g^2\over 2}\phi^2\chi^2$ is small ($g \lesssim 3\times 10^{-4}$), the resonance terminates at the stage when the backreaction of produced particles is unimportant. For larger values of $g^2$ the resonance terminates due to a combined effect of the growth of the effective mass of the inflaton field and rescattering, which in its turn increases the effective mass of $\chi$-particles, making them heavy and hard to produce. We made an estimate of   the number of $\chi$-particles   produced
during preheating and their quantum fluctuations $\langle\chi^2\rangle$ with
all backreaction effects taken into account.

Traditionally, the only purpose of the theory of reheating was to obtain the
value of the reheating temperature. From this point of view the theory of
preheating for the simple  model which we studied in this paper does not
change the situation. For $g \ll 3\times 10^{-4}$ the total energy density of produced particles is exponentially small. Similarly, it remains extremely small even for large $g$ if $\chi$ particles have mass $M$ much greater than about $10 m$. In the case when $M$ is small and $g \gtrsim 3\times 10^{-4}$, the $\chi$-particle production is very efficient. However,  we have found that even in this case after preheating one has many more $\phi$-particles than
$\chi$-particles. If $\chi$-particles are massless, or if they can easily
decay, their contribution to the energy density of the universe rapidly
decreases. Therefore, after preheating the main contribution to the energy
density of the universe is again given by the $\phi$-particles. The only
difference is that prior to preheating these particles constitute the oscillating classical inflaton field $\phi(t)$, whereas after preheating they acquire various spatial momenta and 
become decoherent. Thus, as we already pointed out in \cite{KLS}, it does not
make much sense to calculate the reheating temperature immediately after preheating. One
should study the subsequent decay of the $\phi$-particles . The theory of this
decay is described by the elementary theory of reheating \cite{DL,AFW,PERT}.
So why should one study extremely complicated nonperturbative effects which may happen at the stage
of parametric resonance, if in the end  they will not greatly change   our
old estimates of the reheating temperature?

We believe that the investigation of nonperturbative effects in the very early
universe is worth the trouble. In fact, the complex nature of this process
makes it especially
interesting. Indeed, a few years ago the standard picture of the evolution of the universe
included a remarkable stage of explosive expansion (inflation) in the
vacuum-like state, which is responsible for its large-scale structure,
and a rather dull stage of decay of the inflaton field, which is responsible
for the matter content of the universe. The processes which could happen during
the later stage were typically ignored.

Now we see that the stage of reheating deserves a more detailed
investigation.
Explosive processes far away from thermal equilibrium could  impact
  the further evolution of the universe. As we know, the appearance of baryon
asymmetry requires the absence of thermal equilibrium, so it is only natural to
investigate the possibility of baryogenesis at the stage of reheating, see
e.g. \cite{DL,Fujisaki,Kolb}.

Particles produced by the resonance have   energies which are
determined
by the properties of the resonance bands. Typically this energy is much
smaller
than the temperature which would appear if the particles were instantaneously
thermalized. Meanwhile, the total number of particles produced by parametric
resonance is much greater than the number of particles in thermal equilibrium
with the same energy density.
 Fluctuations associated with these particles can be anomalously large. For
example, we have found that for certain values of coupling constants in our
model $ \sqrt{\langle \chi ^2 \rangle}$ may become of the order of $10^{16}$
GeV, and $\sqrt{\langle \phi ^2 \rangle} $ may become even greater than $
\sqrt{\langle \chi ^2 \rangle}$.
 In   models describing several interacting scalar fields such
anomalously large fluctuations
may
lead to specific nonthermal phase transitions in the early universe on the
scale of $10^{16}$ GeV
\cite{KLSSR,tkachev}. As we pointed out in \cite{KLSSR}, the investigation of
such
phase transitions in the theory of a single self-interacting field $\phi$ is
rather involved because one needs to separate the effects related to the
oscillations from the effects related to the fluctuations of the same field.
Therefore an optimal way to study nonthermal phase transitions is to
investigate the models where the fluctuations produced during preheating
restore symmetry for the field which does not oscillate during the oscillations
of the inflaton field, see Sect. \ref{FIRST}.
We
will return to the discussion of this effect in a separate publication
\cite{PhaseTr}.

Unlike   fluctuations in thermal equilibrium, the nonthermal fluctuations
produced by a parametric resonance often exhibit a nongaussian nature. In
particular, ``fluctuations of fluctuations'' can be very large. This means
that
in some regions of the universe one can find fluctuations at a level much
greater than its average value. This effect may play an important
role in the theory of topological defect production. Indeed, even if the
average level of fluctuations is smaller than the critical level which
leads to monopole production, they may be produced in the rare islands where
the level of the fluctuations is anomalously high. Note that  in
order to avoid cosmological problems and burning of  neutron stars by the
monopole catalysis of  baryon decay, the density
of the
primordial monopoles should be suppressed by 20 to 30 orders of magnitude. It
was easy to achieve such
suppression
 for the usual thermal fluctuations which appear after reheating, but for the
nonthermal fluctuations produced by resonance the situation may be
quite
different.

There is an additional reason which makes the investigation of preheating so
interesting.
The theory of particle production in the early universe was one of the most
challenging problems of theoretical cosmology in the early 70's. However,
powerful methods of investigation developed at that time produced rather
modest results: particle creation could be efficient only near the cosmological
singularity, at   densities comparable with $M_p^4$. This process could not
considerably increase the total number of particles in the universe.

Now we see that in the context of inflationary cosmology {\it all}\, particles
populating our part of the universe have been created due to quantum effects
soon after the end of inflation. The investigation of these effects sometimes
requires the development of new theoretical methods involving quantum field
theory, cosmology, the theory of parametric resonance, the theory of stochastic
processes, and nonequilibrium quantum statistics.

In a situation where nonperturbative effects play an important role, and the
number of produced particles grows exponentially, one could expect that the
only reliable tool for the
investigation of preheating would be numerical simulations. Fortunately,
one can go very far by developing
analytical methods. For sufficiently small values of the coupling constant
($g \lesssim 3\times 10^{-4}$) these methods allow us to make a very detailed
investigation of preheating. For higher values of the coupling constant one can
describe preheating analytically during most of the process. At
the
last stage of preheating the situation becomes too complicated, and numerical
methods become most adequate. Even in these cases analytical methods
 allow us to obtain estimates of the same order of magnitude as the results of
numerical calculations, and sometimes this agreement is even much better.
 Taking into account all of the uncertainties involved in the
analytical investigation of stochastic resonance as well as in the computer
simulations, this agreement looks rather encouraging. It remains a
challenge to develop a complete analytical theory of preheating, and to apply
it to realistic inflationary models with many interacting fields.

\bigskip
\section{Acknowledgments}
The authors are grateful to Igor Tkachev and Sergei Khlebnikov for very
important discussions, to Patrick Greene for assistance
and  useful comments,  and to Juan
Garc\'{\i}a--Bellido for useful comments.
 This work was supported  by NSF grant AST95-29-225.
 The work by A.L. was also supported   by NSF
grant PHY-9219345.
A.S. was supported   by the Russian
Foundation
for Basic Research, grant 96-02-17591.
A.L and A.S. thank the Institute for Astronomy, University of
Hawaii for hospitality.


\begin{thebibliography}{999}
\bibitem{FOOT} In fact, the table in Sec. \ref{STOCHASTIC} might suggest that
the
change of parameters by 10\% can lead to exponentially large changes in the
final results. This is indeed the case for the processes if $g < 3 \times
10^{-4}$, where backreaction is not important. Meanwhile, for $g \gg 3 \times
10^{-4}$ backreaction makes the process terminate at a time which is somewhat
less sensitive to the change of parameters. That is why it becomes possible to
obtain analytical estimates despite the stochastic nature of the process.
\end{thebibliography}
\end{document}